\documentclass{LMCS}
\usepackage{latexsym}

\def\eqalign#1{\null\,\vcenter{\openup\jot\mathsurround=0 pt
  \ialign{\strut\hfil$\displaystyle{##}$&$\displaystyle{{}##}$\hfil
      \crcr#1\crcr}}\,}

\usepackage{amssymb,amsmath,amsfonts,bm}
\usepackage{algorithmic}
\usepackage{subfigure} 
\usepackage{epic} 
\usepackage{eepic} 
\usepackage{enumerate}
\usepackage{vinayak}
\usepackage{graphicx}
\usepackage{verbatim}
\usepackage[mathscr]{eucal}
\usepackage{xspace}

\usepackage{enumerate}

\def\doi{7 (4:08) 2011}
\lmcsheading%
{\doi}
{1--55}
{}
{}
{Feb.~12, 2010}
{Dec.~14, 2011}
{}
\begin{document}

\title{Timed Parity Games: Complexity and Robustness}
\author[K.~Chatterjee]{Krishnendu Chatterjee\rsuper a}
\address{{\lsuper{a,b}}IST Austria (Institute of Science and
Technology, Austria)}
\email{\{krish.chat,tah\}@ist.ac.at}
\author[T.~A.~Henzinger]{Thomas A.~Henzinger\rsuper b}
\address{\vskip-6 pt} 
\author[V.~S.~Prabhu]{Vinayak S.~Prabhu\rsuper c}
\address{{\lsuper c}University of Porto}
\email{vinayak@eecs.berkeley.edu}
\thanks{%
This research was supported in part by the NSF grants
CCR-0132780, CNS-0720884, and CCR-0225610, and
by the European COMBEST project.}

\keywords{Timed Automata, Timed Parity Games, Robust Timed Control}
\subjclass{F.4.1}

\begin{abstract}
We consider two-player games played in real time on game structures with 
clocks where the objectives of players are described using parity
conditions. 
The games are \emph{concurrent} in that at each turn, both players 
independently propose a time delay and an action, and the action with the
shorter delay is chosen.
To prevent a player from winning by blocking time, we restrict each player 
to play strategies that ensure that the player cannot be responsible for causing 
a zeno run.
First, we present an efficient reduction of these games to \emph{turn-based} 
(i.e., not concurrent) \emph{finite-state} (i.e., untimed) parity games.
Our reduction improves the best known complexity for solving timed parity 
games.
Moreover, the rich class of algorithms for classical parity games can now 
be applied to timed parity games. 
The states of the resulting game are based on clock regions of the original 
game, and
the state space of the finite game
is linear in the size of the region graph.

Second, we consider two restricted classes of strategies for the player that 
represents the controller in a real-time synthesis problem, namely, 
\emph{limit-robust} and \emph{bounded-robust} winning strategies.
Using a limit-robust winning strategy, the controller cannot choose an exact 
real-valued time delay but must allow for some nonzero jitter in each of 
its actions.
If there is a given lower bound on the jitter, then the strategy is 
bounded-robust winning.
We show that exact strategies are more powerful than limit-robust 
strategies, which are more powerful than bounded-robust winning 
strategies for any  bound.
For both kinds of robust strategies, we present efficient reductions to 
standard timed automaton games.
These reductions provide algorithms for the synthesis of robust real-time
controllers.
\end{abstract}

\maketitle


\section{Introduction}

Timed automata~\cite{AlurD94} are models of real-time systems in which
states consist of discrete locations and values for real-time clocks.
The transitions between locations are dependent on the clock values.
\emph{Timed automaton games}, introduced
in~\cite{DBLP:conf/stacs/MalerPS95}, and explored further
in~\cite{AFHM+03,DBLP:conf/formats/AdlerAF05,DBLP:conf/concur/CassezDFLL05,DBLP:conf/lics/FaellaTM02,DBLP:conf/vmcai/FaellaTM02}
(amongst others), are played by two players on timed automata, e.g., a
``controller'' and a ``plant'' for modeling real-time controller
synthesis problems.  We consider timed automaton games with
$\omega$-regular objectives specified as \emph{parity conditions}.
The class of $\omega$-regular objectives can express all safety and
liveness specifications that arise in the synthesis and verification
of reactive systems, and parity conditions are a canonical form to
express $\omega$-regular objectives~\cite{Thomas97}.  The construction
of a winning strategy for player~1 in such games corresponds to the
\emph{controller-synthesis problem for real-time systems}
\cite{DBLP:conf/stacs/DSouzaM02,DBLP:conf/stacs/MalerPS95,wongtoi91control}
with respect to achieving a desired $\omega$-regular objective.

Timed automaton games proceed in an infinite sequence of rounds.
In each round, both players simultaneously propose moves, with each move 
consisting of an action and a time delay after which the player wants 
the proposed action to take place.
Of the two proposed moves, the move with the shorter time delay ``wins'' 
the round and determines the next state of the game.
Let a set $\Phi$ of runs be the desired objective for player~1.
Then player~1 has a winning strategy for $\Phi$ if it has a strategy to 
ensure that, no matter what player~2 does, one of the following two 
conditions holds:
(1)~time diverges and the resulting run belongs to $\Phi$, or
(2)~time does not diverge but player-1's moves are chosen only finitely 
often (and thus it is not to be blamed for the convergence of time)
\cite{AFHM+03,HenPra06}.
This definition of winning is equivalent to restricting both players  
to play according to \emph{receptive} strategies 
\cite{AluHen97,SegalaGSL98}, which do not allow a player to win by 
blocking time.

In timed automaton games, there are cases where a player can win by 
proposing a certain strategy of moves, but where moves that deviate 
in the timing by an arbitrarily small amount from the winning strategy 
result in a strategy that does not ensure winning any more. 
If this is the case, then the synthesized controller needs to work with 
infinite precision in order to achieve the control objective.
As this requirement is unrealistic, we propose two notions of 
\emph{robust winning strategies}.
In the first robust model, each move of player~1 (the ``controller'') 
must allow some jitter  when the action of the move is taken.
The jitter may be arbitrarily small, but it must be greater than~0.
We call such strategies \emph{limit-robust}.
In the second robust model, we give a lower bound on the jitter, i.e., 
every move of player~1 must allow for a fixed jitter, which is specified 
as a parameter of the game.
We call these strategies \emph{bounded-robust}.
The strategies of player~2 (the ``plant'') are left unrestricted 
(apart from being receptive).
We show that 
(1)~general strategies are strictly more powerful than limit-robust strategies; 
and 
(2)~limit-robust strategies are strictly more powerful than bounded-robust 
strategies for \emph{any} lower bound on the jitter, i.e., there are games
in which player~1 can win with a limit-robust strategy, but there does
not exist any nonzero bound on the jitter for which player~1 can win with 
a bounded-robust strategy.
The following example illustrates this issue.

\begin{figure}[t]
\strut\centerline{\setlength{\unitlength}{0.00043745in}
\begingroup\makeatletter\ifx\SetFigFont\undefined%
\gdef\SetFigFont#1#2#3#4#5{%
  \reset@font\fontsize{#1}{#2pt}%
  \fontfamily{#3}\fontseries{#4}\fontshape{#5}%
  \selectfont}%
\fi\endgroup%
{\renewcommand{\dashlinestretch}{30}
\begin{picture}(9604,3269)(0,-10)
\put(4694,846){\ellipse{1620}{990}}
\put(7124,2691){\ellipse{1620}{990}}
\put(8786,846){\ellipse{1620}{990}}
\path(4694,1341)(6359,2556)
\path(6248.966,2401.428)(6359.000,2556.000)(6178.229,2498.363)
\path(8249,1251)(7439,2241)
\path(7599.420,2139.682)(7439.000,2241.000)(7506.545,2063.693)
\path(7979,756)(7977,755)(7972,753)
	(7962,749)(7949,743)(7930,735)
	(7906,726)(7879,714)(7848,702)
	(7815,689)(7782,676)(7748,663)
	(7714,651)(7681,639)(7650,628)
	(7619,618)(7590,608)(7561,600)
	(7533,592)(7506,585)(7479,578)
	(7451,572)(7423,566)(7394,561)
	(7370,557)(7346,553)(7321,549)
	(7295,546)(7269,542)(7241,539)
	(7213,536)(7184,533)(7154,530)
	(7123,527)(7091,524)(7059,522)
	(7027,520)(6994,518)(6960,516)
	(6927,514)(6893,513)(6860,511)
	(6827,510)(6794,509)(6761,508)
	(6729,508)(6698,507)(6667,507)
	(6637,507)(6607,507)(6578,507)
	(6550,507)(6522,508)(6494,508)
	(6462,509)(6431,510)(6400,512)
	(6368,513)(6337,515)(6305,517)
	(6273,519)(6242,521)(6210,523)
	(6178,526)(6147,529)(6116,532)
	(6085,535)(6055,538)(6026,542)
	(5998,545)(5971,549)(5944,553)
	(5919,557)(5895,560)(5872,564)
	(5851,568)(5830,572)(5810,576)
	(5792,580)(5774,583)(5750,589)
	(5726,595)(5704,602)(5683,609)
	(5661,616)(5640,624)(5618,633)
	(5595,643)(5573,653)(5550,664)
	(5528,675)(5507,685)(5490,695)(5459,711)
\path(5646.470,681.762)(5459.000,711.000)(5591.433,575.127)
\path(5459,981)(5461,982)(5465,985)
	(5472,990)(5483,997)(5497,1007)
	(5515,1018)(5536,1031)(5560,1045)
	(5585,1060)(5610,1074)(5636,1087)
	(5663,1100)(5689,1112)(5716,1123)
	(5742,1133)(5769,1142)(5797,1150)
	(5826,1157)(5857,1164)(5889,1170)
	(5924,1176)(5949,1180)(5975,1183)
	(6002,1187)(6031,1190)(6061,1194)
	(6092,1197)(6124,1200)(6157,1203)
	(6192,1206)(6228,1208)(6264,1211)
	(6302,1213)(6340,1215)(6380,1217)
	(6419,1219)(6459,1221)(6500,1223)
	(6540,1224)(6580,1226)(6621,1227)
	(6660,1228)(6699,1229)(6738,1229)
	(6776,1230)(6813,1230)(6849,1230)
	(6885,1230)(6919,1230)(6953,1230)
	(6986,1230)(7018,1229)(7049,1228)
	(7087,1228)(7124,1227)(7160,1225)
	(7196,1224)(7231,1222)(7267,1220)
	(7301,1218)(7336,1216)(7370,1213)
	(7403,1210)(7436,1207)(7468,1203)
	(7499,1200)(7529,1196)(7558,1192)
	(7585,1188)(7612,1183)(7637,1179)
	(7661,1174)(7683,1169)(7704,1165)
	(7724,1160)(7742,1154)(7760,1149)
	(7776,1144)(7792,1138)(7815,1129)
	(7837,1119)(7857,1108)(7876,1096)
	(7894,1082)(7912,1067)(7930,1049)
	(7948,1030)(7966,1010)(7982,990)
	(7997,972)(8024,936)
\path(7868.000,1044.000)(8024.000,936.000)(7964.000,1116.000)
\put(4739,2286){\makebox(0,0)[lb]{{\SetFigFont{9}{10.8}{\rmdefault}{\mddefault}{\updefault}$a_2^1, x  > 2$}}}
\put(4424,756){\makebox(0,0)[lb]{{\SetFigFont{9}{10.8}{\rmdefault}{\mddefault}{\updefault}$l^0$}}}
\put(6944,2601){\makebox(0,0)[lb]{{\SetFigFont{9}{10.8}{\rmdefault}{\mddefault}{\updefault}$l^3$}}}
\put(5369,171){\makebox(0,0)[lb]{{\SetFigFont{9}{10.8}{\rmdefault}{\mddefault}{\updefault}$a_1^2, y> 1 \rightarrow y:=0$}}}
\put(8429,756){\makebox(0,0)[lb]{{\SetFigFont{9}{10.8}{\rmdefault}{\mddefault}{\updefault}$l^1$}}}
\put(7799,1971){\makebox(0,0)[lb]{{\SetFigFont{9}{10.8}{\rmdefault}{\mddefault}{\updefault}$a_2^2, y > 2$}}}
\put(5369,1341){\makebox(0,0)[lb]{{\SetFigFont{9}{10.8}{\rmdefault}{\mddefault}{\updefault}$a_1^1, x\leq 1 \rightarrow x:=0$}}}
\put(818,764){\ellipse{1620}{990}}
\path(1589,936)(1591,937)(1596,939)
	(1605,943)(1619,948)(1639,956)
	(1663,966)(1692,977)(1724,990)
	(1760,1004)(1797,1018)(1835,1033)
	(1873,1047)(1910,1061)(1946,1074)
	(1981,1087)(2014,1098)(2046,1109)
	(2077,1118)(2106,1127)(2134,1136)
	(2162,1143)(2189,1150)(2217,1157)
	(2244,1163)(2272,1168)(2297,1174)
	(2324,1178)(2351,1183)(2379,1187)
	(2407,1191)(2437,1195)(2467,1198)
	(2497,1202)(2529,1204)(2561,1207)
	(2594,1209)(2627,1211)(2660,1212)
	(2694,1213)(2727,1213)(2761,1213)
	(2794,1213)(2827,1212)(2860,1210)
	(2892,1208)(2923,1206)(2954,1203)
	(2985,1200)(3014,1196)(3043,1192)
	(3071,1187)(3099,1182)(3127,1176)
	(3154,1170)(3181,1163)(3208,1156)
	(3235,1148)(3263,1138)(3291,1129)
	(3320,1118)(3351,1106)(3383,1092)
	(3416,1078)(3451,1063)(3487,1046)
	(3525,1028)(3564,1010)(3603,990)
	(3643,971)(3682,951)(3720,932)
	(3755,914)(3787,897)(3815,883)
	(3838,871)(3856,861)(3884,846)
\path(3697.000,878.111)(3884.000,846.000)(3753.667,983.889)
\path(3884,756)(3882,755)(3878,753)
	(3870,749)(3857,743)(3840,734)
	(3818,723)(3791,710)(3761,695)
	(3727,679)(3691,662)(3653,644)
	(3615,626)(3577,608)(3539,591)
	(3502,575)(3467,559)(3432,545)
	(3399,532)(3368,519)(3337,508)
	(3307,497)(3278,488)(3249,479)
	(3221,470)(3192,463)(3163,455)
	(3134,448)(3106,442)(3078,437)
	(3049,431)(3019,426)(2989,421)
	(2958,417)(2926,412)(2893,408)
	(2860,405)(2826,401)(2791,398)
	(2757,396)(2721,393)(2686,392)
	(2651,390)(2615,389)(2580,389)
	(2546,388)(2511,389)(2478,389)
	(2445,391)(2413,392)(2382,394)
	(2352,396)(2322,399)(2294,402)
	(2266,406)(2240,410)(2214,414)
	(2189,418)(2159,425)(2129,432)
	(2100,440)(2071,449)(2043,459)
	(2014,470)(1984,482)(1954,496)
	(1923,512)(1890,528)(1857,546)
	(1823,566)(1788,586)(1754,606)
	(1721,626)(1690,646)(1662,663)
	(1639,678)(1620,691)(1589,711)
\path(1772.781,663.835)(1589.000,711.000)(1707.726,562.999)
\path(779,1251)(780,1253)(782,1257)
	(785,1264)(790,1276)(798,1292)
	(807,1313)(820,1340)(834,1370)
	(851,1405)(869,1443)(889,1483)
	(911,1525)(933,1568)(956,1611)
	(980,1654)(1004,1696)(1028,1736)
	(1052,1775)(1076,1812)(1100,1847)
	(1124,1881)(1148,1912)(1172,1943)
	(1197,1971)(1222,1999)(1248,2025)
	(1275,2050)(1303,2074)(1332,2098)
	(1362,2121)(1394,2143)(1420,2161)
	(1447,2179)(1475,2197)(1505,2215)
	(1535,2233)(1566,2250)(1599,2268)
	(1632,2286)(1667,2304)(1703,2322)
	(1740,2339)(1778,2357)(1817,2375)
	(1857,2393)(1898,2411)(1940,2429)
	(1982,2447)(2026,2465)(2070,2483)
	(2114,2501)(2159,2518)(2204,2535)
	(2250,2553)(2296,2569)(2341,2586)
	(2387,2603)(2432,2619)(2478,2635)
	(2523,2650)(2567,2665)(2612,2680)
	(2656,2695)(2699,2709)(2742,2723)
	(2785,2737)(2827,2751)(2868,2764)
	(2910,2778)(2951,2791)(2992,2803)
	(3032,2816)(3073,2829)(3114,2842)
	(3155,2855)(3196,2868)(3237,2880)
	(3279,2893)(3321,2906)(3364,2919)
	(3407,2932)(3450,2944)(3493,2957)
	(3537,2970)(3582,2983)(3626,2995)
	(3670,3008)(3715,3020)(3760,3032)
	(3804,3044)(3849,3056)(3893,3067)
	(3938,3078)(3981,3089)(4025,3100)
	(4068,3110)(4110,3120)(4152,3130)
	(4193,3139)(4234,3147)(4274,3156)
	(4314,3163)(4352,3171)(4390,3178)
	(4428,3185)(4465,3191)(4501,3197)
	(4537,3202)(4572,3207)(4607,3212)
	(4642,3216)(4682,3221)(4722,3225)
	(4763,3228)(4803,3232)(4844,3235)
	(4885,3237)(4926,3239)(4967,3240)
	(5009,3241)(5050,3242)(5092,3242)
	(5134,3242)(5175,3241)(5217,3240)
	(5258,3239)(5299,3237)(5339,3235)
	(5379,3232)(5417,3229)(5456,3226)
	(5493,3222)(5529,3218)(5565,3214)
	(5599,3210)(5632,3205)(5664,3200)
	(5694,3195)(5724,3190)(5752,3185)
	(5780,3179)(5806,3174)(5831,3168)
	(5856,3162)(5879,3156)(5911,3147)
	(5941,3138)(5971,3129)(5999,3118)
	(6026,3107)(6053,3096)(6080,3083)
	(6107,3069)(6135,3054)(6162,3038)
	(6190,3021)(6219,3003)(6247,2985)
	(6274,2966)(6301,2947)(6325,2930)
	(6347,2914)(6366,2900)(6380,2889)(6404,2871)
\path(6224.000,2931.000)(6404.000,2871.000)(6296.000,3027.000)
\put(554,666){\makebox(0,0)[lb]{{\SetFigFont{9}{10.8}{\rmdefault}{\mddefault}{\updefault}$l^2$}}}
\put(1814,1341){\makebox(0,0)[lb]{{\SetFigFont{9}{10.8}{\rmdefault}{\mddefault}{\updefault}$a_1^4, x  < 1$}}}
\put(1769,81){\makebox(0,0)[lb]{{\SetFigFont{9}{10.8}{\rmdefault}{\mddefault}{\updefault}$a_1^3, x  < 1$}}}
\put(1319,2601){\makebox(0,0)[lb]{{\SetFigFont{9}{10.8}{\rmdefault}{\mddefault}{\updefault}$a_2^3, x  > 2$}}}
\end{picture}
}}
\caption{A timed automaton game $\A$.}
\label{figure:jitter}
\end{figure}
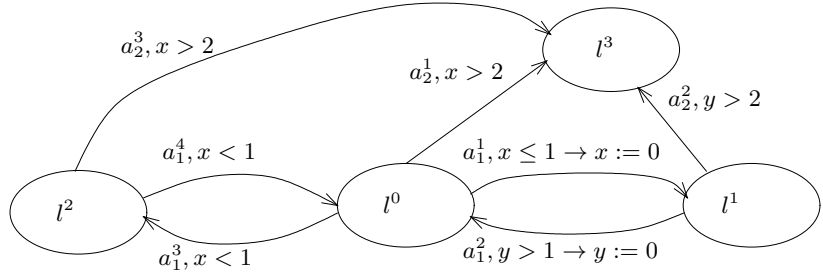

\begin{exa}
Consider the timed automaton $\A$ in Fig.~\ref{figure:jitter}.
The edges denoted $a_1^k$ for $k\in\set{1,2,3,4}$ are controlled by 
player~1, and the edges denoted $a_2^j$ for $j\in\set{1,2,3}$ are controlled
by player~2.
The objective of player~1 is $\Box(\neg l^3)$, i.e., to avoid the location $l^3$.
The important part of the automaton is the cycle $l^0, l^1$.
The only way to avoid $l^3$ in a time divergent run is to cycle  between
$l^0$ and $l^1$ infinitely often.
In addition, player~1 may choose to also cycle  between $l^0$ and $l^2$,
but that does not help (or harm) it.
Due to 
strategies being required to be receptive, player~1 cannot just cycle  between 
$l^0$ and $l^2$ forever, it must also cycle  between $l^0$ and $l^1$;
that is, to satisfy $\Box(\neg l^3)$ player~1 must ensure 
$(\Box\Diamond l^0)\wedge (\Box\Diamond l^1)$, where $\Box\Diamond$ denotes
``infinitely often''.
But note that player~1 may cycle  between $l^0$ and $l^2$ any
finite number of times
as it wants between an $l^0, l^1$ cycle.

In our analysis below, we omit such $l^0,l^2$ cycles for simplicity.
Let the game start from the location $l^0$ at time 0, and let
$l^1$ be visited at time $t^0$ for the first time.
Also, let $\alpha^j$ denote the difference between times when $l^0$ is visited
for the $(j+1)$-th time, and when $l^1$ is visited for the $j$-th time.
We can have at most 1 time unit between two successive visits to $l^0$,
and we must have strictly more than 1 time unit elapse between two
successive visits to $l^1$.
Thus, $\alpha^j$ must be in a strictly decreasing sequence.
Also, for player~1 to cycle between $l^0$ and $l^1$ infinitely often, 
we must have  $\alpha^j \geq 0$ for all $j$ as
the $(j+1)$-th $a_1^1$ transition must always 
happen after the $j$-th
$a_1^2$ transition.
Consider any bounded-robust strategy.
Since the jitter is some fixed $\varjit$, for any strategy of player~1
that tries to cycle between $l^0$ and $l^1$, there are executions
where the transition labeled $a_1^1$ is  taken when $x$ is less
than or equal to $1-\varjit$, and the transition labeled $a_1^2$
 is taken when $y$ is greater than $1$.
This means that there are executions where $\alpha^j$ decreases by at 
least $\varjit$ in each cycle.
But, this implies that we cannot have an infinite decreasing sequence of 
$\alpha^j$'s for any $ \varjit$ and for any starting value of $t^0$.

With a limit-robust strategy, however, player~1 can cycle between the two
locations infinitely often, provided that the starting value of $x$ is
strictly less than 1. 
This is because at each step of the game, player~1 can choose  moves that are
such that the clocks $x$ and $y$ are closer and closer to 1.
A general strategy allows player~1 to win even when the starting value of
$x$ is 1.
The details will be presented later in Example~\ref{example:Jitter} in
Subsection~\ref{subsection:BoundedJitter}.
\qed
\end{exa}

\noindent\textbf{Contributions}.
Our contributions are two-fold: we present improved complexity results to 
solve timed automaton parity games, and we present two notions of robust 
winning in timed automaton parity games and present solutions of them.

\smallskip\noindent\emph{Improved complexity.}
We first show that timed automaton parity games can be reduced to 
classical \emph{turn-based} finite-state (untimed) parity games.
Even though the timed games are \emph{concurrent}, in that in each turn both 
players simultaneously propose moves before one of the moves is chosen, 
our reduction to the untimed finite-state game is turn-based.
The concurrency in timed games is limited (only in proposal of time), and 
we exploit this in the reduction to obtain turn-based games.
In general the reduction of concurrent games to turn-based games is 
only known for B\"uchi and coB\"uchi objectives (and only for
qualitative analysis)~\cite{JurdzinskiKH02}. 
The turn-based game we obtain as result of the reduction has a state space that is
linear in the number of clock regions.
There is a rich literature of algorithms as well much ongoing research to 
solve finite-state turn-based parity games, and our reduction allows 
all these algorithms to be used to solve timed automaton parity games.
A solution for timed automaton games with parity  objectives 
was already presented in~\cite{AFHM+03} and the solution works in 
\[O\left( \left(M\cdot |C|\cdot|A_1|\cdot|A_2|\right)^2 \cdot 
\left(|{S}_{\reg}^{*}| \right)^{d+2} \right)
\] 
time, where $M$ is the maximum constant in the timed automaton; $C$ is
the set of clocks; $A_i$ is the set of player-$i$ edges; $L$ is the
set of locations; $d$ is the number of priorities in the parity index
function; and ${S}_{\reg}^{*}$ is the set of states in the region
graph of the timed automaton expanded to handle receptiveness and we
have $|{S}_{\reg}^{*}| = |S_{\reg}|\cdot 32\cdot |C|\cdot d$, where $
S_{\reg}$ is the set of regions of the timed automaton (which is
bounded by $|L|\cdot\prod_{x\in C}(c_x+1)\cdot |C|!\cdot 4^{|C|}$,
with $c_x$ being the maximum constant that clock $x$ is compared to in
the timed automaton game).  We show that timed automaton games can be
solved in 
\[O\left( \left(|S_{\reg}^*|\cdot(|A_1| + |A_2|)\right)\cdot
\left( |{S}_{\reg}^{*}|\cdot 8 \right)^ {\frac{d+2}{3}+ \frac{1}{2}}
\right)
\]
 time.  Our reduction has two steps: first we show that
certain restrictions can be applied to strategies without changing the
winning set; then we show that the timed game with clocks under the
strategy restrictions can be transformed into a finite-state
turn-based game with $8\cdot |S^{*}_{\reg}|$ states, $O(
|S^{*}_{\reg}|\cdot(|A_1| + |A_2|))$ edges, and
$d+2$ priorities.  Our improved complexity follows from the above
reduction, and the fact that a finite-state turn-based parity game
with $m$ edges, $n$ states and $d$ parity indices can be solved in
$O(m\cdot n^{\frac{d}{3}+\frac{1}{2}})$ time~\cite{Schewe/07/Parity}.
The restriction to receptive strategies is handled by our reduction
with the following two modifications: (1)~the number of regions of the
timed automaton parity game needs to be enlarged by a factor of
$32\cdot |C|\cdot d$, and (2)~the number of indices of the parity
function needs to be increased by 2.  The modifications are similar to
those in~\cite{AFHM+03}.

\smallskip\noindent\emph{Robust winning.}
Second, we show that timed automaton games with limit-robust and 
bounded-robust strategies can be solved by reductions to general 
timed automaton games (with exact strategies).
In the reduction for limit-robust games, the jitter is controlled by
player~1, as the jitter is only required to be greater than 0, with no other
restriction.
For bounded-robust games, the jitter is controlled by
player~2, as there is an uncertainty interval of constant length $\varjit$ (which
is fixed for the game).
The reduction for the limit-robust case is obtained by
changing the winning condition so that moves are only to states where  all the clock
values are non-integral.
The reduction for the  bounded-robust case is by a 
syntactic transformation of the game graph.
The limit-robust game can be solved in time
\[O\left(
 \left(
   |S_{\reg}^*|\cdot(|A_1| + |A_2|)\right)\cdot
    \left(
      |{S}_{\reg}^{*}|\cdot 16
    \right)^ {\frac{d+2}{3}+ \frac{1}{2}}
\right).
\]
Given a rational valued jitter of 
$\varepsilon= \frac{{\varepsilon_n}_{\vphantom{X}}}{\varepsilon_d}$, 
the  bounded-robust game can be
solved in time
\[O\left(
 \left(
   |S_{\reg}^*|\cdot |A_1|^2\cdot |A_2| \cdot |C|\cdot
   \varepsilon_n \cdot \varepsilon_d^{|C|+2}
\right)\cdot
 \left(
     |{S}_{\reg}^{*}|\cdot 32 \cdot |C|\cdot |A_1|\cdot
   \varepsilon_n \cdot \varepsilon_d^{|C|+2}
 \right)^ {\frac{d+2}{3}+ \frac{1}{2}}
\right).
\]
%
%
%
The reductions provide algorithms for synthesizing robust controllers 
for real-time systems, where the controller is guaranteed to achieve the 
control objective even if its time delays are subject to jitter.
The question of the \emph{existence} of a non-zero jitter for 
which a game can be won with a bounded-robust strategy remains open.

\smallskip\noindent\textbf{Comparison to the preliminary
 version~\cite{KCHenPraB08}.}
Our present submission extends and improves upon the results 
of~\cite{KCHenPraB08}.
The state space of
the finite game of
the reduction in~\cite{KCHenPraB08} had size 
$O(|S^{*}_{\reg}|\cdot |C|\cdot M\cdot |A_1|) $; the present
finite-state  game has size $O(|S^{*}_{\reg}|)$.
The correctness proof of the 
present reduction is based on several new non-trivial results that are not 
present in~\cite{KCHenPraB08}.

\smallskip\noindent\textbf{Related work}.
Timed automaton games have been explored before for controller synthesis, e.g, 
in~\cite{maler98controller,DBLP:conf/stacs/DSouzaM02,DBLP:conf/stacs/MalerPS95,wongtoi91control}.
In most  previous work,
 time-divergence has not been handled properly. 
For example,
the formulations 
of~\cite{maler98controller,DBLP:conf/lics/FaellaTM02,JurdzinskiT07,TrivediThesis} 
assume
that the syntactic structure of the game is such that it is not 
possible for the players to block time.
The work of~\cite{AluHen97} looks at safety objectives, and correctly requires 
that a player might not stay safe simply by blocking time. 
It however also requires that the controller achieve its objective even if the opponent
blocks time, and hence is deficient for reachability objectives.
In~\cite{DBLP:conf/stacs/DSouzaM02} the authors require player~1 to always allow player~2 moves,
thus, in particular, player~2 can foil a reachability objective of player~1 by
blocking time.
The strategies of player~1 are also  
assumed to be region strategies  by definition, 
that is, player-1 strategies can only specify regions in its moves 
(and not an exact desired state), and can only be
dependent on the history of the observed regions in a play (and not on the history
of exact states in a play).
A solution for timed automaton games with receptive 
strategies and parity objectives was first
presented in~\cite{AFHM+03}, where the solution is obtained by
first demonstrating that the winning set can be characterized by a 
$\mu$-calculus fixpoint expression, and then showing that only unions
of clock regions arise in its fixpoint iteration.

It has been recognized by many researchers that an important shortcoming
of timed automata is that clock values are assumed to be available as real
numbers with infinite precision.
Our notion of bounded-robustness is closely related to the Almost-ASAP 
semantics of~\cite{WulfDR05}.
The work there is done in a one-player setting where the controller is 
already known, and one wants the know if the composition of the controller
and the system satisfies a safety property in the presence of bounded
jitter and observation delay.
A similar model for hybrid automata is considered in~\cite{AgrawalT04}.
The solution for the existence of bounded jitter and observation delay 
for which a timed system stays safe is presented in~\cite{WulfDMR04}.
Various models of robustness in  timed automata (the one-player case) are also 
considered in~\cite{AlurTM05,BMR-fossacs08,GHJ97,HenRas00}.

\smallskip\noindent\textbf{Outline}.
In Section~\ref{section:TimedGames} we present the definitions of 
timed game structures, objectives, strategies, and review some basic results
for timed automaton games.
In Section~\ref{section:Reduction} we present  a restriction of strategies
of the two players, which does not change the winning set, and which allows
an efficient reduction to finite-state turn-based games.
We also derive the complexity bound obtained from the new reduction to finite-state games.
In Section~\ref{section:Robust} we define limit-robust and  bounded-robust
strategies, and show how winning sets for both can be computed by reductions
to timed automaton games.

\section{Timed Games}
\label{section:TimedGames}
In this section we present the definitions of timed game structures,
runs, objectives, and strategies in timed game structures.

\smallskip\noindent{\bf Timed game structures.} 
A \emph{timed game structure} is a tuple 
$\TG = \tuple{S,\acts_1,\acts_2,\Gamma_1,\Gamma_2,\delta}$
with the following components.
\begin{enumerate}[$\bullet$]
\item 
$S$ is a set of states.
\item 
$\acts_1$ and $\acts_2$ are two disjoint sets of actions for players~1 
and~2, respectively.
We assume that $\set{\bot_1,\bot_2,\bot_*}\cap\acts_i = \emptyset $, and write 
$\acts_1^{\bot}$ for $\acts_1\cup\set{\bot_1,\bot_*}$, and 
$\acts_2^{\bot}$ for $\acts_2\cup\set{\bot_2}$.
The set of \emph{moves} for player $i$ is 
$M_i =\reals_{\geq 0} \times \acts_i^{\bot}$.
Intuitively, a move $\tuple{\Delta,a_i}$ by player $i$ indicates a
waiting period of $\Delta$ time units followed by a discrete
transition labeled with action~$a_i$.
The move  $\tuple{\Delta,\bot_i}$ is used to represent the move where player-$i$
just lets time elapse for $\Delta$ time units without taking any of the discrete 
actions from $\acts_i$.
The action $\bot_*$ is used to represent the fact that player~1 is relinquishing
control to player~2 in the given stage of the game.

\item
$\Gamma_i : S\mapsto 2^{M_i} \setminus \emptyset$ are two move assignments.
At every state~$s$, the set $\Gamma_i(s)$ contains the moves that are 
available to player $i$.
We require that $\tuple{0,\bot_i}\in\Gamma_i(s)$ for all states $s\in S$ 
and $i\in\set{1,2}$.
Intuitively, $\tuple{0,\bot_i}$ is a time-blocking stutter move.
\item
$\delta: S\times (M_1 \cup M_2) \mapsto S$ is the transition function.
We require that for all time delays 
$\Delta,\Delta'\in\reals_{\ge 0}$ with $\Delta'\leq \Delta$,
and all actions $a_i\in \acts_i^{\bot}$,
we have 
\begin{enumerate}[(i)]
\item 
$\tuple{\Delta,a_i}\in\Gamma_i(s)$ iff both
$\tuple{\Delta',\bot_i}\in \Gamma_i(s)$ and 
$\tuple{\Delta -\Delta',a_i}\in\Gamma_i(\delta(s,\tuple{\Delta',\bot_i}))$;
and
\item 
if $\delta(s,\tuple{\Delta',\bot_i})=s'$ and 
$\delta(s',\tuple{\Delta-\Delta',a_i})=s''$, then 
$\delta(s,\tuple{\Delta,a_i}) = s''$.
\end{enumerate}
\end{enumerate}
The game proceeds as follows.
If the current state of the game is~$s$, then both players simultaneously 
propose moves $\tuple{\Delta_1,a_1}\in\Gamma_1(s)$ and 
$\tuple{\Delta_2,a_2}\in\Gamma_2(s)$.
If $a_1\neq \bot_*$, the move with the shorter duration ``wins'' 
in determining the next state of the game. 
If both moves have the same duration, then  the next state is chosen
non-deterministically.
If $a_1=\bot_*$, then the move of player~2 determines the next state,
regardless of $\Delta_1$.
We give this special power to player~1 for modeling convenience 
as (1)~the controller always has the
option of letting the state evolve in a controller-plant framework,
without always having to provide inputs to the plant, and
(2)~it allows a natural model for systems where controller actions are disabled in
certain modes{\footnote{We illustrate the usefulness of $\bot_*$ in
Example~\ref{example:Relinquish}.
The results of this paper do not change if $\bot_*$ is not present in the framework.}}.
Formally, we define the \emph{joint destination function} 
$\delta_{\jd} : S\times M_1\times M_2 \mapsto 2^S$ by
\[
\delta_{\jd}(s,\tuple{\Delta_1,a_1},\tuple{\Delta_2,a_2}) = \left\{
\begin{array}{ll}
\set{\delta(s,\tuple{\Delta_1,a_1})} & \text{ if } \Delta_1 < \Delta_2
\text{ and } a_1\neq \bot_*; \\
\set{\delta(s,\tuple{\Delta_2,a_2})} & \text{ if } \Delta_2 < \Delta_1
\text{ or } a_1=\bot_*;\\
\set{\delta(s,\tuple{\Delta_2,a_2}),
\delta(s,\tuple{\Delta_1,a_1})} & \text{ if } \Delta_2 = \Delta_1
\text{ and } a_1\neq \bot_*.

\end{array}
\right.
\]
The time elapsed when the moves $m_1=\tuple{\Delta_1,a_1}$ and 
$m_2=\tuple{\Delta_2,a_2}$ are proposed is given by 
$$\delay(m_1,m_2) = \left\{
\begin{array}{ll}
\min(\Delta_1,\Delta_2) &  \text{ if } a_1\neq \bot_* \\
\Delta_2 & \text{ if } a_1= \bot_* 
\end{array}
\right.
$$
The boolean predicate $\Blfunc_i(s,m_1,m_2,s')$ indicates whether player~$i$ 
is ``responsible'' for the state change from $s$ to $s'$ when the moves $m_1$ 
and $m_2$ are proposed.
Denoting the opponent of player~$i$ by $\negspaceopp{i} = 3-i$, for $i \in
\set{1,2}$, we define 
 $$\Blfunc_i(s,\tuple{\Delta_1,a_1},\tuple{\Delta_2,a_2},s') =\ 
 \begin{cases}
   \mspace{-2mu}
   \big(\Delta_1 \leq \Delta_2\ \wedge\ 
   \delta(s,\tuple{\Delta_1,a_1}) = s'\big) \ \bigwedge\ 
   (a_1\neq \bot_* ) & \text{if } i=1\\
   \mspace{-2mu}
 \big(\Delta_2 \leq \Delta_1\ \wedge\ 
   \delta(s,\tuple{\Delta_2,a_2}) = s'\big)\ \bigvee\ 
   (a_1 = \bot_*) & \text{if } i=2
\end{cases}
 $$

\smallskip\noindent{\bf Runs.} A \emph{run} $r=\run$ 
of the timed game
structure $\TG$ is an infinite sequence such that $s_k\in S$
and $m_i^k \in \Gamma_i(s_k)$ and $s_{k+1} \in
\delta_{\jd}(s_k,m_1^k,m_2^k)$ for all $k\geq 0$ and $i\in\set{1,2}$.
For $k\ge 0$, let $\runtime(r,k)$ denote the ``time'' at position $k$ of the 
run, namely, 
$\runtime(r,k)=\sum_{j=0}^{k-1}\delay(m_1^j,m_2^j)$ (we let $\runtime(r,0)=0$).
By $r[k]$ we denote the $(k+1)$-th state $s_k$ of~$r$.
The run prefix $r[0..k]$ is the finite prefix of the run $r$ that ends in 
the state~$s_k$.
Let $\iruns$ be the set of all runs of $\TG$, and let $\VRuns$ be the set 
of run prefixes.

\smallskip\noindent{\bf Objectives.}
An \emph{objective} for the timed game structure $\TG$ is a set 
$\Phi\subseteq\iruns$ of runs.
We will be interested in 
parity objectives.
Parity objectives are canonical forms for $\omega$-regular
properties that can express all commonly used specifications
that arise in verification.

%
%
%
Let $\Omega: S\mapsto \set{0,\dots,k-1}$ be a parity index function.
The parity objective for $\Omega$ requires that the 
maximal index visited infinitely often be even.
Formally, let $\infoften(\Omega(r))$ denote the set of indices visited 
infinitely often along a run $r$.
Then the parity objective defines the following set of
runs:
$\parity(\Omega)=\set{r \mid \max(\infoften(\Omega(r))) \text{ is even }}$.
%
A timed game structure $\TG$ together with the index function $\Omega$ 
constitute a  \emph{parity timed game} (of \emph{order} $k$) in which the objective of 
player~1 is $\parity(\Omega)$.

\smallskip\noindent{\bf Strategies.}
A \emph{strategy} for a player is a recipe that specifies
how to extend a run.
Formally, 
a \emph{strategy} $\pi_i$ for player $i\in \set{1,2}$ is a function 
$\pi_i$ that assigns to every run prefix 
$r[0..k]$ a move $m_i$ in 
the set of moves available to  player~$i$ at the state $r[k]$.
%
%
%
For $i\in\set{1,2}$, let $\Pi_i$ be the set of strategies for player~$i$.
Given two strategies $\pi_1\in \Pi_1$ and $\pi_2\in \Pi_2$, the set of 
possible \emph{outcomes} of the game starting from a state $s\in S$ is 
the set of possible runs denoted by $\outcomes(s,\pi_1,\pi_2)$.

\smallskip\noindent{\bf Receptive strategies.}
We will be interested in strategies that are meaningful
(in the sense that they do not block time).
To define them formally we first present  the following two
sets of runs.
\begin{enumerate}[$\bullet$]
\item
A run $r$ is \emph{time-divergent} if 
$\lim_{k\rightarrow\infty}\runtime(r,k) = \infty$.
We denote by $\td$  the set of all time-divergent runs.
\item
The set $\blameless_i\subseteq\iruns$ consists of
the set of runs in which player $i$ is 
responsible only for finitely many transitions.
A run $\run$ belongs to the set $\blameless_i$, for $i=\set{1,2}$, 
if there exists a $k\ge 0$ such that for all $j\ge k$, we have 
$\neg\Blfunc_i(s_j,m_1^{j},m_2^{j}, s_{j+1})$.
\end{enumerate}
A strategy $\pi_i$ is \emph{receptive} if  for
all strategies $\pi_{\opp{i}}$, all states $s\in S$, and all
runs $r\in\outcomes(s,\pi_1,\pi_2)$, either $r\in\td$ or
$r\in\blameless_i$.
Thus, no  matter what the opponent does, a receptive  
strategy of player~$i$ cannot be responsible for blocking time.
Strategies that are not receptive  are not physically meaningful.
A timed game structure $\TG$ is \emph{well-formed} if both players have 
receptive strategies.
We restrict our attention to well-formed timed game structures.
We denote $\Pi_i^R$ to be the set of receptive strategies for player~$i$.
Note that for $\pi_1\in\Pi_1^R, \pi_2\in\Pi_2^R$, we have
$\outcomes(s,\pi_1,\pi_2)\subseteq \td$.

\smallskip\noindent{\bf Winning sets.}
Given an objective $\Phi$, let $\wintimediv_1^{\TG}(\Phi)$ denote the set
of states $s$ in $\TG$ such that player~1 has a receptive strategy
 $\pi_1\in \Pi_1^R$ such that for all receptive strategies 
 $\pi_2\in \Pi_2^R$, we have $\outcomes(s,\pi_1,\pi_2)\subseteq \Phi$.
The strategy $\pi_1$ is said to be a winning strategy.
In computing the winning sets, we shall quantify over \emph{all} strategies,
but modify the objective to take care of time divergence.
Given an objective $\Phi$, let 
\[\timedivbl_1(\Phi) = (\td\cap\ \Phi)\cup 
(\blameless_1\setminus \td)\]
 \emph{i.e.}, $\timedivbl_1(\Phi)$ denotes the
set of runs such that either time diverges and $\Phi$ holds, or else 
time converges and player~1 is not responsible for time to converge.
Let $\win_1^{\TG}(\Phi)$ be the 
set of states in $\TG$ such that for all $s\in \win_1^{\TG}(\Phi)$, 
player~1 has a (possibly non-receptive) strategy $\pi_1\in
\Pi_1$ such that for all (possibly non-receptive) strategies 
$\pi_2\in \Pi_2$, we have
$\outcomes(s,\pi_1,\pi_2)\subseteq\, \Phi$.
The strategy $\pi_1$  is said to be winning for the non-receptive game.
The following result establishes the connection between $\win$ and
$\wintimediv$ sets.

\begin{thm} [\cite{HenPra06}] 
For all well-formed timed game structures
$\TG$, and for all $\omega$-regular objectives $\Phi$, we have
$\win_1^{\TG}(\timedivbl_1(\Phi))= \wintimediv_1^{\TG}(\Phi)$.
\end{thm}

We now define a special class of timed game structures, namely,
timed automaton games.

\smallskip\noindent{\bf Timed automaton games.}
Timed automata~\cite{AlurD94} suggest a finite syntax for specifying
infinite-state timed game structures.
A \emph{timed automaton game} is a tuple 
$\A=\tuple{L,C,\acts_1, \acts_2,E,\inv}$ 
with the following components:
\begin{enumerate}[$\bullet$]	
\item 
$L$ is a finite set of locations.
\item 
$C$ is a finite set of clocks.
\item 
$\acts_1$ and $\acts_2$ are two disjoint sets of actions for players~1 
and~2, respectively.
\item 
$E \subseteq L\times (\acts_1\cup \acts_2)\times \clkcond(C)\times L
  \times 2^C$
is the edge relation, where the set $\clkcond(C)$ of 
\emph{clock constraints} is generated by the grammar 
  $$\theta ::= x\leq d \mid d\leq x\mid \neg\theta \mid 
    \theta_1\wedge\theta_2$$ 
for clock variables $x\in C$ and nonnegative integer constants~$d$.\\
For an edge $e=\tuple{l,a_i,\theta,l',\lambda}$, the clock constraint 
$\theta$ acts as a guard on the clock values which specifies when the 
edge $e$ can be taken, and by taking the edge~$e$, the clocks in the set 
$\lambda\subseteq C$ are reset to~0.
We require that for all edges 
$\tuple{l,a_i,\theta',l',\lambda'} \neq \tuple{l,a_i',\theta'',l'',\lambda''}\in E$,
we have $a_i\neq a_i'$. 
This requirement ensures that a state and a move together uniquely determine 
a successor state.
\item 
$\inv: L\mapsto\clkcond(C)$ is a function that assigns to 
every location an invariant for both players.  
All clocks increase uniformly at the same rate.
When at location~$l$, each player~$i$ must propose a move out of $l$ 
before the invariant $\inv(l)$ expires.
Thus, the game can stay at a location only as long as the invariant is 
satisfied by the clock values.
\end{enumerate}
A \emph{clock valuation} is a function  $\kappa : C\mapsto\reals_{\geq 0}$
that maps every clock to a nonnegative real. 
The set of all clock valuations for $C$ is denoted by $K(C)$.
Given a clock valuation $\kappa\in K(C)$ and a time delay 
$\Delta\in\reals_{\geq 0}$, we write 
$\kappa +\Delta$ for the clock valuation in $K(C)$ defined by 
$(\kappa +\Delta)(x) =\kappa(x) +\Delta$ for all clocks $x\in C$.
For a subset $\lambda\subseteq C$ of the clocks, we write 
$\kappa[\lambda:=0]$ for the clock valuation in $K(C)$ defined by 
$(\kappa[\lambda:=0])(x) = 0$ if $x\in\lambda$, 
and $(\kappa[\lambda:=0])(x)=\kappa(x)$ if $x\not\in\lambda$.
A clock valuation $\kappa\in K(C)$ \emph{satisfies} the clock constraint 
$\theta\in\clkcond(C)$, written $\kappa\models \theta$, if the condition 
$\theta$ holds when all clocks in $C$ take on the values specified 
by~$\kappa$.
A \emph{state} $s=\tuple{l,\kappa}$ of the timed automaton game $\A$ is a 
location $l\in L$ together with a clock valuation $\kappa\in K(C)$ such 
that the invariant at the location is  satisfied, that is,
$\kappa\models\inv(l)$.
We let $S$ be the set of all states of~$\A$.

In a state, each player~$i$ proposes a time delay allowed by the 
invariant map~$\inv$, together either with the action~$\bot_i$ (player~1 can
also propose $\bot_*$), 
or with an action $a_i\in\acts_i$ such that an edge labeled $a_i$ 
is enabled after the proposed time delay.
We require that for all states $s=\tuple{l,\kappa}$, 
either 
\begin{enumerate}[(a)]
\item
$\kappa+\Delta\models\inv(l)$ for all
$\Delta\in\reals_{\geq 0}$, or 
\item
there exist a time delay 
$\Delta\in\reals_{\geq 0}$ and an edge 
$\tuple{l,a_2,\theta,l',\lambda}\in E$ such that
\begin{enumerate}[(1)] 
\item $a_2\in\acts_2$ and
\item 
$\kappa+\Delta\models\theta$ and 
for all $0\le\Delta'\le\Delta$, we have $\kappa+\Delta'\models\inv(l)$, and
\item 
$(\kappa+\Delta)[\lambda:=0]\models\inv(l')$.
\end{enumerate}
\end{enumerate}
Informally, these conditions ensure that, for a legal state, 
either the invariant at the  location is satisfied at all time points in the
future; or there is some time point in the future at which a discrete action
can be taken by the plant (with the location invariant being satisfied up to 
that time point).
This requirement is necessary (but not sufficient) for well-formedness of 
the game.

The timed automaton game $\A$ defines a timed game
structure $\symb{\A} =
\tuple{S,\acts_1,\acts_2,\Gamma_1,\Gamma_2,\delta}$ as follows:
\begin{enumerate}[$\bullet$]
\item
$S = \set{\tuple{l,\kappa} \mid l\in L \text{ and } \kappa(l) \text{ satisfies
} \gamma(l)}$.
\item
For $i\in\set{1,2}$, the set 
$\Gamma_i(\tuple{l,\kappa})$ contains the following elements:
\begin{enumerate}
\item 
$\tuple{\Delta,\bot_i}$ if for all $0\le\Delta'\le\Delta$, we have $\kappa+\Delta'\models\inv(l)$.
\item
$\tuple{\Delta,a_i}$ if  for all $0\le\Delta'\le\Delta$, we have
$\kappa+\Delta'\models\inv(l)$,  $a_i\in\acts_i$, and 
there exists an edge $\tuple{l,a_i,\theta,l',\lambda}\in E$ 
such that $\kappa+\Delta\models\theta$.
\item
$\tuple{0,\bot_*}$ if $i=1$.
\end{enumerate}
\item
The transition function $\delta$ is specified by:
\begin{enumerate}
\item
$\delta(s,\tuple{\Delta,\bot_*})= s$
\item		
$\delta(\tuple{l,\kappa},\tuple{\Delta,\bot_i}) = \tuple{l,\kappa +\Delta}$.
\item 
$\delta(\tuple{l,\kappa},\tuple{\Delta,a_i}) = 
\tuple{l',(\kappa +\Delta)[\lambda:=0]}$ 
for the unique edge $\tuple{l,a_i,\theta,l',\lambda} \in E$ with 
$\kappa+\Delta\models \theta$.
\end{enumerate}
\end{enumerate}
The timed game structure $\symb{\A}$ is not necessarily well-formed, because 
it may contain cycles along which time cannot diverge.
Well-formedness of timed automaton games can be checked in
\EXPTIME~\cite{HenPra06}.
We restrict our focus to well-formed timed automaton games in this paper.

\begin{exa}[Utility of the action $\bot_*$]
\label{example:Relinquish}
Suppose we did not have the action $\bot_*$.
Then in every timed game, we need to require that 
from every location,  there needs to be an outgoing player-1 edge  before the
invariant of the location expires; this requirement needs to be present as now
player~1 cannot simply relinquish active control to player~2.

Consider a timed automaton game $\A$ with a location $l$ in which we want to model
the following.
We want that if a clock condition $\theta$ is met at location $l$, then 
the moves of the controller (player~1)  are disabled for the next $\Delta$ time units 
(the disabled
time interval is right open); and that
player~2 (the plant) is to take a particular action $a_2'$ within these $\Delta$ time units.
Let $x$ be a clock used just for guarding the $\Delta$ condition.
The system  can be modeled as in Figure~\ref{figure:NoRelinquish}.
The incoming player-1 and player-2 edges to $l^1$ have the 
same guard and reset condition.
Without the $\bot_*$ action, we need special player-1 and player-2 actions $a_1^*$
and $a_2^*$ which go to a dummy sink accepting location from
the location $l^1$.
The objective $\Phi$ of player~1 also needs to be modified to 
$\Phi \wedge (\neg \Diamond l^*)$ (where $\Diamond$ is the standard \LTL operator for
reachability).
\begin{figure}[t]
  \begin{minipage}[b]{0.5\linewidth}
    \strut\centerline{\setlength{\unitlength}{0.00043745in}
\begingroup\makeatletter\ifx\SetFigFontNFSS\undefined%
\gdef\SetFigFontNFSS#1#2#3#4#5{%
  \reset@font\fontsize{#1}{#2pt}%
  \fontfamily{#3}\fontseries{#4}\fontshape{#5}%
  \selectfont}%
\fi\endgroup%
{\renewcommand{\dashlinestretch}{30}
\begin{picture}(6240,2982)(0,-10)
\put(4086,522){\ellipse{1620}{990}}
\put(4087,2464){\ellipse{1620}{990}}
\put(818,503){\ellipse{1620}{990}}
\path(1638,511)(3258,511)
\path(3078.000,451.000)(3258.000,511.000)(3078.000,571.000)
\path(3798,961)(3796,964)(3793,972)
	(3787,985)(3778,1004)(3766,1028)
	(3753,1057)(3738,1089)(3723,1122)
	(3708,1156)(3693,1188)(3680,1219)
	(3668,1247)(3658,1274)(3649,1299)
	(3641,1322)(3634,1343)(3628,1364)
	(3622,1384)(3618,1403)(3614,1425)
	(3610,1447)(3607,1469)(3605,1492)
	(3604,1515)(3603,1538)(3603,1561)
	(3604,1584)(3605,1607)(3607,1630)
	(3609,1652)(3613,1673)(3616,1693)
	(3620,1712)(3625,1730)(3630,1747)
	(3635,1763)(3641,1778)(3647,1795)
	(3655,1811)(3663,1826)(3672,1842)
	(3683,1858)(3695,1875)(3709,1893)
	(3723,1911)(3739,1930)(3755,1948)
	(3769,1964)(3798,1996)
\path(3721.586,1822.331)(3798.000,1996.000)(3632.667,1902.914)
\path(4158,1006)(4161,1009)(4168,1016)
	(4178,1028)(4193,1045)(4211,1065)
	(4230,1088)(4250,1111)(4268,1133)
	(4284,1155)(4298,1175)(4311,1194)
	(4322,1213)(4331,1231)(4339,1250)
	(4346,1268)(4351,1286)(4356,1304)
	(4360,1324)(4364,1344)(4367,1365)
	(4369,1387)(4371,1410)(4372,1433)
	(4373,1457)(4372,1481)(4371,1504)
	(4369,1528)(4367,1550)(4364,1572)
	(4360,1593)(4356,1613)(4351,1632)
	(4346,1651)(4339,1669)(4333,1687)
	(4325,1705)(4316,1723)(4306,1742)
	(4294,1762)(4281,1783)(4266,1805)
	(4250,1828)(4233,1852)(4215,1875)
	(4199,1897)(4185,1916)(4158,1951)
\path(4315.452,1845.127)(4158.000,1951.000)(4220.438,1771.831)
\path(4878,511)(6228,511)
\path(6048.000,451.000)(6228.000,511.000)(6048.000,571.000)
\put(5148,196){\makebox(0,0)[lb]{\smash{{\SetFigFontNFSS{9}{10.8}{\familydefault}{\mddefault}{\updefault}$x<\Delta \rightarrow a_2$}}}}
\put(2538,1501){\makebox(0,0)[lb]{\smash{{\SetFigFontNFSS{9}{10.8}{\familydefault}{\mddefault}{\updefault}$x\geq \Delta$}}}}
\put(3123,1141){\makebox(0,0)[lb]{\smash{{\SetFigFontNFSS{9}{10.8}{\familydefault}{\mddefault}{\updefault}$a_1^*$}}}}
\put(4473,1546){\makebox(0,0)[lb]{\smash{{\SetFigFontNFSS{9}{10.8}{\familydefault}{\mddefault}{\updefault}$x\geq \Delta$}}}}
\put(4653,1186){\makebox(0,0)[lb]{\smash{{\SetFigFontNFSS{9}{10.8}{\familydefault}{\mddefault}{\updefault}$a_2^*$}}}}
\put(3978,601){\makebox(0,0)[lb]{\smash{{\SetFigFontNFSS{9}{10.8}{\rmdefault}{\mddefault}{\updefault}$l^1$}}}}
\put(3978,2356){\makebox(0,0)[lb]{\smash{{\SetFigFontNFSS{9}{10.8}{\familydefault}{\mddefault}{\updefault}$l^*$}}}}
\put(738,421){\makebox(0,0)[lb]{\smash{{\SetFigFontNFSS{9}{10.8}{\rmdefault}{\mddefault}{\updefault}$l$}}}}
\put(3618,331){\makebox(0,0)[lb]{\smash{{\SetFigFontNFSS{9}{10.8}{\familydefault}{\mddefault}{\updefault}$x\leq \Delta$}}}}
\put(1773,151){\makebox(0,0)[lb]{\smash{{\SetFigFontNFSS{9}{10.8}{\familydefault}{\mddefault}{\updefault}$\theta \rightarrow x:=0$}}}}
\end{picture}
}}
    \caption{Player-1 actions  disabled for $\Delta$ time units ($\bot_*$ absent 
      in model).}
    \label{figure:NoRelinquish}
  \end{minipage}
  \begin{minipage}[b]{0.5\linewidth}
    \strut\centerline{\setlength{\unitlength}{0.00043745in}
\begingroup\makeatletter\ifx\SetFigFontNFSS\undefined%
\gdef\SetFigFontNFSS#1#2#3#4#5{%
  \reset@font\fontsize{#1}{#2pt}%
  \fontfamily{#3}\fontseries{#4}\fontshape{#5}%
  \selectfont}%
\fi\endgroup%
{\renewcommand{\dashlinestretch}{30}
\begin{picture}(6240,1040)(0,-10)
\put(4086,522){\ellipse{1620}{990}}
\put(818,503){\ellipse{1620}{990}}
\path(1638,511)(3258,511)
\path(3078.000,451.000)(3258.000,511.000)(3078.000,571.000)
\path(4878,511)(6228,511)
\path(6048.000,451.000)(6228.000,511.000)(6048.000,571.000)
\put(5148,196){\makebox(0,0)[lb]{\smash{{\SetFigFontNFSS{9}{10.8}{\familydefault}{\mddefault}{\updefault}$x<\Delta \rightarrow a_2$}}}}
\put(3978,601){\makebox(0,0)[lb]{\smash{{\SetFigFontNFSS{9}{10.8}{\rmdefault}{\mddefault}{\updefault}$l^1$}}}}
\put(738,421){\makebox(0,0)[lb]{\smash{{\SetFigFontNFSS{9}{10.8}{\rmdefault}{\mddefault}{\updefault}$l$}}}}
\put(3618,331){\makebox(0,0)[lb]{\smash{{\SetFigFontNFSS{9}{10.8}{\familydefault}{\mddefault}{\updefault}$x < \Delta$}}}}
\put(1773,151){\makebox(0,0)[lb]{\smash{{\SetFigFontNFSS{9}{10.8}{\familydefault}{\mddefault}{\updefault}$\theta \rightarrow x:=0$}}}}
\end{picture}
}}
    \caption{Player-1 actions  disabled for $\Delta$ time units ($\bot_*$ present
      in model).}
    \label{figure:Relinquish}
  \end{minipage}
\end{figure}
With the implicit presence of the
relinquishing action $\bot_*$, the situation can be modeled more naturally as
in Figure~\ref{figure:Relinquish} without having to add dummy sink locations and 
edges; or having to change the objective.
Note that replicating the $a_2'$ action to a similar player-1 action from $l_1$
\emph{does not} work in a $\bot_*$-less model, as we want the time at
which the $a_2'$ action is taken to be under the control of player~2.
\qed
\end{exa}

\smallskip\noindent{\bf Clock regions.}
Timed automaton games can be solved using a region construction from 
the theory of timed automata~\cite{AlurD94}.
For a real $t\ge 0$, let  $\fractional(t)=t-\floor{t}$ denote the 
fractional part of~$t$.
Given a timed automaton game $\A$, for each clock $x\in C$, let $c_x$
denote the largest integer constant that appears in any clock
constraint involving $x$ in~$\A$ (let $c_x=1$ if there is no clock
constraint involving~$x$).
Two states $\tuple{l_1,\kappa_1}$ and $\tuple{l_1,\kappa_1}$ are said to be
\emph{region equivalent} if all the following conditions are satisfied:
(a)~ $l_1 = l_2$,
(b)~ for all clocks $x$, $\kappa_1(x) \leq c_x $ iff $\kappa_2(x) \leq c_x $,
(c)~for all  clocks $x$ with $\kappa_1(x) \leq c_x $,
$\floor{\kappa_1(x)}=\floor{\kappa_2(x)}$,
(d)~for all  clocks $x,y$ with $\kappa_1(x) \leq c_x $ and $\kappa_1(y) \leq c_y $,
$\fractional(\kappa_1(x)) \leq \fractional(\kappa_1(y))$ iff
$\fractional(\kappa_2(x)) \leq \fractional(\kappa_2(y))$, and
(e)~for all  clocks $x$ with $\kappa_1(x) \leq c_x $, 
$\fractional(\kappa_1(x))=0$ iff $\fractional(\kappa_2(x))=0$.
A \emph{region} is an equivalence class of states with respect to the region equivalence
relation.
There are finitely many clock regions;
more precisely, the number of clock regions is bounded by 
$|L|\cdot\prod_{x\in C}(c_x+1)\cdot |C|!\cdot 2^{2|C|}$.

\smallskip\noindent\textbf{Representing regions}.
A region  of a timed automaton game $\A$ can be represented as  a tuple
$R=\tuple{l,h,\parti(C)}$ where
(a)~$l$ is a location of $\A$;
(b)~$h$ is a function which specifies the integral part of
clocks $ h : C \rightarrow (\nat\cap [0,M])$
($M$ is the largest constant in $\A$); and
(c)~$\parti(C)$ is a ordered disjoint partition of the clocks
 $\tuple{C_{-1},C_0,\dots C_n} $ such that $\uplus C_i = C$, with 
$C_i\neq\emptyset$
 for  $i>0$.
Then, a state $s$ with clock valuation $\kappa$ is in the region corresponding to
 $R$ when all
the following conditions hold:
(a)~the location of $s$ corresponds to the location of $R$;
(b)~for all  clocks $x$ with $\kappa(x) \leq c_x $,
$\lfloor \kappa(x) \rfloor = h(x)$;
(c)~for $\kappa(x) >  c_x$, $h(x) =c_x$;
(d)~for all pair of clocks $(x,y)$, with $\kappa(x) \leq c_x$ and
$\kappa(y)\leq c_y$, we have
$\fractional(\kappa(x)) < \fractional(\kappa(y))$ iff
$ x\in C_i \text{ and } y\in C_j \text{ with } 0\leq i<j$
(so, $x,y \in C_k$ with $k\geq 0$ implies
 $\fractional(\kappa(x)) = \fractional(\kappa(y))$);
(e)~for $\kappa(x) \leq c_x$, $\fractional(\kappa(x)) = 0$ iff $ x\in C_0$;
and
(f)~$x\in C_{-1}$ iff $\kappa(x) > c_x$.

\smallskip\noindent{\bf Region strategies and objectives.}
For a state $s\in S$, we write $\reg(s)\subseteq S$ for the clock region 
containing~$s$.
For a run $r$, we let the \emph{region sequence} $\reg(r)= \reg(r[0]),\reg(r[1]),\cdots$.
Two runs $r,r'$ are region equivalent if their region sequences are the same.
An $\omega$-regular objective $\Phi$ is a region objective if for all region-equivalent runs 
$r,r'$, we have $r\in \Phi$ iff $r'\in \Phi$.
A strategy $\pi_1$ is a \emph{region strategy}, if for all runs
$r_1$ and $r_2$ and all $k\geq 0$
such that $\reg(r_1[0..k])=\reg(r_2[0..k])$, we have that
if $\pi_1(r_1[0..k]) = \tuple{\Delta,a_1}$, then
$\pi_1(r_2[0..k]) = \tuple{\Delta',a_1}$ with 
$\reg(r_1[k]+\Delta)= \reg(r_2[k]+\Delta')$.
The definition for player~2 strategies is analogous.
Two region strategies $\pi_1$ and $\pi_1'$ are region-equivalent if for all 
runs $r$ and all $k\geq 0$ we have that if 
$\pi_1(r[0..k]) = \tuple{\Delta, a_1}$, then 
$\pi_1'(r[0..k]) = \tuple{\Delta', a_1}$ with
$\reg(r[k]+\Delta)= \reg(r[k]+\Delta')$.
A parity index function $\Omega$ is a region (resp. location)  
parity index function if
$\Omega(s_1)=\Omega(s_2)$ whenever $\reg(s_1) = \reg(s_2)$ 
(resp. $s_1,s_2$ have the same location).
Henceforth, we shall restrict our attention to region and location objectives.


\smallskip\noindent{\bf Encoding time-divergence by enlarging the game structure.}
 Given a timed automaton game $\A$, consider the enlarged game structure
$\w{\A}$ (based mostly on the construction in~\cite{AFHM+03}) 
with the state space $S^{\w{\A}} \subseteq S \times
\reals_{[0,1)}\times\set{\true,\false}^2$,
and an augmented transition relation $\delta^{\w{\A}}:
S^{\w{\A}}\times (M_1 \cup M_2) \mapsto S^{\w{\A}}$.  In an
augmented state $\tuple{s,\z,\tick,\bl_1} \in S^{\w{\A}}$, the
component $s\in S$ is a state of the original game structure
$\symb{\A}$, $\z$ is the value of a fictitious clock $z$ which gets reset to 0
every time it hits 1, 
$\tick$ is true 
iff $z$ hit 1 during the 
 last transition, and $\bl_1$ is true if player~1 is to blame 
for the last transition (i.e., $\Blfunc_1$ is true for the last transition).
Note that any strategy $\pi_i$ in $\symb{\A}$, can be considered a strategy in
$\w{\A}$.
The values of the clock $z$, $\tick$ and $\bl_1$ correspond to the values
each player keeps in memory in constructing his strategy.
Given any initial value of $\z=\z^*,\tick=\tick^*,\bl_1=\bl_1^*$;
any run $r$ in $\A$ has a corresponding unique run $\w{r}$ in 
$\w{\A}$ with $\w{r}[0]=\tuple{r[0],\z^*,\tick^*,\bl_1^*}$ such 
 that $r$ is a projection of $\w{r}$ onto $\A$. 
For an objective $\Phi$, we can now encode time-divergence as the 
objective:  
\[
\timedivbl_1(\Phi)=(\Box\Diamond \tick \rightarrow \Phi)\ \wedge\ 
(\neg\Box\Diamond\tick \rightarrow \Diamond\Box \neg\bl_1)
\]
 where
$\Box$ and $\Diamond$ are the standard LTL modalities (``always'' and
``eventually'' respectively), the combinations $\Box\Diamond$ and
$\Diamond\Box$ denoting
``infinitely often'' and ``all but for a finite number of steps''  
respectively.
This is formalized in the following  proposition.


\begin{prop}
\label{proposition:ExpandedGame}
Let $\A$ be a timed automaton game and $\w{\A}$ be the corresponding
enlarged game structure.  Let $\Phi$ be an objective on $\A$.
Consider a run $r\!=\!s^0,\tuple{m_1^0,m_2^0},
s^1,\tuple{m_1^1,m_2^1}, \dots$ in $\A$.  Let $\w{r}$ denote the
corresponding run in $\w{\A}$ such that
$\z^0=0, \tick^0=\false,\bl_1^0=\false$ and 
$\w{r}= \tuple{s^0,\z^0,\tick^0,\bl^0_1},\tuple{m_1^0,m_2^0}, 
\tuple{s^1,\z^1,\tick^1,\bl_1^1},\tuple{m_1^1,m_2^1}, \dots$.
%

\noindent Then, 
$r \in  \timedivbl_1(\Phi)\quad\hbox{iff}\quad \w{r}\in
\left((\Box\Diamond \tick \rightarrow \Phi)\ \wedge\ 
(\neg\Box\Diamond\tick \rightarrow \Diamond\Box \neg\bl_1)\right)$.
\end{prop}
\proof
Time diverges in the run $r$ iff it diverges in the corresponding run $\w{r}$.
Moreover, time diverges in $\w{r}$ iff time crosses integer boundaries 
infinitely often, i.e., $\Box\Diamond\tick$ holds.
Also, the run $\w{r}$ belongs to $\blameless_1$ iff player~1 is blamed only
finitely often, i.e., $\Diamond\Box \neg\bl_1$ holds.
\qed

The following lemma states that because of the correspondence between
$\A$ and $\w{\A}$, we can obtain the winning sets of  $\A$ by obtaining
the winning sets in $\w{\A}$.

\begin{lem}
\label{lemma:ExpandedGame}
Let $\A$ be a timed automaton game  and
$\w{\A}$ be the corresponding enlarged game
structure.  
Let $\Phi$ be an objective on $\A$. 
Any state $s$ of $\A$ satisfies
%
$s\in\win_1^{\A}(\timedivbl_1(\Phi))$ iff $\tuple{s, 0, \false,\false}\in  
 \win_1^{\w{\A}}\left((\Box\Diamond \tick \rightarrow \Phi)\ \wedge\ 
 (\neg\Box\Diamond\tick \rightarrow \Diamond\Box \neg\bl_1)\right)$.
\end{lem}
\proof
Consider a state $s$ of $\A$, and a corresponding state 
$\tuple{s, 0, \false,\false}$ of $\w{\A}$.
The variables $\z,\tick$ and $\bl_1$ only ``observe'' properties in $\w{\A}$, they
do not restrict transitions. 
Thus, given a run $r$ of $\A$ from $s$, there is a unique run $\w{r}$ of $\w{\A}$ 
from  $\tuple{s, 0, \false,\false}$ and vice versa.
Similarly, any player-$i$ strategy $\pi_i$ in $\A$ corresponds to a strategy 
$\w{\pi}_i$ in $\w{\A}$; and any strategy $\w{\pi}_i$ in $\w{\A}$ corresponds
to a strategy $\pi_i$ in $\A$ such that both strategies propose the same moves for
corresponding runs.
The result then follows from Proposition~\ref{proposition:ExpandedGame}.
\qed

\smallskip\noindent{\bf Encoding $\mathbf{\timedivbl_1(\parity(\Omega))}$ 
as  a parity
objective.}
If $\Phi$ is a parity objective, then $\timedivbl_1(\Phi)$ can be specified
as a parity objective in a related game structure $\wtd{\A}$.
The following encoding is based on a construction in~\cite{AFHM+03}.
Given $\Phi=\parity(\Omega)$ where $\Omega$ is a parity index function
of order $d$, the structure $\wtd{\A}$ has the state 
space $S^{\wtd{\A}} \subseteq S \times
\reals_{[0,1)}\times\set{\true,\false}^2\times \set{0,1,\dots,d-1}$.
Given a state $\w{s} = \tuple{s,\z,\tick,\bl_1,p}$, the set of
available moves $\Gamma_i^{\wtd{\A}}(\w{s})$ is equal to
$\Gamma_i^{\A}(s)$.
The transition relation $\delta^{\wtd{\A}}$ is specified as follows.
For $\tuple{\Delta,a_i} \in \Gamma_i^{\A}(s)$, we have
$\delta^{\wtd{\A}}(\tuple{s,\z,\tick,\bl_1,p}, \tuple{\Delta,a_i}) =$
$\tuple{s', \z',\tick', \bl_1',p'}$ where
\begin{enumerate}[$\bullet$]
\item
$s'=\delta^{\A}(s,\tuple{\Delta,a_i})$.
\item
$\z'= \fractional(\z+\Delta)$.
\item
$\tick'=\true$ iff $\z+\Delta \geq 1$.
\item
$\bl_1'=\true$ iff $i=1$ (i.e., its a player-1 move).
\item
$p'=
\left\{
\begin{array}{ll}
\max(p,\Omega(s')) & \text{if } \tick=\false \\
\Omega(s') & \text{if } \tick=\true 
\end{array}
\right.$

\end{enumerate}

The following lemma states that a parity index function $\otd$ can be defined on
$\wtd{\A}$ such that the time divergence conditions are encoded.
The proof of the lemma is technical and is presented in the appendix.
\begin{lem}
\label{lemma:ParityTimeDiv}
Let $\A$ be a timed automaton game, $\parity(\Omega)$ 
an objective on $\A$,
and  $\wtd{\A}$  the corresponding enlarged game
structure.
Consider the parity index function $\otd$ for $\wtd{\A}$ 
defined as 
\[\otd(\tuple{s,\z,\tick,\bl_1,p}) =
\left\{
\begin{array}{ll}
0 & \text{if } \tick=\bl_1=\false \\
1 & \text{if } \tick=\false, \bl_1=\true \\
p+2 & \text{if } \tick=\true
\end{array}
\right.
\]
Extend the  parity function $\Omega$ to states of  $\wtd{\A}$, 
such that the parity
 $\Omega(\tuple{s,\z,\tick,\bl_1,p})$ is the same as the parity $\Omega(s)$ in
$\A$.
Then in the game structure $\wtd{\A}$, we have
\[\timedivbl_1(\parity(\Omega)) =
\left((\Box\Diamond \tick \rightarrow \parity(\Omega))\ \wedge\ 
(\neg\Box\Diamond\tick \rightarrow \Diamond\Box \neg\bl_1)\right) = 
\parity(\otd).
\]
\qed
\end{lem}

The next lemma states that we can consider games on $\wtd{\A}$ with the parity
index function $\otd$ to obtain the winning states of $\A$ for the objective
$\parity(\Omega)$.
The proof of the lemma follows from the results of Lemma~\ref{lemma:ExpandedGame}
(the result of the  lemma also holds for the structure $\wtd{\A}$) 
and Lemma~\ref{lemma:ParityTimeDiv}.

\begin{lem}
Let $\A$ be a timed automaton game,  
$\Omega$  a parity index function on states of $\A$,
$\wtd{\A}$  the corresponding enlarged game
structure, and 
$\otd$ the parity index function on states of $\wtd{\A}$ defined
in Lemma~\ref{lemma:ParityTimeDiv}.
 Let $\Omega$ be extended to states of $\wtd{\A}$ such that
$\Omega(\tuple{s,\z,\tick,\bl_1,p}) = \Omega(s)$.
Any state $s$ of $\A$ satisfies
$s\in\win_1^{\A}(\timedivbl_1(\parity(\Omega)))$\\
 iff 
$\tuple{s, 0, \false,\false,0}\in 
 \win_1^{\wtd{\A}}\left((\Box\Diamond \tick \rightarrow \parity(\Omega))\ \wedge\ 
 (\neg\Box\Diamond\tick \rightarrow \Diamond\Box \neg\bl_1)\right)$\\
iff
$\tuple{s, 0, \false,\false,0}\in \win_1^{\wtd{\A}}(\parity(\otd))$.
\qed
\end{lem}

Let $\w{\kappa}$ be a valuation for the clocks in $\w{C}=C\cup\set{z}$.
A state of $\wtd{\A}$ can then be considered as 
$\tuple{\tuple{l,\w{\kappa}},\tick,\bl_1,p}$.
We extend the clock equivalence relation to these expanded states: 
$\tuple{\tuple{l,\w{\kappa}}, \tick,\bl_1,p}\cong 
\tuple{\tuple{l',\w{\kappa}'},\tick',\bl_1',p'}$ 
iff $l=l', \tick=\tick', \bl_1=\bl_1', p=p'$ and $\w{\kappa}\cong\w{\kappa}'$.
We let $\tuple{l,\tick,\bl_1,p}$ be the ``locations'' in $\wtd{\A}$.
If $\Omega$ is a location parity index function for $\A$, we have $\otd$ to 
be a location parity index function for $\wtd{\A}$


\medskip\noindent\textbf{A $\mathbf{\mu}$-calculus formulation for
describing the winning set.}
  A $\mu$-calculus formula $\varphi$ to describe the winning set 
  $\win_1^{\wtd{\A}}(\parity(\otd))$ is given in~\cite{AFHM+03}.
  The $\mu$-calculus formula uses the 
  \emph{controllable predecessor} operator for player~1,
  $\CPre_1 : 2^{\w{S}}\mapsto 2^{\w{S}}$ (where $\w{S}=S^{\wtd{\A}}$), 
  defined formally
  by
\[\w{s}\in \CPre_1(Z)\quad\hbox{iff}\quad
  \exists m_1\in\Gamma^{\wtd{\A}}_1(\w{s})\;
  \forall m_2\in\Gamma^{\wtd{\A}}_2(\w{s})\,.\, \delta^{\wtd{\A}}_{\jd}
  (\w{s},m_1,m_2) \subseteq Z.
\]
  Informally, $\CPre_1(Z)$ consists of the set of states from which player~1
  can ensure that the next state will be in $Z$,  no matter what player~2 does.
  The operator  $\CPre_1$ preserves regions of $\wtd{\A}$ 
  (this  follows from the results of Lemma~\ref{lemma:RegionsBeatRegions}).
  It follows from~\cite{AFHM+03} that
  given a parity index function $\w{\Omega}: \w{S} \mapsto \set{0,1,\dots,d-1}$,
  the winning set
  $\win_1^{\wtd{\A}}(\parity(\w{\Omega}))$ can be  described by the following
  $\mu$-calculus formula in case $d-1$ is odd:
  $$
  \mu X_{d-1} \cdot \nu Y_{d-2}\cdot \mu X_{d-3} \dots \nu Y_2 \cdot
  \mu X_1 \cdot \nu Y_0 \left[
    \begin{array}{c}
    {\w{\Omega}}^{-1}(0) \cap \CPre_1(Y_0) \\
    \cup\\
    {\w{\Omega}}^{-1}(1) \cap \CPre_1(X_1)\\
    \cup\\
    {\w{\Omega}}^{-1}(2) \cap \CPre_1(Y_2)\\
    \vdots\\
     {\w{\Omega}}^{-1}(d-2) \cap \CPre_1(Y_{d-2})\\
     \cup\\
     {\w{\Omega}}^{-1}(d-1) \cap \CPre_1(X_{d-1})
   \end{array}
   \right]
  $$
 where  $\mu$ and $\nu$ denote the
 least fixpoint and the greatest fixpoint operators respectively, and 
 ${\w{\Omega}}^{-1}(j)$ denotes the set of states of parity $j$.
  In case $d-1$ is even, the $\mu$-calculus formula is
  $$
  \nu Y_{d-1} \cdot \mu X_{d-2}\cdot \nu X_{d-3} \dots \mu Y_2 \cdot
  \nu X_1 \cdot \mu Y_0 \left[
    \begin{array}{c}
    {\w{\Omega}}^{-1}(0) \cap \CPre_1(Y_0) \\
    \cup\\
    {\w{\Omega}}^{-1}(1) \cap \CPre_1(X_1)\\
    \cup\\
    {\w{\Omega}}^{-1}(2) \cap \CPre_1(Y_2)\\
    \vdots\\
     {\w{\Omega}}^{-1}(d-2) \cap \CPre_1(X_{d-2})\\
     \cup\\
     {\w{\Omega}}^{-1}(d-1) \cap \CPre_1(Y_{d-1})
   \end{array}
   \right]
   $$

We now present a lemma which states that in the structure $\wtd{\A}$,
for location $\omega$-regular objectives,
(1)~memoryless region strategies suffice for winning, and 
(2)~from states in the winning set there exists a winning memoryless 
region strategy $\pi_1$  such that all strategies
region-equivalent to $\pi_1$ are also 
winning.
The proof of the lemma can be found in the appendix.

\begin{lem}
\label{lemma:RegionStrategies} 
Let $\A$ be a timed automaton game, $\Omega$ a location parity index
function on states of $\A$,  and
$\wtd{\A}$  the corresponding enlarged game
structure with the parity index function $\otd$. 
Then, (1)~there exists a memoryless region winning strategy $\pi_1$ for 
$\parity(\otd)$ from $\win_1^{\wtd{\A}}(\parity(\otd))$, and
(2)~if $\pi_1'$ is 
a strategy that is region-equivalent to  $\pi_1$,
then $\pi_1'$ is a winning strategy for 
$\parity(\otd)$ from 
$\win_1^{\wtd{\A}}(\parity(\otd))$.
\qed
\end{lem}

We say a strategy $\pi_i$ is \emph{move-independent} if for any two runs $r,r'$
such that $r[k]=r'[k]$ for all $k\geq 0$ we have 
$\pi_i(r[0..j])= \pi(r'[0..j])$ for all $j\geq 0$.
A  move-independent region strategy is a region strategy that is 
move-independent.
The following corollary follows from Lemma~\ref{lemma:RegionStrategies} observing
that  a memoryless strategy is a move independent strategy.
%
%
%
A memoryless
 strategy in $\wtd{\A}$ does  not always have a 
corresponding  memoryless strategy in $\A$.
It may not even have a move-independent strategy in $\A$.
This is because to infer the values of $\z,\tick,\bl_1$ in $\A$, we need the values
of the previous moves taken. 
The following proposition states that if player~1 has access to a global clock
then move-independent strategies suffice in $\A$ (the proof can be found in the 
appendix).

\begin{prop}
\label{proposition:MoveIndependent}
Let $\A$ be a timed automaton game such that the set of clocks includes a 
global clock $z$ that is never reset and
let $\Phi=\parity(\Omega)$ be an $\omega$-regular location objective of $\A$.
Then, move-independent strategies in $\A$ suffice for winning $\timedivbl_1(\Phi)$.
\qed
\end{prop}

The next lemma states that memoryless strategies of player~2 suffice as spoiling
strategies (the proof is in the appendix).

\begin{lem}
\label{lemma:ExpandedMemorylessPlayerTwo}
Let $\A$ be a timed automaton game, $\Omega$ a location parity index
function on states of $\A$,  and
$\wtd{\A}$  the corresponding enlarged game
structure with the parity index function $\otd$. 
Then, memoryless  region strategies of player~2 in $\wtd{\A}$ 
suffice 
for preventing player~1 from winning 
$\parity(\otd)$ from a state $\w{s}\notin\win_1^{\wtd{\A}}(\parity(\otd))$ .
\qed
\end{lem}

\section{Exact Winning of Timed Parity Games}
\label{section:Reduction}
In this section we will present a reduction of infinite-state 
timed automaton games with parity objectives to finite-state
turn-based games with parity objectives.
The reduction gives us several results related to complexity 
and algorithms to solve timed automata parity games:
(a) we obtain algorithms to solve timed automaton parity 
games with better time complexity than the algorithm presented in~\cite{AFHM+03};
(b) our reduction allows us to use the rich literature on algorithms
for finite-state parity games for solving timed automaton parity games.

\medskip\noindent\textbf{Finite-state  turn-based  games}.
 A finite-state  turn-based  game $G$ consists of the tuple 
$\tuple{(S,E),(S_1,S_2)}$, where $(S_1,S_2)$ forms a partition of the finite 
set $S$ of states,
$E$ is the set of edges, $S_1$ is the set of states from which only player~1
can make a move to choose an outgoing edge, and $S_2$ is the set of states 
from which only player~2 can make a move.
The game is bipartite if every outgoing edge from a player-1 state leads
to a player-2 state and vice-versa.

\medskip\noindent{\bf First idea of the reduction.}
Let $\A$ be a timed automaton game, and let $\wtd{\A}$ be the corresponding
enlarged timed game structure that encodes time divergence.
We shall construct a finite-state  turn-based  game structure $\A^f$
based on the regions
of $\wtd{\A}$ which can be used to compute winning states for parity
objectives for the timed automaton game $\A$.
In this finite-state  game, 
first player~1 proposes a destination
region $\w{R}_1$ together with a discrete action $a_1$
(intuitively, this can be taken to mean that in the game $\wtd{\A}$,
player~1 wants to first let
 time elapse to get to the region $\w{R}_1$, and then take the discrete action
$a_1$).
The finite-state  game then moves to an intermediate state which remembers 
the proposed
action of player~1 in the game $\wtd{\A}$.
Let us denote this intermediate state in $\A^f$ which specifies the
desired destination region and action of player~1 in  $\wtd{\A}$
 by the tuple  $\tuple{\w{R},\w{R}_1,a_1}$.
From this state in $\A^f$,  player~2 similarly also proposes a move 
consisting of a region $\w{R}_2$ together
with a discrete action $a_2$.
These two moves in $\A^f$ signify that in the game in $\wtd{\A}$ 
player~$i$ proposed a move $\tuple{\Delta_i,a_i}$
 from a state $\w{s}\in \w{R}$ such that 
$\w{s}+\Delta_i \in \w{R}_i$.
Depending on the move $\tuple{\Delta_2,a_2}$, the game in $\A^f$ will then
 proceed from $\tuple{\w{R},\w{R}_1,a_1}$
to   destination states 
$\reg\left(\delta^{\wtd{\A}}\left(\w{s},\tuple{\Delta_1,a_1}\right)\right)$, or 
$\reg\left(\delta^{\wtd{\A}}\left(\w{s},\tuple{\Delta_2,a_2}\right)\right)$, or 
both, after the move of player~2 depending on whether
$\Delta_1 <\Delta_2$, or, $\Delta_1 > \Delta_2$, or, $\Delta_1=\Delta_2$,
respectively.
The following lemma indicates that only the regions of $\w{s}+\Delta_i$ are
important in determining  whether the move of player~1 or player~2 determines
the successors after this two-step process.
The proof is technical and is presented in the appendix.

\begin{lem}
  \label{lemma:RegionsBeatRegions}
Let $\A$ be a timed automaton game, $\Omega$ a parity index function on
states of $\A$ and let $Y,Y_1',Y_2'$ be
regions in the enlarged timed game structure
$\wtd{\A}$.
Suppose player-$i$ has a move $\tuple{\Delta_i,\bot_i}$ 
from some $\w{s}\in Y$
to $\w{s}_{i}\in Y_i'$, for $i\in\set{1,2}$.
Then,  for all states $\widehat{s}\in Y$ and for all  player-1 moves
  $m_1^{\widehat{s}} = \tuple{\Delta_1,a_1}$ with
  $\widehat{s} + \Delta_1 \in Y_1'$ and $a_1\neq \bot_*$, 
one of the following cases must hold.
\begin{enumerate}[\em(1)]
\item 
  $Y_1'\neq Y_2'$ and
  for all moves $m_2^{\widehat{s}}=\tuple{\Delta_2,a_2}$  of player-2 with
  $\widehat{s} + \Delta_2 \in Y_2'$, we have $\Delta_1 < \Delta_2$ 
  (and hence $\Blfunc_1(\widehat{s},m_1^{\widehat{s}},m_2^{\widehat{s}},\widehat{\delta}
(\widehat{s},m_1^{\widehat{s}}))=\true$ and 
$\Blfunc_2(\widehat{s},m_1^{\widehat{s}},m_2^{\widehat{s}},
\widehat{\delta}(\widehat{s},m_2^{\widehat{s}}))=\false$).
\item 
  $Y_1'\neq Y_2'$ and
  for all 
  player-2  moves $m_2^{\widehat{s}}= \tuple{\Delta_2,a_2}$ with
  $\widehat{s} + \Delta_2\in Y_2'$, we have $\Delta_2 < \Delta_1$
  (and hence $\Blfunc_2(\widehat{s},m_1^{\widehat{s}},m_2^{\widehat{s}},
  \widehat{\delta}(\widehat{s},m_2^{\widehat{s}}))=\true$ and
$\Blfunc_1(\widehat{s},m_1^{\widehat{s}},m_2^{\widehat{s}},\widehat{\delta}
(\widehat{s},m_1^{\widehat{s}}))=\false$).
\item
  $Y_1'=Y_2'$ and
  there exists a  player~2 move $m_2^{\widehat{s}}= \tuple{\Delta_2,a_2}$
  with
  $\widehat{s} + \Delta_2\in Y_2'$ 
  such that $\Delta_1 = \Delta_2$ 
  (and hence 
$\Blfunc_1(\widehat{s},m_1^{\widehat{s}},m_2^{\widehat{s}},\widehat{\delta}
(\widehat{s},m_1^{\widehat{s}}))=\true$ and 
$\Blfunc_2(\widehat{s},m_1^{\widehat{s}},m_2^{\widehat{s}},
\widehat{\delta}(\widehat{s},m_2^{\widehat{s}}))=\true$).
\qed
\end{enumerate}
\end{lem}

\noindent Lemma~\ref{lemma:RegionsBeatRegions} states that given an initial state in $\w{R}$, 
for moves of both players to some fixed $\w{R}_1,\w{R}_2$, either 
the move of player~1 is always chosen, or player~2 can always pick 
a move such that player-1's move is foiled.

Let $S^{\wtd{\A}}_{\reg} = \set{ X\mid X \text{ is a region of } \wtd{\A}}$, and
let 
$S^{\dagger} = S^{\wtd{\A}}_{\reg} \times S^{\wtd{\A}}_{\reg} \times A_1^{\bot}$.
Using Lemma~\ref{lemma:RegionsBeatRegions}, a bipartite turn-based  finite game 
 $\A^f= \tuple{(S^f, E^f),(S^{\wtd{\A}}_{\reg}\times\set{1}, 
{S}^{\dagger}\times\set{2})}$ 
can be constructed to capture the timed game $\A$ as follows.
\begin{enumerate}[$\bullet$]
\item
  The state space $S^f$  is equal to
  $S^{\wtd{\A}}_{\reg}\times\set{1} \, \cup\, {S}^{\dagger}\times\set{2}$.
\item 
  Each $\tuple{\w{R},1}\in S^{\wtd{\A}}_{\reg}\times\set{1}$  encodes 
  states in the 
  timed game $\wtd{\A}$ that belongs to the region 
  $\w{R}$.
  $S^{\wtd{\A}}_{\reg}\times\set{1}$ are player-1 states.
\item 
  Each $\tuple{Y,2}\in {S}^{\dagger}\times\set{2}$  encodes the 
  following information:
  (a)~the previous state of $\A^f$ (which corresponds to a region $\w{R}$
  of $\wtd{\A}$), 
  (b)~a region $\w{R}'$ of $\wtd{\A}$ (representing an intermediate state
  which results from  time passage  in $\wtd{\A}$
  from the state in the  previous region $\w{R}$ to a state in $\w{R}'$), and
  (c)~the desired discrete action of player~1 to be taken from the
  intermediate state in $\w{R}'$.
  ${S}^{\dagger}\times\set{2}$ are player-2 states.
\item
  An edge from $\tuple{\w{R},1}$ to $\tuple{Y,2}=\tuple{\w{R},\w{R}',a_1,2}$ 
  would represent  the fact that 
  in the timed game $\wtd{\A}$, from some state $\w{s} \in \w{R}$, 
  player~1 has a move $\tuple{\Delta,a_1}$ such that $\w{s}+\Delta$ is in
  the intermediate region component $\w{R}'$ of $\tuple{Y,2}$, with $a_1$ 
  being the desired final discrete action.
  From the state $\tuple{Y,2}$, player~2 would have  moves to 
  $S^{\wtd{\A}}_{\reg}\times\set{1}$
  depending on what moves of player~2 in the timed game $\wtd{\A}$ can
  beat the player-1 moves from $\w{R}$ to $\w{R}'$ according to
  Lemma~\ref{lemma:RegionsBeatRegions}.
\end{enumerate}
The construction alluded to above requires having a state space that has size
roughly $|S^{\wtd{\A}}_{\reg}|^2\cdot |A_1|$, where $|S^{\wtd{\A}}_{\reg}|$ is the number
of regions of $\wtd{\A}$.
An optimized construction was presented in~\cite{KCHenPraB08} such that the size
of the state space was roughly 
$O(|S^{\wtd{\A}}_{\reg}|\cdot M\cdot |C|\cdot |A_1|)$ where $M$ is the largest
constant and $C$ the set of clocks in $\wtd{A}$.
The optimization was due to the observation that a state in a region $\w{R}$ does
 not have transitions to every region, but only to a few restricted ones.

\medskip\noindent\textbf{Second idea of the reduction.}
We will show  in subsection~\ref{subsection:FiniteStateReduction}
that the strategies of both players in the timed game $\wtd{\A}$ can be 
be restricted so that  from
any state $\w{s}$, each
player can only propose  moves $\tuple{\Delta_i,a_i}$ such that 
the discrete action $a_i$ 
is taken  either from the current region, or from the two following
successor regions.
That is, the cardinality of the set 
$\set{\reg(\w{s}+\Delta) \mid 0\leq \Delta \leq \Delta_i}$ is at most 3.
The winning set remains the set with strategies being restricted in this manner.
This result
allows us to restrict the size of the state space in the region pair
construction mentioned above to
$O(|S^{\wtd{\A}}_{\reg}|\cdot |A_1|)$.
A further optimization allows us to have a state space \emph{linear} in
$|S^{\wtd{\A}}_{\reg}|$.

\medskip\noindent\textbf{Outline of Section~\ref{section:Reduction}}.
In subsection~\ref{subsection:ThreeRegionReduction}, we show that the strategies of
both players can be restricted so that the edges of the timed automaton game are
taken from the current
region, or from the two following successor regions;  with the restriction
not changing the winning set
$\win_1^{\wtd{\A}}(\parity(\otd))$.
We call these restricted games \emph{3-region timed parity games}.
Then in subsection~\ref{subsection:FiniteStateReduction}, we reduce these 3-region
timed parity games to finite-state  turn-based  games, the state space of the 
turn-based game being linear in the number of regions of $\wtd{\A}$.

\subsection{Reduction to 3-Region Timed Parity Games}
\label{subsection:ThreeRegionReduction}
\subsubsection{3-Region strategies.}
We define the boolean functions 
$\succr_j^{\wtd{\A}} : S^{\wtd{\A}}\times\reals_{\geq 0}\mapsto \set{\true,\false}$
for $j\in\set{2,3}$ as
\[
\succr_j^{\wtd{\A}}(\w{s},\Delta) = \left\{
  \begin{array}{ll}
    \true & \text{ if } 
    |\set{\reg(\w{s}+\Delta') \mid 0\leq \Delta' \leq \Delta}| \leq j \\
    \false & \text{ otherwise}
  \end{array}
  \right.
\]
%
%
%
For player $i$, with $i\in \set{1,2}$, we say a strategy $\pi_i$ is a
\emph{3-region strategy} of $\wtd{\A}$ if for any run prefix $r[0..k]$, 
we have that $\pi_i$ plays a move of duration $\Delta$ such that $r[k]+\Delta$
is at most in the second following region.
Formally,
$\pi_i(r[0..k]) = \tuple{\Delta_i^k, a_i^k}$ with
$\succr_3^{\wtd{\A}}(r[k], \Delta_i^k) = \true$.
We wish to show that in the game $\wtd{\A}$, we can restrict both players to 
using only 3-region strategies.
Consider  3-region strategies of player~1. 
Informally, they suffice for winning as a)~3-region strategies allow time to 
diverge, and b)~if $\pi_1$ is a player-1 winning strategy, then we can 
obtain a winning 3-region strategy $\pi_1^*$ that plays the same moves as
$\pi_1$ whenever $\pi_1$ proposes moves within two successor regions; and plays
simple time moves to  the second
successor  region whenever $\pi_1$ plays a move to outside the second following
region.
The strategy $\pi_1^*$ works in general because for a 
run $\w{r}[0..k]$ such that $\pi_1$ proposes moves outside two
successor regions, a player-2 strategy $\pi_2$
can counter $\pi_1$ by playing  similar pure time move as $\pi_1^*$.
Unfortunately this argument does not formally work, as a player-1 move makes
the $\bl_1$ component true, and a player-2 move makes $\bl_1$ false, that is, 
 $\delta(r[k],\tuple{\Delta,\bot_1}) \neq 
\delta(r[k],\tuple{\Delta,\bot_2})$, the only difference being in the
  $\bl_1$ components.
We get around this roadblock by working in another expanded game structure 
where 
the $\bl_1$ component is true only if a move of player~1 is chosen, \emph{and
the move is either to the originating region, or to the immediately succeeding
region}.
It turns out that this modification does not change the time divergence
condition.
We present this new game structure  next.

\subsubsection{The expanded game structure $\wtdbl{\A}$} 
Analogous to the definition of $\wtd{\A}$, let $\wtdbl{\A}$ be a similar enlarged
game structure, the only difference from $\wtd{\A}$ being in the $\bl_1$ 
component (we will refer to the new $\bl_1$ component in the new game
structure as $\tbl$).
We denote the transition relation by $\delta^{\wtdbl{\A}}$, and the joint transition
relation by $\delta^{\wtdbl{\A}}_{\jd}$.
In an
augmented state $\tuple{s,\z,\tick,\tbl_1,p} \in S^{\wtdbl{\A}}$, the
component 
$\tbl_1$ is true only if  both of the following conditions are
satisfied:
\begin{enumerate}[(a)]
\item player~1 is to blame for the last
transition, and
\item
 $
\begin{array}{ll}
\text{if} &  \delta^{\wtdbl{\A}}(\tuple{s^*,\z^*, \tick^*, \tbl_1^*, p^*}, \tuple{\Delta,a_1 }) =
\tuple{s,\z,\tick,\tbl_1,p }\\
 \text{then} &
\succr_2^{\wtdbl{\A}}(\tuple{s^*,\z^*, \tick^*, \tbl_1^*,p},\Delta)=
\true.
\end{array}$

\end{enumerate}
A run $r$ of $\A$ has  corresponding unique run $\w{r}$ in $\wtd{\A}$ and
$\w{r}_3$ in 
$\wtdbl{\A}$ such 
 that $r$ is a projection of $\w{r}$ and $\w{r}_3$ onto $\A$,
given the starting values of $\z,\tick,\bl_1, \tbl_1$ and $p$.
The reverse also holds --- for any run in the  expanded game structures
starting from $\tuple{s, \z, \tick, \bl_1, p}$,
we have a corresponding unique run in $\A$ from $s$.
Observe that the available moves are the same in all game structures, and we
may view the additional components of the states in the expanded game structures
as being kept in memory by the two players in $\A$.
A similar correspondence between strategies also holds, thus a strategy $\pi_i$
in $\A$ has corresponding  matching strategies in $\wtd{\A}$ and in 
$\wtdbl{\A}$; and vice versa.

The next lemma states that time diverges in a run $\w{r}$ of $\wtdbl{\A}$ 
when the run has infinitely many winning moves  such that the
moves allow  a time elapse to a region farther than the immediate successor
region.

\begin{lem}
\label{lemma:ThreeRegionsTimeDiverge}
Consider a run $r\in \A$ such that $r=\runs$ and the corresponding
runs $\w{r}$ and $\w{r}_3$ in $\wtd{\A}$ and $\wtdbl{\A}$
respectively.  Suppose for infinitely many $k$
 we have $\succr_2^{\wtd{\A}}(\w{r}[k],\delay(m_1^k, m_2^k)) $
to be $\false$
$\bigl($or 
 $\succr_2^{\wtd{\A}}(\w{r}_3[k],\delay(m_1^k, m_2^k))$
to be $\false\bigr)$.
Then time diverges in the run $r$ (and hence also in  $\w{r},\w{r}_3$).
\end{lem}
\proof
We have that for all $k\geq 0$, 
except for the blame component,  
the remaining four  corresponding components of the 5-tuples $\w{r}[k]$ and of
$\w{r}_3[k]$ match.
Thus, 
\[\succr_2^{\wtd{\A}}(\w{r}[k],\delay(m_1^k, m_2^k)) = 
\succr_2^{\wtd{\A}}(\w{r}_3[k],\delay(m_1^k, m_2^k)).
\]
A region of $\wtd{\A}$ can be represented
as a tuple
$\w{R}=\tuple{l, \tick, \bl_1, p, \w{h}, \parti(\w{C)}}$  (similar to clock regions in $\A$) 
where (a)~$h$ is a function which specifies the integer values of
clocks  $ h : \w{C} \rightarrow (\nat\cap [0,M])$
($M$ is the largest constant in $\A$); and
(b)~$\parti(\w{C})$ is a disjoint partition of the clocks
 $\tuple{\w{C}_{-1},\w{C}_0,\dots \w{C}_n}$ such that  $ \uplus \w{C}_i = \w{C}$, 
and 
$\w{\w{C}}_i\neq\emptyset$ for   $i>0 $ 
(see Section~\ref{section:TimedGames} for details on regions and this
representation).

Consider the clock partitions $\parti^k(\w{C}) =
\tuple{\w{C}_{-1}^k,\w{C}_{0}^k,\dots \w{C}_{n^k}^k}$ of the regions
$\reg(\w{r}[k])$.
Suppose $\w{C}_{-1}^k \neq \emptyset$. 
Then, the immediate time successor of the region will have
$\w{C}_{-1}=\emptyset$, and $\w{C}_j= \w{C}^{k}_{j-1}$ for 
$n^k+1\geq j\geq 0$.
Suppose $\w{C}_{-1}^k = \emptyset$, then the immediate time successor of the 
region will have
$\w{C}_{-1} \neq \emptyset$; this happens because all the clock values
in $\w{C}_{n^k}^k$ reach an integer boundary.
Suppose $\succr_2^{\wtd{\A}}(\w{r}[k],\delay(m_1^k, m_2^k)) = \false$.
This means that $\reg(\w{r}[k] + \delay(m_1^k, m_2^k))$ is at least two
region successors away from $\reg(\w{r}[k])$.
Thus, some clock must be crossing an integer boundary (greater than 0) 
between $\runtime(\w{r},k)$ and $\runtime(\w{r}, k+1)$.
Since $\succr_2^{\wtd{\A}}(\w{r}[k],\delay(m_1^k, m_2^k)) = \false$ for
infinitely many $k$, we must have that some clock crosses an
integer boundary greater than 1 infinitely often.
Observing that at least one time unit must pass between any two such crossings,
we have that time diverges in the run $\w{r}$ (and hence also in $r, \w{r}_3$).
\qed

We next show that the change in  the $bl_1$ component in $\wtdbl{\A}$ does
not change the time divergence condition.
Given an objective $\Phi$, let $\timedivtbl_1(\Phi)$ denote the objective
$(\Box\Diamond \tick \rightarrow \Phi)\ \wedge\ 
(\neg\Box\Diamond\tick \rightarrow \Diamond\Box \neg\tbl_1)$.

\begin{lem}
\label{lemma:ThreeBlameGames}
Let $\A$ be a timed automaton game, $\Omega$ a location parity index
function on states of $\A$,  and
$\wtd{\A}$  the corresponding enlarged game.
Consider a run $r\in \A$ and the corresponding runs $\w{r}$ and $\w{r}_3$ in
$\wtd{\A}$ and $\wtdbl{\A}$ respectively.
The run $\w{r}$ belongs to the  objective
$\timedivbl_1(\parity(\Omega))=(\Box\Diamond \tick \rightarrow \parity(\Omega))\ \wedge\ 
(\neg\Box\Diamond\tick \rightarrow \Diamond\Box \neg\bl_1)$
iff the run $\w{r}_3$ belongs to  the objective
$\timedivtbl_1(\parity(\Omega))=(\Box\Diamond \tick \rightarrow \parity(\Omega))\ \wedge\ 
(\neg\Box\Diamond\tick \rightarrow \Diamond\Box \neg\tbl_1)$.
\end{lem}
\proof
We have that for all $k\geq 0$, except for the blame component,  
the remaining four  corresponding components of the 5-tuples $\w{r}[k]$ and of
$\w{r}_3[k]$ match.
Hence the membership of the run $\w{r}$ in the objective $\timedivbl_1(\parity(\Omega))$ 
differs from the membership of the run $\w{r}_3$ in the objective 
$\timedivtbl_1(\parity(\Omega))$ only when both of the following conditions
hold:
(a)~time converges on both runs (i.e., $(\Box\Diamond\tick) = \false$), and
(b)~$\bl_1$ is true infinitely often in $\w{r}$ and $\tbl_1$ is true
only finitely often in $\w{r}_3$; or vice versa.
Observe that as the first four  components of $\w{r}[k]$ and in $\w{r}_3[k]$
match, and both runs correspond to a run $r$ in $\A$, we have $\bl_1^k$ to 
be true whenever 
$\tbl_1^k $ is true.
Thus, we cannot have  $\tbl_1$ to be true infinitely often in 
$\w{r}_3$ and $\bl_1$ to be true only finitely often in $\w{r}$.
Thus, we can restrict our attention to the case where $\bl_1$ is true 
infinitely often, but $\tbl_1$ is only true  finitely often.
The $\bl_1^k$ component of $\w{r}[k]$ differs from the $\tbl_1^k$ component
in $\w{r}_3[k]$ for $k\geq 1$ only when a move 
$\tuple{\Delta^{k-1},a_1^{k-1}}$ 
of player~1 is chosen from the state $r[k-1]$ (and correspondingly from
$\w{r}[k]$ and $\w{r}_3[k]$), and we have
$\succr_2^{\wtdbl{\A}}(\w{r}_3[k-1],\Delta^{k-1}) = \false$ 
(note that
$\succr_2^{\wtdbl{\A}}(\w{r}_3[k-1],\Delta^{k-1}) =
\succr_2^{\wtd{\A}}(\w{r}[k-1],\Delta^{k-1})$).
Thus, we must have that  player-1 moves 
$\tuple{\Delta^k,a_1^k}$ were chosen from the state  $\w{r}_3[k]$ 
such that $\succr_2^{\wtd{\A}}(\w{r}_3[k],\Delta^k) = \false$ for infinitely many $k$.
Thus, from Lemma~\ref{lemma:ThreeRegionsTimeDiverge}, we have the runs 
$\w{r}_3$ and $\w{r}$
to be time divergent.
Hence, the membership of the run $\w{r}$ in the objective $\timedivbl_1(\parity(\Omega))$ 
must be the same as the  membership of the run $\w{r}_3$ in the objective 
$\timedivtbl_1(\parity(\Omega))$,
\qed

As in Lemma~\ref{lemma:ParityTimeDiv}, the objective 
$\timedivtbl_1(\parity(\Omega))$ can be expressed as a parity objective
$\parity(\otdbl)$.
Proposition~\ref{proposition:ExpandedGame} and 
Lemma~\ref{lemma:ThreeBlameGames} give us the following proposition which
states that we can consider games in $\wtdbl{\A}$ to compute winning sets in 
$\A$.

\begin{prop}
\label{proposition:ExpandedThreeRegionGame}
Consider a timed automaton game $\A$ and a location parity objective
$\Omega$ on states of $\A$ with $\wtdbl{\A}$ being the
corresponding expanded game structure.
Consider a run $r=s^0,\tuple{m_1^0,m_2^0}, s^1,\tuple{m_1^1,m_2^1}, \dots$ 
in $\A$.
Let $\w{r}_3$ be the corresponding run in $\wtdbl{\A}$ such that
$ \w{r}_3=\tuple{s^0,\z^0,\tick^0,\bl^0_1,p^0},\tuple{m_1^0,m_2^0}, 
\tuple{s^1,\z^1,\tick^1,\bl_1^1, p^1},\tuple{m_1^1,m_2^1}$ with
$\z^0=0, \tick^0=\false$, $\bl_1^0=\false$, $p^0=0$.
%
Then 
$r \in  \timedivbl_1(\parity(\Omega))$
iff $\w{r}_3\in 
 \timedivtbl_1(\parity(\Omega))$
(i.e., $\w{r}_3\in =\parity(\otdbl)$, or equivalently,
$\w{r}_3\in 
 (\Box\Diamond \tick \rightarrow \parity(\Omega))\ \wedge\ 
 (\neg\Box\Diamond\tick \rightarrow \Diamond\Box \neg\tbl_1)$.\qed
%
\end{prop}

\subsubsection{Sufficiency of 3-region strategies.}
We next show that we can restrict both players to use only 3-region
strategies in the game structure $\wtdbl{\A}$ without changing the winning
sets.
We first consider 3-region strategies of player~1.
The following lemma is a key lemma of the paper and an example is presented after
the  proof to help illustrate its workings.
\begin{lem}
\label{lemma:ThreeRegionWinningStrategies}
Consider a timed automaton game $\A$ with $\Omega$ a location parity function 
on states of $\A$ and $\wtdbl{\A}$  the
corresponding expanded game structure.
Then, 3-region memoryless  region strategies  of player~1 suffice 
for winning from $\win_1^{\wtdbl{\A}}(\timedivtbl_1(\parity(\Omega)))$.
\end{lem}
\proof
As in Lemma~\ref{lemma:RegionStrategies}, let $\pi_1$ be a player-1 winning memoryless  region strategy
 in $\wtdbl{\A}$ for
$\timedivtbl_1(\parity(\Omega))$ 
 (the results of Lemma~\ref{lemma:RegionStrategies},
 also hold for the structure $\wtdbl{\A}$).
We construct a 3-region memoryless region strategy 
$\pi_1^*$ of player~1 which wins against
all  strategies of player~2.
Intuitively, from a state $\w{s}$, the strategy  $\pi_1^*$ prescribes 
(a)~the
 same move as 
$\pi_1$ when $\pi_1$ 
prescribes a  move to either the current region or  to the two immediately succeeding
 regions
(b)~prescribes a relinquishing move $\bot_*$ when $\pi$ prescribes a relinquishing move, and
(c)~prescribes a time blocking
move to the second region following  $\reg(\w{s})$ whenever $\pi_1$ proposes a move
to outside the second succeeding region.
Formally, let $\w{r}_3$ be any run in $\wtdbl{\A}$.
The strategy $\pi_1^*$ is specified by:
\[
\pi_1^*(\w{r}_3[0..k]) = \left\{
\begin{array}{ll}
\tuple{\Delta,a_1} & \text{ if } \pi_1(\w{r}_3[0..k]) = \tuple{\Delta,a_1} 
\text{ and } \succr_3^{\wtdbl{\A}}(\w{r}_3[k], \Delta) = \true, \\
&\text{ and } a_1\neq\bot_*\\
\tuple{0,\bot_*} & \text{ if } \pi_1(\w{r}_3[0..k]) = \tuple{\Delta,\bot_*}\\
\tuple{\Delta,\bot_1} & \text{ if } 
\pi_1(\w{r}_3[0..k]) = \tuple{\Delta_1,a_1},\ a_1\neq\bot_*, 
\text{ and }\\
& \succr_3^{\wtdbl{\A}}(\w{r}_3[k], \Delta_1)= \false; 
	 \text{ where } \Delta  
	\text{ is any real number}\\
&        \text{such that }
	|\set{\reg(\w{r}_3[k]+\Delta') \mid 0\leq \Delta' \leq \Delta}| = 3.

\end{array}
\right.
\]
Note that $\pi_1^*$ is a memoryless region strategy as
$\pi_1$ is a memoryless  region strategy.

We claim $\pi_1^*$ wins against all strategies of player~2 for
every state $\w{s}\in\win_1^{\wtdbl{\A}}$.
Suppose this is not true.
Let $\pi_2^*$ be a spoiling move-independent region  strategy of player~2 
against $\pi_1^*$
from $\w{s}\in\win_1^{\wtdbl{\A}}$ (move-independent region strategies suffice
as spoiling strategies by a lemma corresponding to 
Lemma~\ref{lemma:ExpandedMemorylessPlayerTwo} for the structure 
$\wtdbl{\A}$).
Suppose $\w{r}_3^*\in\outcomes(\w{s}, \pi_1^*,\pi_2^*)$ and
$\w{r}_3^*\notin \timedivtbl_1(\parity(\Omega))$.
We show that in that case we can construct a player-2 spoiling strategy for
$\pi_1$, contrary to the assumption that $\pi_1$ was a winning strategy.
Intuitively, given a  finite run $\w{r}_3[0..k]$ the strategy $\pi_2$:
\begin{enumerate}[(1)]
\item
  Acts like 
  $\pi_2^*$ when $\pi_1$ proposes a relinquishing move.
\item
  Acts like 
  $\pi_2^*$ when $\pi_1$ proposes moves within three regions.
\item
  Acts like $\pi_2^*$ when $\pi_1$ proposes moves outside three
  regions and $\pi_2^*$ proposes moves of shorter duration than
  $\pi_1^*$ (observe that $\pi_1^*$ moves are pure time moves in case
  $\pi_1$ proposes moves outside three regions).
\item
  Proposes the same time delay moves as $\pi_1^*$ when $\pi_1$ proposes moves 
  outside three regions, and $\pi_2^*$ proposes moves longer than $\pi_1^*$; 
\item
  Proposes the same moves as $\pi_2^*$ when 
  $\pi_1$ proposes moves outside three
  regions, $\pi_2^*$ proposes moves of exactly the same duration as
  $\pi_1^*$, and the  move of $\pi_2^*$ is chosen at  $\w{r}^*_3[k+1]$.
\item
  Proposes the same time delay moves as $\pi_1^*$ 
  when 
  $\pi_1$ proposes moves outside three
  regions and when $\pi_2^*$ proposes moves of exactly the same duration as
  $\pi_1^*$ if the move of $\pi_1^*$ is chosen at $\w{r}^*_3[k+1]$.
\end{enumerate}
Formally, the player-2 strategy $\pi_2$ (dependent on the strategies
$\pi_1, \pi_1^*$ and $\pi_2^*$)  is defined to be:
\[
\pi_2(\w{r}_3[0..k]) = \left\{
\begin{array}{ll}
\pi_2^*(\w{r}_3[0..k]) & \text{ if } \pi_1(\w{r}_3[0..k]) = 
\tuple{\Delta,a_1}, \text{ and } \\
 &  a_1=\bot_*\\
\\
\pi_2^*(\w{r}_3[0..k]) & \text{ if } \pi_1(\w{r}_3[0..k]) = 
\tuple{\Delta,a_1}, \text{ and } \\
 &  \succr_3^{\wtd{\A}}(\w{r}_3[k], \Delta) = \true\\
\\
 \pi_2^*(\w{r}_3[0..k]) & \text{ if } \pi_1(\w{r}_3[0..k]) = 
 \tuple{\Delta,a_1}, a_1\neq \bot_*,  \text{ and }  \\
 & \succr_3^{\wtd{\A}}(\w{r}_3[k], \Delta) = \false; \\
 & \pi_1^*(\w{r}_3[0..k]) = \tuple{\Delta',\bot_1}, \text{ and }\\
 & \pi_2^*(\w{r}_3[0..k]) = \tuple{\Delta^*,a_2^*} \text{ with } 
 \Delta^* < \Delta' \\
\\
 \tuple{\Delta', \bot_2} & \text{ if } \pi_1(\w{r}_3[0..k]) = 
 \tuple{\Delta,a_1}, a_1\neq \bot_*, \text{ and } \\
 & \succr_3^{\wtd{\A}}(\w{r}_3[k], \Delta) = \false; \\
 & \pi_1^*(\w{r}_3[0..k]) = \tuple{\Delta',\bot_1}, \text{ and }\\
 & \pi_2^*(\w{r}_3[0..k]) = \tuple{\Delta^*,a_2^*} \text{ with } 
 \Delta^* > \Delta' \\
\\
 \pi_2^*(\w{r}_3[0..k]) &
 \text{ if } \pi_1(\w{r}_3[0..k]) = 
 \tuple{\Delta,a_1}, a_1\neq \bot_*, \text{ and}\\
 & \succr_3^{\wtd{\A}}(\w{r}_3[k], \Delta) = \false;\\
 & \pi_1^*(\w{r}_3[0..k]) = \tuple{\Delta',\bot_1}, \text{ and }\\
 & \pi_2^*(\w{r}_3[0..k]) = \tuple{\Delta^*,a_2^*} \text{ with } 
 \Delta^* = \Delta'; \text{ and} \\
 &\delta^{\wtdbl{\A}}(\w{r}_3^*[k], \pi_2^*(\w{r}^*_3[0..k])) = 
 \w{r}_3^*[k+1], \text{ and}\\
 &\delta^{\wtdbl{\A}}(\w{r}_3^*[k], \pi_1^*(\w{r}^*_3[0..k])) 
 \neq \w{r}_3^*[k+1]\\
%
%
\\
 \tuple{\Delta',\bot_2} &
 \text{ if } \pi_1(\w{r}_3[0..k]) = \tuple{\Delta,a_1}, a_1\neq \bot_*,
 \text{ and } \\
 & \succr_3^{\wtd{\A}}(\w{r}_3[k], \Delta) = \false;\\
 & \pi_1^*(\w{r}_3[0..k]) = \tuple{\Delta',\bot_1}, \text{ and }\\
 & \pi_2^*(\w{r}_3[0..k]) = \tuple{\Delta^*,a_2^*} \text{ with } 
 \Delta^* = \Delta'; \text{ and}\\
 &\delta^{\wtdbl{\A}}(\w{r}_3^*[k], \pi_1^*(\w{r}^*_3[0..k])) =
 \w{r}_3^*[k+1]
%

\end{array}
\right.
\]

Note that $\pi_2$ is a memoryless strategy (as $\pi_1, \pi_1^*, \pi_2^*$
are all memoryless).
We show that $\pi_2$ is a spoiling strategy for 
$\pi_1$ from  $\w{s}\in\win_1^{\wtdbl{\A}}$.
This is a contradiction since $\pi_1$ was assumed to be a player-1 
winning strategy.
We show there exists a run $\w{r}_3 \in \outcomes(\w{s},\pi_1, \pi_2)$ such that
$\w{r}_3[k] = \w{r}_3^*[k]$ for all $ k\geq 0$ (recall that $\w{r}_3^*$ is the
run used in defining $\pi_2$, and is such that
$\w{r}_3^*\in\outcomes(\w{s}, \pi_1^*,\pi_2^*)$ and
$\w{r}_3^*\notin \timedivtbl_1(\parity(\Omega))$).

We proceed by induction on $k$.
For $k=0$ the claim is trivially true.
Suppose the claim is true for all $j\leq k$.
Thus, we have a run $\w{r}_3$ such that $\w{r}_3[j] = \w{r}_3^*[j]$ for all 
$j\leq k$.
We show that the run $\w{r}_3[0..k]$ can be extended to $\w{r}_3[0..k+1]$ according
to $\pi_1,\pi_2$ such that $\w{r}_3[k+1]=\w{r}_3^*[k+1]$.
Informally, we have the following cases (we can ignore the moves taken
and only focus on the states in the runs $\w{r}_3[0..k]$, which is the same as
the states in the run $\w{r}^*_3[0..k]$ since $\pi_1,\pi_2,\pi_1^*,\pi_2^*$ are
all memoryless and hence move independent):
\begin{enumerate}[(1)]
\item
  The strategies $\pi_1$ and $\pi_1^*$ 
  propose a relinquishing move for the state sequence of
  $\w{r}_3[0..k]$ (which is the same as the state sequence $\w{r}^*_3[0..k]$
  by the inductive hypothesis).
  In this case the move proposed by player~2 is the same for $\pi_2$ and for 
  $\pi_2^*$,
  so the next state $\w{r}_3[k+1]$ is the same as $\w{r}^*_3[k+1]$.
\item
  For the state 
  sequence $\w{r}_3[0..k]$,
  the strategies $\pi_1$ and $\pi_1^*$ propose the same
  non-relinquish\-ing  move to either
  the current region, or the next two immediately succeeding regions.
  In this case 
  the move proposed by player~2 is the same for $\pi_2$ and for $\pi_2^*$,
  so the next state $\w{r}_3[k+1]$ is the same as $\w{r}^*_3[k+1]$.
\item
  The strategy $\pi_1$ proposes a move outside the second succeeding region,
  $\pi_1^*$ proposes a delay move to the second succeeding region, and 
  $\pi_2^*$ proposes
  a move shorter than $\pi_1^*$.
  In this case, the strategy $\pi_2$ proposes the same move as $\pi_2^*$, 
  and hence the same
  move of player~2 determines both $\w{r}_3[k+1]$ and $\w{r}^*_3[k+1]$.
  Thus, the two states are equal.
\item
  The strategy $\pi_1$ proposes a move outside the second succeeding region,
  $\pi_1^*$ proposes a delay move to the second succeeding region, and 
  $\pi_2^*$ proposes
  a move longer than $\pi_1^*$.
  In this case, the strategy $\pi_2$ proposes the same delay move as $\pi_1^*$, and 
  hence the state $\w{r}_3[k+1]$ is the same as $\w{r}^*_3[k+1]$ as
  both states are determined by delay moves of equal duration.
  The $\tbl_1$ component remains 
  $\false$ in both $\w{r}_3[k+1]$ and $\w{r}^*_3[k+1]$
  as the transition is outside the immediately succeeding region.
\item
  The strategy $\pi_1$ proposes a move outside the second succeeding region,
  $\pi_1^*$ proposes a delay move to the second succeeding region,  $\pi_2^*$ 
  proposes
  a move of exactly the 
  same duration as  $\pi_1^*$, and the move of player~2 according
  to the strategy $\pi_2^*$ determines $\w{r}^*_3[k+1]$.
  In this case $\pi_2$ behaves like $\pi_2^*$, and hence the 
  state $\w{r}_3[k+1]$ is the same as $\w{r}^*_3[k+1]$ as both are determined by 
  the same move of player~2.
\item
  The strategy $\pi_1$ proposes a move outside the second succeeding region,
  $\pi_1^*$ proposes a delay move to the second succeeding region,  $\pi_2^*$ 
  proposes
  a move of exactly the same duration as  $\pi_1^*$, and the delay
  move of player~1 
  according to $\pi_1^*$ determines the next state $\w{r}^*_3[k+1]$.
  In this case $\pi_2$ proposes a delay move of the same duration as $\pi_1^*$
  and hence the 
  state $\w{r}_3[k+1]$ is the same as $\w{r}^*_3[k+1]$ as both are determined by
  delay moves of equal duration.
  The $\tbl_1$ component 
  remains $\false$ in both $\w{r}_3[k+1]$ and $\w{r}^*_3[k+1]$
  as the transition is outside the immediately succeeding region.
\end{enumerate}
The full details can be found in the appendix. 

Thus, in all cases, we have that $\w{r}_3[0..k]$ can be extended to 
$\w{r}_3[0..k+1]$ according to $\pi_1,\pi_2$ such that 
$\w{r}_3[0..k+1]=\w{r}_3^*[0..k+1]$.
Hence, we have $\w{r}_3\in \outcomes(\w{s},\pi_1,\pi_2)$ and 
$\w{r}_3\notin\timedivtbl_1(\parity(\Omega))$ as 
$\w{r}^*_3\notin\timedivtbl_1(\parity(\Omega))$, 
a contradiction since $\pi_1$ was assumed to be a winning strategy.
Hence, we cannot have the existence of the strategy $\pi_2^*$ from
which $\w{r}_3^*$ and $\pi_2$ were derived, i.e., 
$\pi_1^*$ is a winning strategy for player~1 from $\w{s}$.
\qed

The next example illustrates the above lemma and the usefulness of the structure
$\wtdbl{\A}$.

\begin{exa}
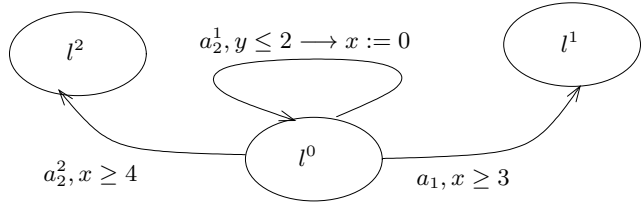
\begin{figure}[t]
\strut\centerline{\setlength{\unitlength}{0.00043745in}
\begingroup\makeatletter\ifx\SetFigFontNFSS\undefined%
\gdef\SetFigFontNFSS#1#2#3#4#5{%
  \reset@font\fontsize{#1}{#2pt}%
  \fontfamily{#3}\fontseries{#4}\fontshape{#5}%
  \selectfont}%
\fi\endgroup%
{\renewcommand{\dashlinestretch}{30}
\begin{picture}(7486,2378)(0,-10)
\put(818,1749){\ellipse{1620}{990}}
\put(6668,1860){\ellipse{1620}{990}}
\put(3612,503){\ellipse{1620}{990}}
\path(4419,509)(4421,509)(4425,509)
	(4433,509)(4445,509)(4462,510)
	(4485,510)(4513,511)(4548,512)
	(4589,513)(4635,514)(4686,515)
	(4741,517)(4800,518)(4862,520)
	(4926,523)(4990,525)(5056,528)
	(5120,531)(5184,534)(5247,537)
	(5308,540)(5367,544)(5423,548)
	(5477,552)(5529,556)(5578,561)
	(5625,566)(5669,571)(5712,576)
	(5752,582)(5790,588)(5826,594)
	(5860,601)(5893,608)(5924,616)
	(5955,624)(5984,633)(6012,642)
	(6039,651)(6074,665)(6108,679)
	(6140,695)(6172,712)(6203,730)
	(6233,750)(6263,772)(6292,795)
	(6322,820)(6352,848)(6382,877)
	(6412,909)(6443,943)(6473,979)
	(6505,1016)(6535,1054)(6566,1094)
	(6596,1133)(6624,1171)(6651,1208)
	(6675,1242)(6697,1272)(6715,1299)
	(6730,1321)(6742,1338)(6759,1364)
\path(6710.713,1180.511)(6759.000,1364.000)(6610.277,1246.180)
\path(2799,554)(2797,554)(2793,554)
	(2785,554)(2772,554)(2754,555)
	(2731,555)(2701,556)(2665,557)
	(2624,558)(2577,559)(2526,560)
	(2471,562)(2413,564)(2353,566)
	(2291,568)(2229,570)(2167,573)
	(2106,576)(2046,579)(1988,582)
	(1932,586)(1878,589)(1827,593)
	(1778,598)(1731,602)(1687,607)
	(1645,612)(1606,617)(1568,622)
	(1532,628)(1498,635)(1466,641)
	(1435,648)(1405,656)(1376,664)
	(1348,673)(1322,681)(1288,694)
	(1256,707)(1225,720)(1195,736)
	(1165,752)(1136,770)(1106,789)
	(1077,810)(1048,833)(1018,857)
	(988,884)(957,912)(926,942)
	(894,974)(862,1008)(829,1042)
	(798,1077)(766,1112)(737,1147)
	(709,1179)(683,1210)(660,1237)
	(641,1261)(625,1280)(613,1295)(594,1319)
\path(752.769,1215.114)(594.000,1319.000)(658.684,1140.630)
\path(3879,1004)(3881,1005)(3887,1007)
	(3897,1011)(3912,1018)(3933,1027)
	(3960,1038)(3992,1052)(4029,1068)
	(4069,1085)(4111,1103)(4154,1122)
	(4198,1142)(4240,1161)(4281,1180)
	(4320,1198)(4357,1215)(4391,1232)
	(4422,1248)(4450,1263)(4476,1277)
	(4500,1291)(4521,1304)(4540,1316)
	(4557,1328)(4573,1340)(4587,1352)
	(4599,1364)(4608,1374)(4617,1383)
	(4624,1393)(4631,1403)(4637,1413)
	(4642,1423)(4645,1433)(4648,1443)
	(4650,1453)(4650,1464)(4649,1474)
	(4647,1484)(4643,1495)(4638,1505)
	(4632,1515)(4624,1525)(4615,1535)
	(4605,1544)(4592,1554)(4579,1563)
	(4564,1572)(4548,1580)(4530,1589)
	(4511,1597)(4490,1604)(4468,1612)
	(4445,1618)(4421,1625)(4395,1631)
	(4368,1637)(4340,1642)(4310,1647)
	(4279,1652)(4247,1656)(4216,1660)
	(4184,1664)(4151,1667)(4116,1670)
	(4080,1672)(4043,1675)(4004,1677)
	(3964,1679)(3922,1681)(3880,1682)
	(3836,1684)(3791,1685)(3745,1685)
	(3698,1686)(3651,1686)(3603,1686)
	(3555,1685)(3506,1685)(3458,1684)
	(3409,1683)(3361,1681)(3313,1679)
	(3266,1677)(3219,1675)(3174,1672)
	(3129,1669)(3086,1666)(3044,1663)
	(3003,1660)(2963,1656)(2925,1652)
	(2888,1648)(2853,1644)(2820,1639)
	(2788,1634)(2758,1629)(2729,1624)
	(2702,1619)(2676,1613)(2651,1608)
	(2628,1602)(2606,1595)(2585,1589)
	(2566,1582)(2548,1575)(2531,1568)
	(2515,1560)(2501,1552)(2487,1544)
	(2475,1535)(2464,1527)(2455,1518)
	(2446,1509)(2439,1499)(2433,1490)
	(2428,1480)(2425,1470)(2423,1460)
	(2422,1450)(2422,1440)(2423,1430)
	(2425,1420)(2428,1410)(2432,1400)
	(2438,1391)(2444,1381)(2451,1371)
	(2458,1362)(2467,1352)(2476,1343)
	(2486,1333)(2496,1324)(2507,1315)
	(2519,1307)(2531,1298)(2544,1289)
	(2561,1278)(2579,1267)(2598,1256)
	(2619,1245)(2641,1234)(2665,1223)
	(2691,1211)(2718,1199)(2749,1186)
	(2781,1173)(2816,1159)(2853,1145)
	(2893,1130)(2935,1115)(2978,1099)
	(3023,1083)(3069,1067)(3115,1051)
	(3160,1035)(3202,1020)(3242,1007)
	(3278,995)(3308,984)(3334,976)
	(3353,969)(3384,959)
\path(3194.272,957.158)(3384.000,959.000)(3231.113,1071.363)
\put(3429,419){\makebox(0,0)[lb]{\smash{{\SetFigFontNFSS{9}{10.8}{\rmdefault}{\mddefault}{\updefault}$l^0$}}}}
\put(4824,194){\makebox(0,0)[lb]{\smash{{\SetFigFontNFSS{9}{10.8}{\rmdefault}{\mddefault}{\updefault}$a_1, x  \geq 3$}}}}
\put(414,239){\makebox(0,0)[lb]{\smash{{\SetFigFontNFSS{9}{10.8}{\rmdefault}{\mddefault}{\updefault}$a_2^2, x  \geq  4$}}}}
\put(2259,1814){\makebox(0,0)[lb]{\smash{{\SetFigFontNFSS{9}{10.8}{\rmdefault}{\mddefault}{\updefault}$a_2^1, y  \leq 2 \longrightarrow x:=0$}}}}
\put(6534,1769){\makebox(0,0)[lb]{\smash{{\SetFigFontNFSS{9}{10.8}{\rmdefault}{\mddefault}{\updefault}$l^1$}}}}
\put(684,1679){\makebox(0,0)[lb]{\smash{{\SetFigFontNFSS{9}{10.8}{\rmdefault}{\mddefault}{\updefault}$l^2$}}}}
\end{picture}
}}
\caption{A timed automaton game $\A$.}
\label{figure:ExampleOneLemma}
\end{figure}

Consider the timed automaton $\A$ in Figure~\ref{figure:ExampleOneLemma}.
Suppose the objective of player~1 is to reach the location $l^1$ starting
from $l^0$. 
Player~1 controls only the edge $a_1$.
Player~2 controls the other edges $a_2^1$ and $a_2^2$.
Player~1 wins from $l^0$ so long as $x <4$.
Let $r[0..k]$ be a run such that $\alpha_x < 4$ where 
$\alpha_x$ is the value of clock $x$ in state $r[k]$.
A winning strategy for player~1 is given by 
$$\pi_1(r[0..k]) = 
\begin{array}{ll}
  \tuple{3- \alpha_x, a_1} & \text{ if } \alpha_x \leq 3\\
  \tuple{0,a_1}  & \text{ otherwise}
  \end{array}
$$
Based on $\pi_1$, a 3-region winning strategy $\pi_1^*$ can be obtained as
in Lemma~\ref{lemma:ThreeRegionWinningStrategies}.
Intuitively, from a state $\w{s}$, the strategy  $\pi_1^*$ prescribes 
(a)~the
 same move as 
$\pi_1$ when $\pi_1$ 
prescribes a  move to either the current region or  to the two immediately succeeding
 regions, and
(b)~prescribes a time elapsing
move to the second region following  $\reg(\w{s})$ whenever $\pi_1$ proposes a move
to outside the second succeeding region.

Let the initial state be $\w{s}_0
=\tuple{l^0, z=x=y=0, \tick=\false, \tbl_1=\false}$ (we skip the parity
component for simplicity).
We consider a sample play from this state.
Initially, $\pi_1^*$ proposes the move $\tuple{1,\bot_1}$ from $\w{s}_0$.
Suppose it is allowed by player~2.
The resulting state is then 
$\w{s}_1=\tuple{l^0, z=0,x=y=1, \tick=\true, \tbl_1=\false}$.
In the second step, $\pi_1^*$ again proposes the 
move $\tuple{1,\bot_1}$ from $\w{s}_1$.
the resulting state is 
$\w{s}_2=\tuple{l^0, z=0,x=y=2, \tick=\true, \tbl_1=\false}$.
In the third step, $\pi_1^*$ 
 proposes the move $\tuple{1,a_1}$ which is blocked by the
player-2 move $\tuple{0,a_2^1}$, leading to the state
$\w{s}_3=\tuple{l^0, z=0,x=0, y=2, \tick=\false, \tbl_1=\false}$.
Player~2 can then take the action $a_2^1$ a finite number of times, but eventually
it must let time elapse. 
When it does, player~1 is able to take the action $a_1$ in three steps.

Now we show that there exists a player-2 strategy $\pi_2$  such that $\pi_1$ and
$\pi_2$ result in the same sequence of states
$\w{s}_0, \w{s}_1, \w{s}_2, \dots$.
The strategy $\pi_2$ proposes a time elapsing move $\tuple{1,\bot_2}$ from
$\w{s}_0$ and then again from $\w{s}_1$.
It then proposes $\tuple{0,a_2^1}$ from $\w{s}_2$ 
as against $\pi_1^*$, and then again
$\tuple{1,\bot_2}$ for one more step.
It can be verified that this leads to the same sequence of states as with $\pi_1^*$.

Suppose we had been working in the structure $\wtd{\A}$.
Then the states $\w{s}_1$ and $\w{s}_2$ would have had $\bl_1=\true$.
Thus, in reaching $\l^1$, the strategy $\pi_1^*$ would have resulted in some
states having $\bl_1=\true$, in contrast to the strategy $\pi_1$ which
manages to have $\bl_1=\false$ until $l^1$ is reached.
Thus $\pi_1$ and $\pi_1^*$ would on the surface appear to be inherently incompatible
with respect to objectives.
In this example however, it can be seen that we may as well work with $\tbl_1$
as player-1 time-elapsing moves can be  chosen only finitely often, thus
$\tbl_1$ and $\bl_1$ differ only finitely often.
\qed


\end{exa}

Next, we show that 3-region memoryless strategies of player~2 suffice as 
spoiling strategies.
An example will be presented after the lemma to illustrate the sufficiency of
3-region strategies.
\begin{lem}
\label{lemma:ThreeRegionSpoilingStrategies}
Consider a timed automaton game $\A$ and a location parity objective
$\Omega$ on states of $\A$ with $\wtdbl{\A}$ being the
corresponding expanded game structure.
Suppose 
$\w{s}\notin \win_1^{\wtdbl{\A}}(\timedivtbl_1(\parity(\Omega)))$.
Then, 3-region memoryless region strategies of player~2 suffice for
preventing  player~1 from satisfying the objective
$ \timedivtbl_1(\parity(\Omega))$ from $\w{s}$.
\end{lem}
\proof
  The objective $\timedivtbl_1(\parity(\Omega))$ 
  corresponds to the parity objective $\parity(\otd)$ by 
  Lemma~\ref{lemma:ParityTimeDiv} (replacing $\bl_1$ by $\tbl_1$).
  As mentioned in Section~\ref{section:TimedGames}, 
  there exists a game
  $\mu$-calculus formula which precisely characterizes the winning set
  $\win_1^{\wtdbl{\A}}(\parity(\otd))$.
  The winning sets can be obtained by a $\mu$-calculus iteration.
  The  iteration uses the 
  \emph{controllable predecessor} operator for player~1,
  $\CPre_1 : 2^{\w{S}}\mapsto 2^{\w{S}}$ (where $\w{S}$ is the state space of
  $\wtdbl{\A}$), defined formally
  by $ \w{s}\in \CPre_1(Z)$ iff
  $\exists m_1\in\Gamma^{\wtdbl{\A}}_1(\w{s})\;
  \forall m_2\in\Gamma_2^{\wtdbl{\A}}(\w{s})\,.\, \delta^{\wtdbl{\A}}_{\jd}
  (\w{s},m_1,m_2) \subseteq Z$.
  Informally, $\CPre_1(Z)$ consists of the set of states from which player~1
  can ensure that the next state will be in $Z$,  no matter what player~2 does.
  It can be shown that $\CPre_1$ preserves regions of $\wtdbl{\A}$ using 
  Lemma~\ref{lemma:RegionsBeatRegions}.
  The iteration also suggests winning strategies for player~1 based on the sets
  that arise in the iteration.
  The sets that arise  depend on the parity labeling, the 
  $\CPre_1$ and the fixpoint operators.
  If we have a location parity objective, it can be shown that only unions of
  regions arise as sets in the fixpoint iteration.
  To prove that 3-region memoryless region strategies of player~2 suffice
  as spoiling strategies, it hence suffices to show that the $\CPre_1$ sets of
  unions of regions remains unchanged if we restrict player-2 strategies to be
  3-region memoryless region strategies.


 Let $\CPre_{1,3}(Z)$ denote the set of states from which player~1
 can ensure that the next state will be in $Z$,  no matter what move
 player~2 takes within 3 regions.
 Also, clearly $\CPre_{1,3}(Z)$ depends only on $Z$, and not on the history of
 the game (hence we shall have memoryless 3-region spoiling strategies).
 We show $\CPre_{1,3}(Z) = \CPre_{1}(Z)$ for $Z$ a union of regions.
 Clearly $\CPre_{1}(Z) \subseteq \CPre_{1,3}(Z)$ as player~2 has fewer moves
 to counter with in $\CPre_{1,3}(Z)$.
 To prove the other direction, we show if $\w{s}\in \CPre_{1,3}(Z)$, then
 $\w{s}\in \CPre_{1}(Z)$.
 We first characterize the $\CPre_1$ sets.
 A state $\w{s}  \in \CPre_1(Z)$ iff \emph{either} one of the following
 conditions is met:
 \begin{enumerate}
   \item
     $\set{\delta(\w{s}, \tuple{\Delta,a_2}) \mid \tuple{\Delta,a_2} \in
       \Gamma_2(\w{s})} \subseteq Z$.
   \item
     There exists $\tuple{\Delta,a_1}\in  \Gamma_1(\w{s})$ with 
     $a_1\neq \bot_*$ such that
     \begin{enumerate}
     \item
       $\delta(\w{s},\tuple{\Delta,a_1})\in Z$, and 
     \item
       $\set{\delta(\w{s},\tuple{\Delta',a_2}) \mid \Delta' \leq \Delta,
         \text{ and }\tuple{\Delta',a_2} \in
         \Gamma_2(\w{s})}
       \subseteq Z $.
     \end{enumerate}

   \end{enumerate}
   The first condition corresponds to the case when player~1 proposes a move
   $\tuple{\Delta,\bot_*}$ from $\w{s}$.
   In this case, the move of player~2 will be chosen, no matter the move.
   Thus, we must have that no matter the move of player~2, the resultant state
   must be in $Z$.
   The second condition corresponds to the case when player~1 proposes a move
   $\tuple{\Delta,a_1}$ from $\w{s}$ with $a_1\neq \bot_*$.
   In this case, every move $\tuple{\Delta',a_1}$ of player~2 with 
   $\Delta'\leq \Delta$ must lead to $Z$.
   A similar characterization exists for $\CPre_{1,3}(Z)$, the only difference
   being that player-2 moves are restricted to be within 3 regions: 
   a state $\w{s}  \in \CPre_{1,3}(Z)$ iff \emph{either} one of the following
   conditions is met:
   \begin{enumerate}
   \item
     $\set{\delta(\w{s}, \tuple{\Delta,a_2}) \mid \tuple{\Delta,a_2} \in
       \Gamma_2(\w{s}) \text{ and } \succr_3(\w{s},\Delta)=\true}
     \subseteq Z$.
   \item
     There exists $\tuple{\Delta,a_1}\in  \Gamma_1(\w{s})$ with 
     $a_1\neq \bot_*$ such that
     \begin{enumerate}
     \item
       $\delta(\w{s},\tuple{\Delta,a_1})\in Z$, and 
     \item
       $\set{\delta(\w{s},\tuple{\Delta',a_2}) \mid \Delta' \leq \Delta,\ 
         \succr_3(\w{s}^*,\Delta')=\true, 
         \text{ and }\tuple{\Delta',a_2} \in
         \Gamma_2(\w{s})}
       \subseteq Z $.
     \end{enumerate}
   \end{enumerate}
   Now we show that if $\w{s}\in \CPre_{1,3}(Z)$, then
   $\w{s}\in \CPre_{1}(Z)$.
   Suppose $\w{s}\in  \CPre_{1,3}(Z)$.
   Informally, we can have the following cases (the formal details are in the
   appendix):
   \begin{itemize}
   \item
     The available moves of player~2 are only until the second immediately 
     succeeding region.
     In this case, the restriction of player-2 strategies to be 3-region
     strategies has no effect, hence $\w{s} \in \CPre_1(Z)$.
   \item
     Player~2 has a move from $\w{s}$ to
     a state that is farther than the second succeeding region.
     This means that it has a delay move to the second succeeding region.
     In this case, player~1 can play the same
     delay move to the second succeeding region.
     The resulting state will be the same as 
     due to the delay move of player~2 to the
     second succeeding region, because $\tbl_1$ will be false in both cases.
     Moreover, all moves of player~2 which lie in the current or the next 
     two regions
     result in a state in $Z$ by assumption.
     In particular, the delay move of player~1 will also result in a state in $Z$.
     Thus, $\w{s} \in \CPre_1(Z)$.
   \end{itemize}
   Thus, in all cases $\w{s}\in\CPre_1(Z)$ whenever $\w{s}\in\CPre_{1,3}(Z)$.
\qed

\begin{exa}
Consider the timed automaton game in Figure~\ref{figure:ExampleOneLemma} and
the following regions.
\begin{align*} 
\w{R}_5 & =  \set{
\tuple{l^1, z=0, x=3, y, \tick = \true, \tbl_1=\false}\mid y >2}\\
 \w{R}_4 & = \set{\tuple{l^0, z=0, x=y=1, \tick = \true, \tbl_1=\false}}\\
\w{R}_3 & = \set{\tuple{l^0, z=x=y, \tick = \false, \tbl_1=\false} \mid
   0< x <1}\\
\w{R}_2 &= \set{\tuple{l^0, z=0, x=0, y=1, \tick = \true, \tbl_1=\false}}\\
\w{R}_1  & = \set{\tuple{l^0, z=y, x=0, \tick = \false, \tbl_1=\false} \mid
   0< y <1}\\
 \w{R}_0 & =  \set{\tuple{l^0, z=0, x=y=0, \tick = \false, \tbl_1=\false}}
\end{align*}
The regions can be seen as the thick lines and dots in 
Figure~\ref{figure:ExampleTwoLemma} (the clock $z$ is not shown for simplicity).
Let $Z= \bigcup_{j=0}^5 \w{R}_j$.
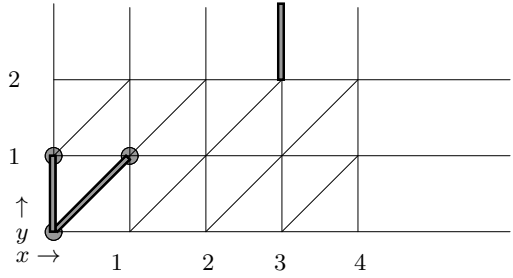
\begin{figure}[t]
\strut\centerline{\setlength{\unitlength}{0.00043745in}
\begingroup\makeatletter\ifx\SetFigFontNFSS\undefined%
\gdef\SetFigFontNFSS#1#2#3#4#5{%
  \reset@font\fontsize{#1}{#2pt}%
  \fontfamily{#3}\fontseries{#4}\fontshape{#5}%
  \selectfont}%
\fi\endgroup%
{\renewcommand{\dashlinestretch}{30}
\begin{picture}(5967,3285)(0,-10)
\texture{55888888 88555555 5522a222 a2555555 55888888 88555555 552a2a2a 2a555555 
	55888888 88555555 55a222a2 22555555 55888888 88555555 552a2a2a 2a555555 
	55888888 88555555 5522a222 a2555555 55888888 88555555 552a2a2a 2a555555 
	55888888 88555555 55a222a2 22555555 55888888 88555555 552a2a2a 2a555555 }
\put(555,537){\shade\ellipse{200}{200}}
\put(555,537){\ellipse{200}{200}}
\put(1455,1437){\shade\ellipse{200}{200}}
\put(1455,1437){\ellipse{200}{200}}
\put(555,1437){\shade\ellipse{200}{200}}
\put(555,1437){\ellipse{200}{200}}
\path(555,2337)(5955,2337)
\path(1455,3192)(1455,537)
\path(2355,3192)(2355,537)
\path(1455,537)(3255,2337)
\path(2355,537)(4155,2337)
\path(3255,537)(4155,1437)
\path(4155,3192)(4155,537)
\path(555,1437)(1455,2337)
\path(555,1437)(5955,1437)
\path(555,537)(2355,2337)
\thicklines
\shade\path(555,582)(1410,1437)(1455,1392)
	(600,537)(555,582)(555,582)
\path(555,582)(1410,1437)(1455,1392)
	(600,537)(555,582)(555,582)
\thinlines
\path(555,3237)(555,537)(5955,537)
\thicklines
\shade\path(510,1437)(577,1437)(577,537)
	(510,537)(510,1437)
\path(510,1437)(577,1437)(577,537)
	(510,537)(510,1437)
\thinlines
\path(3255,3192)(3255,537)
\thicklines
\shade\path(3210,3237)(3278,3237)(3278,2337)
	(3210,2337)(3210,3237)
\path(3210,3237)(3278,3237)(3278,2337)
	(3210,2337)(3210,3237)
\put(105,762){\makebox(0,0)[lb]{\smash{{\SetFigFontNFSS{9}{10.8}{\familydefault}{\mddefault}{\updefault}$\uparrow$}}}}
\put(2310,87){\makebox(0,0)[lb]{\smash{{\SetFigFontNFSS{9}{10.8}{\familydefault}{\mddefault}{\updefault}$2$}}}}
\put(3165,87){\makebox(0,0)[lb]{\smash{{\SetFigFontNFSS{9}{10.8}{\familydefault}{\mddefault}{\updefault}$3$}}}}
\put(15,2247){\makebox(0,0)[lb]{\smash{{\SetFigFontNFSS{9}{10.8}{\familydefault}{\mddefault}{\updefault}$2$}}}}
\put(15,1347){\makebox(0,0)[lb]{\smash{{\SetFigFontNFSS{9}{10.8}{\familydefault}{\mddefault}{\updefault}$1$}}}}
\put(4110,87){\makebox(0,0)[lb]{\smash{{\SetFigFontNFSS{9}{10.8}{\familydefault}{\mddefault}{\updefault}$4$}}}}
\put(1230,87){\makebox(0,0)[lb]{\smash{{\SetFigFontNFSS{9}{10.8}{\familydefault}{\mddefault}{\updefault}$1$}}}}
\put(105,177){\makebox(0,0)[lb]{\smash{{\SetFigFontNFSS{9}{10.8}{\familydefault}{\mddefault}{\updefault}$x\rightarrow$}}}}
\put(105,447){\makebox(0,0)[lb]{\smash{{\SetFigFontNFSS{9}{10.8}{\familydefault}{\mddefault}{\updefault}$y$}}}}
\end{picture}
}}
\caption{Regions of the  timed automaton game $\A$ of 
Figure~\ref{figure:ExampleOneLemma}.}
\label{figure:ExampleTwoLemma}
\end{figure}

The state $\w{s}_0 = \tuple{l^0, z=0, x=y=0, \tick = \false, \tbl_1=\false}$
belongs to $\CPre_{1,3}(Z)$.
This can be seen as follows.
From state $\w{s}_0$, player~1 proposes the move $\tuple{3,a_1}$.
Player~2 can only propose a counter move $\tuple{\Delta, a^*}$ such that
$0\leq\Delta\leq 1$ and $a^*\in \set{\bot_2, a_2^1}$.
Now observe that $\w{s}_0$ also belongs to $\CPre_{1}(Z)$.
To see this, we have to consider a different player-1 move, namely 
$\tuple{1,\bot_1}$.
If player~2 allows this move of player~1, the next state will be in $\w{R}_4$
(the variable $\tbl_1$ remains false as the allowed move of player~1 is not in
the immediate successor region of $\w{s}_0$).
If player~2 instead proposes a shorter (or equal) duration move, the next state will
be in $\bigcup_{j=0}^4 \w{R}_j$.

Let $Z^*$ be any set of states which includes $\w{R}_5$.
Suppose $\w{s}_0 \in \CPre_{1,3}(Z^*)$ or $\w{s}_0 \in \CPre_{1}(Z^*)$.
Then it can be seen that $Z^*$ must contain the states from $Z$.
Hence, we have $\w{s}_0 \in \CPre_{1,3}(Z^*)$ iff $\w{s}_0 \in \CPre_{1}(Z^*)$.
In general, 3-region strategies of player~2 suffice as if there exists a move of
player~2 beyond the two immediately succeeding regions, then \emph{both} player~1 and 
player~2 have simple time passage moves to the second succeeding region, with
the variable $\tbl_1$ remaining false if any of these two moves is chosen.
To mimic a general player-2 counter move, a 
3-region strategy of player~2 can play simple time passage moves to the 
second successor regions till it gets to a state from which a desired 
player-2 edge lies within two successor regions.
\qed
\end{exa}

Given an objective $\parity(\Omega)$, let $\win_{1}^{3,\widehat{\A}_3}
(\parity(\Omega))$ denote the states in the expanded game structure
$\widehat{\A}_3$ such that for all states
$\w{s}\in\win_{1}^{3,\widehat{\A}_3} (\parity(\Omega))$, player~1 has
a 3-region strategy $\pi_1$ such that for all 3-region strategies
$\pi_2$ of player~2, we have $\outcomes(\w{s},\pi_1,\pi_2) \subseteq
\parity(\Omega)$.  The following theorem, which follows from
Lemmas~\ref{lemma:ThreeRegionWinningStrategies},
\ref{lemma:ThreeRegionSpoilingStrategies}, and
\ref{lemma:ThreeBlameGames} states that to solve for
$\timedivtbl_1(\parity(\Omega))$ in $\wtdbl{\A}$, we can restrict both
players to use only 3-region strategies.  We call such games
\emph{3-region parity games}.

\begin{thm}
\label{theorem:ThreeRegionGamesSuffice}
Consider a timed automaton game $\A$ and a location parity objective
$\Omega$ on states of $\A$ with $\wtdbl{\A}$ being the
corresponding expanded game structure.
We have 
\begin{flalign*}
\win_1^{3, \wtdbl{\A}}(\timedivtbl_1(\parity(\Omega))) &= \ 
\win_1^{\wtdbl{\A}}(\timedivtbl_1(\parity(\Omega))) &\\
&=\ 
\win_1^{\wtdbl{\A}}(\timedivbl_1(\parity(\Omega))).  &
\end{flalign*}
\qed
\end{thm}

\subsection{Reduction from 3-Region Parity Games to Finite-State 
Turn-Based  Parity Games}
\label{subsection:FiniteStateReduction}

We gave a rough construction 
of a bipartite turn-based  game to capture the winning sets
of timed parity games at the beginning of this section .
The idea of the construction being that first player~1 proposes a move
representing its intention to let time pass to get to an
intermediate region of $\wtdbl{\A}$ 
and then take a discrete action from that region.
This intermediate region together with the action and the originating region
 can be encoded as a 
player-2 state in the finite game as $\tuple{\w{R},\w{R}', a_1}$ with
$\w{R}$ denoting the originating region, $\w{R}'$ representing the
intermediate region to let time pass to, and $a_1$ the player-1 action to take
from $\w{R}'$.
Note that this player-2 state $\tuple{\w{R},\w{R}', a_1}$ corresponds
to an intermediate step in the game  $\wtdbl{\A}$, where from a state
$\w{s}\in\w{R}$ player~1 has just \emph{proposed} its move (say 
$\tuple{\Delta_1,a_1}$ with the time delay $\Delta$
 leading to the region $\w{R}'$ from which the discrete action $a_1$ will be
taken.
The game of $\wtdbl{\A}$ is still in the state $\w{s}$, waiting for the move
from player~2.
In the finite-state  game, from the player-2 state 
$\tuple{\w{R},\w{R}', a_1}$,
player~2 then takes an action
depending on whether it would allow the previously
proposed player-1 move from the originating region $\w{R}$ 
in $\wtdbl{\A}$ or not.
This corresponds to the move of player~2 in $\wtdbl{\A}$ where it allows or 
disallows the player-1 move $\tuple{\Delta_1,a_1}$ from $\w{s}$ (hence the
element of ``surprise'' from~\cite{AFHM+03} remains in this finite-state game).
From Theorem~\ref{theorem:ThreeRegionGamesSuffice}, we can consider parity
games where each player is restricted to use only 3-region strategies.
This allows us to restrict the pairs of regions that can occur as player-2
states in the finite game.

\medskip\noindent\textbf{Outline of 
  subsection~\ref{subsection:FiniteStateReduction}.}
We first present a finite-state  turn-based  game $\A^f_{\Omega}$  which can be used 
to compute the winning sets of $\A$.
The size of the state space will be $O(|S_{\reg}^{\wtdbl{\A}}|\cdot |A_1|)$ where
$S_{\reg}^{\wtdbl{\A}}$ is the set of regions in $\wtdbl{\A}$ and $A_1$ is the
set of actions of player~1 in $\A$.
We then present a finite-state  turn-based  game ${\A^f_{\Omega}}^*$ such that 
(1)~the 
winning set of ${\A^f_{\Omega}}^*$ corresponds to that for $\A^f_{\Omega}$, and
(2)~the size of the state space of  ${\A^f_{\Omega}}^*$ is 
$O(|S_{\reg}^{\wtdbl{\A}}|)$.

\subsubsection{Construction of the finite turn-based  bipartite game $\A^f_{\Omega}$.}
Given a timed automaton $\A$, and a location parity index function $\Omega$
on states of $\A$, the bipartite finite turn-based  game $\A^f_{\Omega}$ 
consists of a tuple 
$\tuple{S^f, E^f,S_1^f,S_2^f}$ 
where,
\begin{enumerate}[$\bullet$]
\item
  $S^f = S_1^f\,\cup\,S_2^f$ is the state space.
  The states in $S_i^f$ are controlled by player-$i$ for $i\in\set{1,2}$.
\item
  $S_1^f= \w{S}_{\reg}\times\set{1}$, where
  $\w{S}_{\reg}$ is the set of regions in $\wtdbl{\A}$.
\item
  $S_2^f=\w{S}_{\tup}\times\set{2}$, where
  $\w{S}_{\tup} = \w{S}_{\reg}\times\set{0,1,2}\times A_1^{\bot}$.
\item
  $E^f$ contains the following edges:
  \begin{enumerate}[$-$]
    \item
      For $s_1^f = \tuple{\w{R},1} \in S_1^f$, the set $E^f$ contains
      the outgoing  edges 
      $(\tuple{\w{R},1}, \tuple{\w{R}, j, a_1,2} )$ such that
      in the structure $\wtdbl{\A}$, there exists $\w{s}\in \w{R}$ and
      $\tuple{\Delta,a_1}\in\Gamma_1^{\wtdbl{\A}}(\w{s})$ with 
      $|\set{\reg({\w{s}+\Delta'})\mid \Delta'\leq \Delta}| = j+1$.
      These types of edges encode in $\A^f$ the fact that in the game structure
      $\wtdbl{\A}$ there exists a 
      state in $\w{R}$ such that player~1 has a move such that time
      passes to the $j$-th following region from where the discrete action 
      $a_1$ is taken.
    \item
      For $s_2^f = \tuple{\w{R}, j, a_1,2}$ in
      $S_2^f$, the set $E^f$ contains the outgoing edges
      $(\tuple{\w{R}, j, a_1,2},\tuple{\w{R}',1})$ such that in
      the structure $\wtdbl{\A}$, there exists $\w{s}\in \w{R}$,
      $\tuple{\Delta_1,a_1}\in\Gamma_1(\w{s})$ and 
      $\tuple{\Delta_2,a_2}\in\Gamma_2(\w{s})$ with 
      $|\set{\reg(\w{s}+\Delta')\mid \Delta'\leq \Delta_1}| = j+1$
      such that either
      \begin{enumerate}[(1)]
      \item
        $a_1 \neq \bot_*$, \;
        $|\set{\reg(\w{s}+\Delta')\mid \Delta'\leq \Delta_2}| \leq j+1$ and
        $\delta(\w{s},\tuple{\Delta_2, a_2}) \in \w{R}'$.\\
        This edge corresponds to the case when player~2 does not allow a 
        player-1 move $\tuple{\Delta_1,a_1}$ from a state $\w{s}\in\w{R}$ by
        proposing its own move $\tuple{\Delta_2,a_2}$ such that
        $\Delta_2\leq \Delta_1$.
      \item
        $a_1 \neq \bot_*$, \;
        $3\geq |\set{\reg(\w{s}+\Delta')\mid \Delta'\leq \Delta_2}| \geq j+1$
        and $\delta(\w{s}, \tuple{\Delta_1,a_1}) \in \w{R}'$.\\
        This edge corresponds to the case when player~2  allows a 
        player-1 move $\tuple{\Delta_1,a_1}$ from a state $\w{s}\in\w{R}$ by
        proposing a move of duration
        $\Delta_2 \geq \Delta_1$.
      \item
        $a_1=\bot_*$ and
        $\delta(\w{s},\tuple{\Delta_2, a_2}) \in \w{R}'$.\\
        This edge corresponds to the case when player~1 plays a $\bot_*$ move
        so the move of player~2 is chosen.
        
      \end{enumerate}

  \end{enumerate}
      Note that we can pick any states in $\w{R}$ and $\w{R}'$ 
      to check the satisfiability
      of the above conditions by Lemma~\ref{lemma:RegionsBeatRegions}.

\end{enumerate}

\noindent In the construction for $\A^f_{\Omega}$ above, the states in 
$\w{S}_{\tup}\times\set{2}$ contain
player-1 actions as components.
The actions are used only for determining the destination states, so instead of
player-1 actions as components of states in $\w{S}_{\tup}$ we can as well use 
a tuple from $L\times 2^C$, with $\tuple{l,\alpha}$ denoting an action
in $\A$ such that the destination location is $l$, and which resets the clocks
in $\alpha$.

Each $s^f\in S^f$ is  a tuple, with the first component being a region
of $\A$.
Given the location parity index function $\otdbl$  on $\wtdbl{\A}$, 
we let $\Omega^f$ be
the parity index function on $\A^f_{\Omega}$ such that $
\Omega^f(\tuple{\w{R},\cdot})
=\otdbl(\w{s})$ where $\w{s}$ is any state in the region $\w{R}$ (all states
in a region have the same parity as $\otdbl$ is a location parity index
function on $\wtdbl{\A}$).
Note that if the parity index function $\Omega$ is of order $d$, then
$\otdbl$ and $\Omega^f$ are of order $d+2$.
Given a set $X=X_1\times\set{1}\, \cup\, X_2\times\set{2}\subseteq S^f$, we let
$\regstates(X)=\set{\w{s}\in S^{\wtdbl{\A}} \mid \reg(\w{s})\in X_1}$.
Theorem~\ref{theorem:Reduction} shows that the turn-based  game
$\A^f_{\Omega}$ captures the  game $\wtdbl{\A}$.
\begin{thm}
\label{theorem:Reduction}
Let $\A$ be a timed automaton game, $\Omega$ a location parity index
function on states of $\A$, with
$\wtdbl{\A}$ being  the corresponding enlarged game,
and  $\A^f$  the 
corresponding finite game structure with the parity index function $\Omega^f$.
Then, we have
\[\win_1^{3,\wtdbl{\A}}(\timedivtbl_1(\parity(\Omega))) = 
\regstates(\win_1^{\A^f}(\parity(\Omega^f))).
\]
\end{thm}
\proof
Recall the $\mu$-calculus formulation for obtaining the winning set 
\[\win_1^{3,\wtdbl{\A}}(\timedivtbl_1(\parity(\Omega)))\] in $\wtdbl{\A}$.
We use the $\Pre_1$ operator in turn-based  games:
\[\Pre_1(X) = 
\begin{array}{c}
  \set{s\in \w{S}_{\reg}\times\set{1} \mid \exists s'\in X
    \text{ such that } 
    (s,s')\in E^f}\\
  \cup\\ 
  \set{s\in \w{S}_{\tup}\times\set{2} \mid \forall  (s,s')\in E^f
    \text{ we have } s'\in X}
\end{array}
\]
Given $X=X_1\times\set{1}\cup X_2\times\set{2}\subseteq S^f$, we have
\begin{equation}
  \label{equation:PreOne}
\regstates\left(\Pre_1^{\A^f}\left(\Pre_1^{\A^f}\left(X\right)\right)\right) = 
 \regstates\left(\Pre_1^{\A^f}\left(\Pre_1^{\A^f}\left(X_1\times \set{1}\right)\right)\right) 
 \end{equation}
 This is because
the structure of the games $\A^f_{\Omega}$ is such that
a state with $1$ as the last component of the tuple can reach 
$X=X_1\times\set{1}\,\cup\,X_2\times\set{2}$ in exactly two
steps iff it can reach $X_1\times\set{1}$ in exactly two steps (the game is 
bipartite).

From  Lemma~\ref{lemma:RegionsBeatRegions}, it  follows that
\begin{equation}
\label{equation:PreCPre}
\regstates\left(\Pre_1^{\A^f}\left(\Pre_1^{\A^f}\left(X_1\times \set{1}\right)\right)\right) 
=\CPre_1^{\wtdbl{\A}}\left(\regstates\left(X_1\times\set{1}\right)\right)
\end{equation}
Let $\phi_c$ be the $\mu$-calculus formula using the $\CPre_1$ operator
describing the  
winning set for $\parity(\otdbl)=\timedivtbl_1(\parity(\Omega))$ .
Let $\phi_t$ be the $\mu$-calculus formula using the $\Pre_1$ operator
in a turn-based  game describing the  
winning set for $\parity({\Omega}^f)$ .
As $\Omega^f$ and $\otdbl$ contain the same number of priorities,
the formula $\phi_t$ can be obtained from $\phi_c$ by syntactically 
replacing every $\CPre_1$ by $\Pre_1$ (see~\cite{AHM01}).
Let the winning set for $\parity(\Omega^f)$ in $\A^f_{\Omega}$ be 
$W_1\times\set{1} \, \cup\, W_2\times\set{2}$.
It is described by $\phi_t$.
The game in $\A^f$ proceeds in a bipartite fashion --- player~1 and player~2
alternate moves, 
with the state resulting from the move of player~1 having the
same parity index as the originating state.
Note that the objective $\parity(\Omega^f)$ depends
only on the infinitely often occurring indices in the trace.
Thus, $W_1\times\set{1}$ can be also be described by the
$\mu$-calculus formula $\phi_t'$ obtained by
replacing  each $\Pre_1$ in $\phi_t$ with $\Pre_1\circ\Pre_1$, and taking 
states of the form $s\times\set{1}$ in the result.
Consider the fixpoint iteration $\mathcal{I}_{\phi_t'}
$ for computing the set $\phi_t'$.
Since we are only interested in the set $W_1\times\set{1}$,
and since we have a bipartite game,
the set $W_1\times\set{1}$ can also be described by an iteration
$\mathcal{I}^*$
in which each set in the iteration $\mathcal{I}_{\phi_t'}$
is intersected with 
$\w{S}_{\reg}\times\set{1}$.
This is because  each step in the iteration 
$\mathcal{I}_{\phi_t'}$
applies $\Pre_1\circ\Pre_1$,
and if $X_1\times\set{1}\cup X_2\times\set{2} =
\Pre_1\circ\Pre_1 (Z_1\times\set{1}\cup Z_2\times\set{2})$, then
$X_1\times\set{1} = \Pre_1\circ\Pre_1(Z_1\times\set{1})$.
Now, this new iteration $\mathcal{I}^*$ describes the winning sets of
 the $\mu$-calculus formula $\phi_t''$ obtained from $\phi_t'$
by intersecting every variable with $\w{S}_{\reg}\times\set{1}$.
Using the identity~\ref{equation:PreCPre}, we have that the sets in the
fixpoint iteration computation of $\phi_t''$ correspond to the
sets in the fixpoint iteration computation of $\phi_c$, that is,
if $X\times\set{1}$ occurs in the computation of $\phi_t''$ at stage $j$,
then $\regstates(X)$ occurs in the computation of $\phi_c$ at the same
stage $j$.
This implies that the sets are the same on termination for both $\phi_t''$
and $\phi_c$.
Thus, 
$\win_1^{3,\wtdbl{\A}}(\timedivtbl_1(\parity(\Omega))) = 
\regstates(\win_1^{\A^f}(\parity(\Omega^f)))$.
\qed

From Theorem~\ref{theorem:Reduction}, we can solve the finite-state  game
$\A^f$ to compute winning sets for all $\omega$-regular region parity
objectives $\parity(\Omega)$
for a timed automaton game $\A$, using \emph{any} algorithm for finite
state turn-based  games, e.g., 
strategy improvement, small-progress algorithms~\cite{VJ00,Jur00}.

\subsubsection{Construction of the finite turn-based  game ${\A^f_{\Omega}}^*$.}
We now present the turn-based  game ${\A^f_{\Omega}}^*$ which can be used to
compute the sets $\regstates(\win_1^{\A^f_{\Omega}}(\parity(\Omega^f)))$ of
$\A^f_{\Omega}$, with the state space of ${\A^f_{\Omega}}^*$ being linear
in the number of regions of $\wtdbl{\A}$.
The idea behind this construction is that when player~1 proposes a move
$\tuple{\Delta,a_1}$ in $\wtdbl{\A}$ from state $\w{s}$, player~2 can 
ignore the discrete component
$a_1$ of the move.
Intuitively this is because player~2 should only care about the moves it can make
before time $\Delta$ from state $\w{s}$.
If there is a move it would like to make before time $\Delta$, then player~2 should 
preempt the move of player~1.
If there is no such move, then it should allow the move of player~1, irrespective of 
$a_1$.
The exact value of $a_1$ (other than whether it is $\bot_*$ or not) 
 is irrelevant in determining the moves for which player~2 
``wins'' in a round.
Since the value of $a_1$ is irrelevant, there is no need to explicitly remember the
discrete component of the moves of player~1 as was done in the game structure
$\A^f_{\Omega}$.
This allows us to bring the state space down, so that it is linear in
the number of regions of $\wtdbl{\A}$.
In the game $\A^f_{\Omega}$, we broke a step of the game in $\wtdbl{\A}$ into
two steps.
In the game ${\A^f_{\Omega}}^*$, we break a step of $\wtdbl{\A}$ into
\emph{three} steps:
first player~1  makes a move, then, player~2,
then player~1 and then the sequence repeats, i.e., the sequence of  players
making moves is $(121)^{\omega}$.
Note that this game is not bipartite as player~1 is making two consecutive
moves.
Informally, the three steps are as follows.
\begin{enumerate}[(1)]
\item
  First, player~1 proposes either a
  \begin{enumerate}[$\bullet$]
    \item
      Relinquishing move, or
    \item
      A move which corresponds to
      time elapsing  to a desired region.
      The discrete part of the move is left for later.
    \end{enumerate}
  \item
    Then, player~2 moves according to one of the following:
    \begin{enumerate}[$\bullet$]
    \item
      If player~1 had relinquished in the previous step, then it transitions
      to a player~1 state such that the region component is the new region.
      In addition, this is a \emph{special} player~1 state (we shall see what this
      means in a while).
    \item
      If player~1 had proposed a non-relinquishing move to a region, then player~2
      either
      \begin{enumerate}[$-$]
      \item
        Blocks the player~1 move by proposing a move that corresponds to a shorter
        duration move in $\wtdbl{\A}$.
        With this move, it transitions
        to a player~1 state such that the region component is the new region.
        In addition, this is again a  \emph{special} player~1 state.
      \item
        Decides to allow the player~1 move.
        But remember that we only know the time elapse part of the player~1 move
        till now.
        Thus, this move of player~2 transitions the game into an
        \emph{intermediate} state from which player~1 can only propose the
        \emph{remaining discrete part of the move}.
      \end{enumerate}
    \end{enumerate}
    \item
      In the third stage of the game, if the state is not special, then
      player~1 proposes the remaining discrete part
      of its original move.
      If the state is special, then it only has one dummy move, which makes
      the special state non-special without changing the region component.
      The special states thus allow us to maintain a three stage game.

\end{enumerate}

\noindent Formally, given a timed automaton $\A$, and a location parity index function $\Omega$
on states of $\A$, the  finite turn-based  game ${\A^f_{\Omega}}^*$ 
consists of a tuple 
$\tuple{{S^f}^*, {E^f}^*,{S_1^f}^*,{S_2^f}^*}$ 
where,
\begin{enumerate}[$\bullet$]
\item
  ${S^f}^* = {S_1^f}^*\,\cup\,{S_2^f}^*$ is the state space.
  The states in ${S_i^f}^*$ are controlled by player-$i$ for $i\in\set{1,2}$.
\item
  ${S_1^f}^*= \w{S}_{\reg}\times\set{1_a, 1^{\dum}_c}\, \cup\, 
  \w{S}_{\reg}\times\set{0,1,2}\times\set{1_c}$, where
  $\w{S}_{\reg}$ is the set of regions in $\wtdbl{\A}$.
  Informally, the sequence of  the players
  making moves is $(121)^{\omega}$;
  the moves from states labelled with $1_a$ as the last component will correspond
  to the first ``1'' and from states labelled with $1_c^{\dum}$ or $1_c$ 
  as the last component will 
  correspond to the second ``1'' in $(121)^{\omega}$.
\item
  ${S_2^f}^*=\w{S}_{\reg}\times\set{2}\,\cup\,  
  \w{S}_{\reg}\times\set{0,1,2}\times\set{2}$.
\item
  ${E^f}^*$ contains the following edges:
  \begin{enumerate}[$-$]
  \item
    For $s^f_1 = \tuple{\w{R},1_a}$, the set ${E^f}^*$ contains the following
    outgoing edges 
    \begin{enumerate}[(1)]
    \item
      $(\tuple{\w{R},1_a}, \tuple{\w{R},2})$.
      These edges correspond to  player~1  playing a relinquishing
      move from  states in $\w{R}$ in the game structure $\wtdbl{\A}$.
    \item
      $(\tuple{\w{R},1_a}, \tuple{\w{R},j,2})$
      such that
      in the structure $\wtdbl{\A}$, there exists $\w{s}\in \w{R}$ and
      $\tuple{\Delta,a_1}\in\Gamma_1^{\wtdbl{\A}}(\w{s})$ with 
      $|\set{\reg(\w{s}+\Delta')\mid \Delta'\leq \Delta}| = j+1$.
      These types of edges encode in ${\A^f}^*$ the fact that in the
      game structure $\wtdbl{\A}$ there exists a 
      state in $\w{R}$ such that player~1 has a move such that time
      passes to the $j$-th following region from where the discrete action 
      $a_1$ is taken.
    \end{enumerate}
  \item
    For $s^f_2 = \tuple{\w{R},2}$, the set ${E^f}^*$ contains the
    outgoing edges  $(\tuple{\w{R},2}, \tuple{\w{R}',1_c^{\dum}})$ 
    such that in the structure $\wtdbl{\A}$, there exists a state
    $\w{s}\in \w{R}$ and
    $\tuple{\Delta,a_2}\in\Gamma_2^{\wtdbl{\A}}(\w{s})$ with
    $\delta(\w{s},\tuple{\Delta,a_2})\in\w{R}\,'$ and
    $\succr_3(\w{s},\Delta) = \true$.
    These edges correspond to the fact in the game structure $\wtdbl{\A}$,
    player~1 had played a relinquishing move from some state $\w{s}$ in 
    $\w{R}$, and player~2
    had played a move according to its 3-region strategy.
  \item
    For $s^f_2 = \tuple{\w{R},j,2}$, the set ${E^f}^*$ contains the
    following
    outgoing edges:
    \begin{enumerate}[(1)]
    \item
      $(\tuple{\w{R},j, 2}, \tuple{\w{R}',1_c^{\dum}})$ 
      such that in the structure $\wtdbl{\A}$, there exists a state
      $\w{s}\in \w{R}$ and
      $\tuple{\Delta,a_2}\in\Gamma_2^{\wtdbl{\A}}(\w{s})$ with
      $\delta(\w{s},\tuple{\Delta,a_2})\in\w{R}\,'$ and
      $|\set{\reg(\w{s}+\Delta')\mid \Delta'\leq \Delta}| \leq j+1$.
      These types of edges encode in ${\A^f}^*$ the fact that in the
      game structure $\wtdbl{\A}$ there exists a 
      state in $\w{R}$ such that from $\w{s}$ player~1 proposes 
      a move such that time
      would pass  to the $j$-th following region from where some discrete action 
      would be taken, and player~2 is countering this player~1
      move with a move of shorter(or equal) delay.
    \item
      $(\tuple{\w{R},j, 2}, \tuple{\w{R},j, 1_c^{\dum}})$.
      This edge encode in ${\A^f}^*$ the fact that in the
      game structure $\wtdbl{\A}$ there exists a 
      state in $\w{R}$ such that from $\w{s}$ player~1 proposes 
      a move such that time
      would pass  to the $j$-th following region from where some discrete action 
      would be taken, and player~2 is ``allowing'' this move of player~1.
    \end{enumerate}
  \item
    For $s^f_1 = \tuple{\w{R},j,1_c}$, the set ${E^f}^*$ contains the
    outgoing edges $(\tuple{\w{R},j,1_c}, \tuple{\w{R}\,',1})$ such that
    in the structure $\wtdbl{\A}$, there exists a state
    $\w{s}\in \w{R}$ and
    $\tuple{\Delta,a_1}\in\Gamma_1^{\wtdbl{\A}}(\w{s})$ with
    $\delta(\w{s},\tuple{\Delta,a_1})\in\w{R}\,'$ and
    $|\set{\reg(\w{s}+\Delta')\mid \Delta'\leq \Delta}| = j+1$.
    This edge corresponds to the fact that in the game 
    $\wtdbl{\A}$ player~2 has allowed player~1 to take a move
    from a state $\w{s}$ in $\w{R}$ such that time passes to the $j$-th following
    region from where
    the discrete action $a_1$ is taken.
 \item
    For $s^f_1 = \tuple{\w{R},1_c^{\dum}}$, the set ${E^f}^*$ contains the
    single outgoing edge $( \tuple{\w{R},1_c^{\dum}},  \tuple{\w{R},1_a})$.
    This is just a dummy edge, to preserve the $(121)^{\omega}$ move pattern.
  \end{enumerate}
\end{enumerate}

Given a set 
$X=X_{1_a}\times\set{1_a}\, \cup\, X_{1_c}\times\set{1_c, 1_c^{\dum}}\, \cup\,
X_2\times\set{2}\subseteq {S^f}^*$, we let
$\regstates(X)=\set{\w{s}\in S^{\wtdbl{\A}} \mid \reg(\w{s})\in X_{1_a}}$.
Given the location parity index function $\otdbl$  on $\wtdbl{\A}$, 
we let ${\Omega^f}^*$ be
the parity index function on $\A^f$ such that ${\Omega^f}^*(\tuple{\w{R},\cdot})
=\otdbl(\w{s})$ where $\w{s}$ is any state in the region $\w{R}$ (all states
in a region have the same parity as $\otdbl$ is a location parity index
function on $\wtdbl{\A}$).

The following lemma shows that two steps in the bipartite game $\A^f_{\Omega}$
correspond to three steps in the game ${\A^f_{\Omega}}^*$.
The proof is presented in the appendix.
\begin{lem}
\label{lemma:TwoToThreeGame}
Let $\A$ be a timed automaton game, $\Omega$ a location parity index
function on states of $\A$, with
$\wtdbl{\A}$ being  the corresponding enlarged game,
and  $\A^f$, ${\A^f}^*$ being  the 
corresponding finite game structures with the parity index functions $\Omega^f$
and ${\Omega^f}^*$ respectively.
Suppose $X\subseteq S^f$ and $X^*\subseteq S^f$ are such that
$\regstates(X)=\regstates(X^*)$.
Then,
$\regstates\left(\Pre_1^{\A^f}\left(\Pre_1^{\A^f}\left(X\right)\right)\right) =
\regstates\left(\Pre_1^{{\A^f}^*}\left(\Pre_1^{{\A^f}^*}
\left(\Pre_1^{{\A^f}^*}\left(X^*\right)\right)\right)
\right)$.
\qed
\end{lem}

Using the fact that two steps in the bipartite game $\A^f_{\Omega}$
correspond to three steps in the game ${\A^f_{\Omega}}^*$, we have the following
theorem which shows that we can use the game structure ${\A^f_{\Omega}}^*$ to
compute the winning sets for $\A$.
\begin{thm}
\label{theorem:FinalFiniteReduction}
Let $\A$ be a timed automaton game, $\Omega$ a location parity index
function on states of $\A$, with
$\wtdbl{\A}$ being  the corresponding enlarged game,
and  $\A^f$, ${\A^f}^*$ being  the 
corresponding finite game structures with the parity index functions $\Omega^f$
and ${\Omega^f}^*$ respectively.
Then, we have
$\regstates(\win_1^{\A^f}(\parity(\Omega^f))) =
\regstates(\win_1^{{\A^f}^*}(\parity({\Omega^f}^*)))=
\win_1^{3,\wtdbl{\A}}(\timedivtbl_1(\parity(\Omega)))$.
\end{thm}
\proof
Let the winning set in $\A^f_{\Omega}$ be 
$W_1\times\set{1} \, \cup\, W_2\times\set{2}$; and the 
winning set in ${\A^f_{\Omega}}^*$ be 
${W_{1_a}}^*\times\set{1_a} \, \cup\, {W_{1_c}}^*\times\set{1_c,1_c^{\dum}}\,
\cup\, {W_2}^*\times\set{2}$.
As proved in the proof of Theorem~\ref{theorem:Reduction}, 
the set $W_1\times\set{1}$ is also described by the $\mu$-calculus
formula $\phi''_t$ obtained from the  $\mu$-calculus formula $\phi_c$,
which describes the winning set for $\parity(\otdbl)$, by 
(1)~syntactically 
replacing each occurrence of $\CPre_1$ in $\phi_c$
 by $\Pre_1\circ\Pre_1$, and,
(2)~intersecting
each variable in $\phi_c$ 
with $\w{S}_{\reg}$ (where $\w{S}_{\reg}$ is the set of regions of
$\wtdbl{\A}$).
Using a similar argument, the set ${W_{1_a}}^*\times\set{1_a}$ can
be described by a $\mu$-calculus formula $\phi_t^*$ obtained from
$\phi_c$ by
(1)~syntactically 
replacing each occurrence of $\CPre_1$ in $\phi_c$
 by $\Pre_1\circ\Pre_1\circ\Pre_1$, and,
(2)~intersecting
each variable in $\phi_c$ 
with $\w{S}_{\reg}$.
By Lemma~\ref{lemma:TwoToThreeGame}, we have that the sets in the
fixpoint iteration computation of $\phi_t''$ correspond to the
sets in the fixpoint iteration computation of $\phi_t^*$, that is,
if $X\times\set{1}$ occurs in the computation of $\phi_t''$ at stage $j$,
then $X\times\set{1_a}$ occurs in the computation of $\phi_t^*$ at the same
stage $j$.
This implies that the sets are the same on termination for both $\phi_t''$
and $\phi_t^*$.
Thus, $W_1={W_{1_a}}^*$, and hence
$\regstates(\win_1^{\A^f}(\parity(\Omega^f))) =
\regstates(\win_1^{{\A^f}^*}(\parity({\Omega^f}^*)))=
\win_1^{3,\wtdbl{\A}}(\timedivtbl_1(\parity(\Omega)))
$.
\qed

\smallskip\noindent\textbf{Eliminating redundant states in $\mathbf{{\A^f}^*}$.}
The states with $1_c^{\dum}$ as the last component are redundant and were
introduced just to have  a three stage game to prove 
Theorem~\ref{theorem:FinalFiniteReduction} using Lemma~\ref{lemma:TwoToThreeGame}.
These redundant states can be removed as follows.
The objective $\parity({\Omega^f}^*)$ depends only on the infinitely often occurring
indices in traces.
Consider the states in ${\A^f}^*$ of the form $\tuple{\w{R},1_c^{\dum}}$.
Player~1 has only one move from such a state, to $\tuple{\w{R},1_a}$.
The parity of $\tuple{\w{R},1_c^{\dum}}$ is the same as the parity of 
$\tuple{\w{R},1_a}$.
Thus, these kinds of states can be removed, and the incoming edges made to point
to the destination of these states without changing the part of the 
winning set with $1_a$ as the
last component.
That is, given a state $\tuple{\w{R},1_c^{\dum}}$ with $m$
incoming edges $({s^f}^*_j, \tuple{\w{R},1_c^{\dum}})$ for $1\leq j\leq m$,
 we can remove the
state $\tuple{\w{R},1_c^{\dum}}$ and add $m$ edges
$({s^f}^*_j, \tuple{\w{R},1_a})$  for $1\leq j\leq m$
 without changing the part of the  winning set with 
$1_a$ as the
last component.

\subsubsection{Complexity of reduction.}
Recall that for a timed automaton game $\A$,
$A_i$ is the set of actions for player~$i$ and
$C$ is the set of clocks.
Let $|A_1|^*=1+\min\left\{|A_1|+1,\, |L|\cdot 2^{|C|}\right\}$ (the +1 outside is
due to the relinquishing move of player~1, the min is because we can either
use player-1 actions, including pure time moves;
or use tuples of locations with clock
reset sets as actions), and let
$|A_2|^*=\min\left\{|A_2|+1,\, |L|\cdot 2^{|C|}\right\}$.
Let  $|\A_{\clkcond}|$ denote the length of the clock constraints in 
$\A$ and $d$ be the order of the parity index function
$\Omega$.
The size of the state space of ${\A^f}^*$
(with the $1_c^{\dum}$ states eliminated) is bounded by
$8\cdot |S^{\wtdbl{\A}}_{\reg}|$ where
$S^{\wtdbl{\A}}_{\reg}$ is the set of regions of $\wtdbl{\A}$.
The number of edges originating from player-1 states is at most
$|S^{\wtd{\A}}_{\reg}|\cdot(4+3|A_1|^*)$.
The number of edges  originating from player-2 states is at most
$|S^{\wtd{\A}}_{\reg}|\cdot 6 \cdot |A_2|^*$.
The total number of edges is thus at most
$|S^{\wtd{\A}}_{\reg}|\cdot(4 + 3|A_1|^* + 6|A_2|^*)$.
We also have $|S^{\wtd{\A}}_{\reg}|$ to be bounded by
$32\cdot d\cdot |L|\cdot \prod_{x\in C}(c_x+1) \cdot |C+1|! \cdot 4^{|C|}$.



\begin{thm}
\label{theorem:complexity}
Let $\A$ be a timed automaton game, and let $\Omega$ be a location
parity index function of order $d$.
The set $\wintimediv_1^{\A}(\parity(\Omega))$ can be computed in
time 
\[
O\left(
  |S^{\wtdbl{\A}}_{\reg}|\cdot|\A_{\clkcond}|  + 
  \left[   
    \left(|A_1|^* + |A_2|^*\right) \cdot
    \left(8\cdot |S^{\wtdbl{\A}}_{\reg}|\right)^{\frac{d+2}{3}+
      \frac{3}{2}} 
  \right]
  \right)
\]
which equals
\[
O\left(
  |S^{\A}_{\reg}|\cdot d\cdot |C|\cdot|\A_{\clkcond}|  + 
  \left[
    (|A_1|^* + |A_2|^*)\cdot
    \left(
      |S^{\A}_{\reg}|\cdot d\cdot 256 \cdot |C|
    \right)^ {\frac{d+2}{3}+ \frac{3}{2}}
  \right]
\right)
\]
 where 
\begin{enumerate}[$\bullet$]
\item
  $S^{\A}_{\reg}$ is the set of regions of $\A$, with
  $|S^{\A}_{\reg}| =
  |L|\cdot \prod_{x\in C}(c_x+1) \cdot |C|! \cdot 4^{|C|}$,
\item
  $S^{\wtdbl{\A}}_{\reg}$ is the set of regions of
  $\wtdbl{\A}$, with 
  $|S^{\wtdbl{\A}}_{\reg}| = 32\cdot (|C|+1)\cdot d\cdot |S^{\A}_{\reg}|$,
\item
  $|\A_{\clkcond}|$ is the length of the clock constraints in 
  $\A$, 
\item
  $|A_1|^* =  1+\min\left\{|A_1|+1,\, |L|\cdot 2^{|C|}\right\}$ and 
  $|A_2|^* =  \min\left\{|A_2|+1,\, |L|\cdot 2^{|C|}\right\}$ with  
  $A_i$ being the set 
  of discrete actions of player~$i$ for $i\in\set{1,2}$, 
\item  
  $L$ is the set of 
  locations, and $C$ is the set of clocks in $\A$.
\end{enumerate}
\end{thm}
\proof
To solve for $\wintimediv_1^{\A}(\parity(\Omega))$, we solve
the turn-based  game $A^f$ by Theorems~\ref{theorem:ThreeRegionGamesSuffice} 
and~\ref{theorem:FinalFiniteReduction}, and
 Lemma~\ref{lemma:ThreeBlameGames}.
For  constructing  ${\A^f}^*$, we need to check which regions satisfy
clock constraints from $\A$.
For this, we build a list of regions with valid invariants
together with  edge constraints satisfied at the region.
This takes $O(|S^{\wtdbl{\A}}_{\reg}|\cdot |\A_{\clkcond}|)$ time 
(we assume a region can be represented in constant space in our analysis).
From \cite{Schewe/07/Parity}, we have that a turn-based  parity game with $m$
edges, $n$ states and $d$ parity indices can be solved in 
$O(m\cdot n^{\frac{d}{3}+\frac{1}{2}})$ time.
The turn-based  game $A^f$ can hence be solved in time
$$O\left(
\left[  |S^{\wtdbl{\A}}_{\reg}|  \cdot  (4 + 3|A_1|^* + 6|A_2|^*)    \right]
\cdot
\left[   \left(8\cdot |S^{\wtdbl{\A}}_{\reg}|\right)^{\frac{d+2}{3}+
     \frac{1}{2}} \right]\right)
$$
which is equal to
$$O\left( \left(|A_1|^* + |A_2|^*\right) \cdot
\left[   \left(8\cdot |S^{\wtdbl{\A}}_{\reg}|\right)^{\frac{d+2}{3}+
     \frac{3}{2}} \right] \right)\eqno{\qEd}
$$

\begin{rem}[Effects of receptive formulation]
In the complexity,
 we  multiply $|S^{\A}_{\reg}|$ by $d$ as we go from $\parity(\Omega)$ to
$\parity(\otd) = \timedivbl_1(\parity(\Omega))$ (this also increases the value
of the exponent from $\frac{d}{3}+\frac{3}{2}$ to 
 $\frac{d+2}{3}+\frac{3}{2}$).
Multiplication of $|S^{\A}_{\reg}|$ by $32\cdot |C|$ comes in due to the extra clock
$z$ which measures global time, and introduction of the variables
$\tick$ and $\tbl_1$.
\end{rem}

\smallskip\noindent\textbf{Usefulness
 of the finite turn-based  game $\mathbf{{\A^f}^*}$.}
The reduction allows us to solve the finite-state  game
${\A^f}^*$ to compute winning sets for all $\omega$-regular region parity
objectives $\parity(\Omega)$
for a timed automaton game $\A$, using \emph{any} algorithm for finite
state turn-based  games, e.g., 
strategy improvement, small-progress algorithms~\cite{VJ00,Jur00}.
It also allows us to leverage any future improvements in algorithms for
finite-state  games for solving timed parity games.
The reduction to a finite-state  game that is linear in the number
of regions of $\wtdbl{\A}$ also shows that we do not pay any complexity 
penalty due to the
concurrent nature of the timed game where both players simultaneously propose
moves.


%


\section{Robust Winning of Timed Parity Games}
\label{section:Robust}
In this section we study restrictions on player-1 strategies to
model robust winning, and show how
the winning sets can be obtained by reductions to general timed automaton
games.
The results of Section~\ref{section:Reduction} can then be used to obtain
algorithms for computing the robust winning sets.

There is inherent uncertainty in real-time systems.
In a physical system,  an action may be prescribed by a 
controller, but the controller
can never prescribe a single timepoint where that action will be taken
with probability 1.
There is usually some \emph{jitter} when the specified action is taken, the
jitter being non-deterministic.
The model of general timed automaton games, where player~1 can specify 
exact moves of the form $\tuple{\Delta,a_1}$ consisting of an action together 
with a precise delay $\Delta$, assumes that the jitter is 0.
In subsection~\ref{subsection:Jitter}, we obtain robust 
winning sets for player~1
in the presence of  non-zero jitter (which are assumed to be 
arbitrarily small) for each of its proposed moves.
In subsection~\ref{subsection:BoundedJitter}, we assume the  jitter
to be some fixed $\varjit \geq 0$ for every move that is known.
The strategies of player~2 are left unrestricted.
In the case of lower-bounded jitter, we also introduce a \emph{response time} for
player-1 strategies.
The response time is the minimum delay between a discrete action, and a 
discrete action of the controller.
We note that the set of moves with a jitter of $\varjit >0$ around
$\tuple{\Delta, a_1}$ contains the set of moves with a jitter of 
$\varjit/2$  and 
a response time of  $\varjit/2$ around $\tuple{\Delta, a_1}$
(as $\set{\tuple{\Delta',a_1} \mid \Delta+\varjit/2
\leq \Delta' \leq 
\Delta+\varjit} \subseteq
\set{\tuple{\Delta',a_1} \mid \Delta\leq \Delta' \leq 
\Delta+\varjit}$.
Thus, the strategies of subsection~\ref{subsection:Jitter} 
may be considered to have  a response time greater than 0.
The winning sets in both sections are hence robust towards the presence
of jitter and response times.

\subsection{Winning in the Presence of Jitter}
\label{subsection:Jitter}

In this subsection, we model games where the jitter is assumed to be 
greater than 0, but arbitrarily small in each round of the game. 

Given a state $s$,  a \emph{limit-robust move} for player~$1$ is either 
 the move $\tuple{\Delta,\bot_*}$; 
or it is a tuple $\tuple{[\alpha,\beta], a_1}$ for some
$0\leq\alpha < \beta$ such that for every $\Delta \in [\alpha,\beta]$
we have $\tuple{\Delta, a_1}\in \Gamma_1(s)$
\footnote{We can alternatively have an open, or half-open time interval, 
the results do not change.} ($|\beta-\alpha|$  does not have a lower
bound, it just needs to be greater than 0 --- hence the term limit-robust).
Note that a  move $\tuple{\Delta,\bot_*}$ for player~1 implies that 
it is relinquishing the current round to player~2, as the move
of player~2 will always be chosen, and hence we allow a singleton
time move.
Given a limit-robust move $\mrob_1$ for player~$1$, and a move $m_2$ for player~2,
the set of possible outcomes is the set 
\[
\left\{\delta_{\jd}(s,m_1,m_2)  \left\vert \text{ either }
\begin{array}{l}
\text{(a) }
\mrob_1=\tuple{\Delta,\bot_*} \text{ and }
m_1=\mrob_1;  \text{ or} \\
\text{(b)~} \mrob_1=\tuple{[\alpha,\beta], a_1} \text{ and }
m_1=\tuple{\Delta,a_1} \text{ with } \Delta\in [\alpha,\beta]
\end{array}
\right.
\right\}.
\]

 A \emph{limit-robust strategy} $\pi_1^{\rob}$ for player~1 prescribes 
limit-robust moves to finite run prefixes.
A limit-robust strategy $\pi_1^{\rob}$ is \emph{receptive} if  for
all player-2 strategies $\pi_2$, all states $s\in S$, and all
runs $r\in\outcomes(s,\pi_1^{\rob},\pi_2)$, either $r\in\td$ or
$r\in\blameless_i$.
We let $\Pi_1^{\rob}$ denote the set of limit-robust strategies for player-$1$
and $\Pi_1^{\rob,R}$  the set of limit-robust receptive strategies.
Given an objective $\Phi$, let $\robwintimediv_1^{\A}(\Phi)$ denote the set
of states $s$ in $\A$ such that player~1 has a limit-robust receptive strategy
 $\pi_1^{\rob}\in \Pi_1^{\rob,R}$ such that for all receptive strategies 
 $\pi_2\in \Pi_2^R$, we have $\outcomes(s,\pi_1^{\rob},\pi_2)\subseteq \Phi$.
We say a limit-robust strategy $\pi_1^{\rob}$ is region equivalent to a 
(non-robust) strategy
$\pi_1$ if for all runs $r$ and for all $k \geq 0$, 
the following conditions hold:
\begin{enumerate}[(i)]
\item if $\pi_1(r[0..k])=\tuple{\Delta,\bot_*}$, then $\pi_1^{\rob}(r[0..k])=
\tuple{\Delta',\bot_*}$ with $\reg(r[k]+\Delta)= \reg(r[k]+\Delta')$; and
\item if $\pi_1(r[0..k])=\tuple{\Delta,a_1}$ with $a_1\neq\bot_*$, 
then $\pi_1^{\rob}(r[0..k])= \tuple{[\alpha,\beta], a_1}$ with
$\reg(r[k]+\Delta)= \reg(r[k]+\Delta')$ for all $\Delta'\in [\alpha,\beta]$.
\end{enumerate}
Note that for any limit-robust move $\tuple{[\alpha,\beta], a_1}$ with 
$a_1\neq \bot_*$ from a state $s$, we must have that the set
$\set{s+\Delta \mid \Delta\in [\alpha,\beta]}$ contains an open region of $\A$.


We first present an extension to Lemma~\ref{lemma:RegionStrategies}.
\begin{lem}
\label{lemma:RegionStrategiesExtension}
Let $\A$ be a timed automaton game  and
$\w{\A}$ be the corresponding enlarged game
structure.
Let $\w{\Phi}$ be an $\omega$-regular region objective of $\w{\A}$.
If $\pi_1$ is a region strategy that is  winning for
$\w{\Phi}$ from $\win_1^{\w{\A}}(\w{\Phi})$ and $\pi_1^{\rob}$ is
a robust strategy that is region-equivalent to $\pi_1$,
then $\pi_1^{\rob}$ is a winning strategy for
$\w{\Phi}$ from
$\win_1^{\w{\A}}(\w{\Phi})$.
\end{lem}
\proof
Consider any strategy $\pi_2$ for player~2, and a state 
$\w{s}\in \win_1^{\w{\A}}(\w{\Phi})$. 
We observe that the set $\outcomes(s,\pi_1^{\rob},\pi_2) $ consists of runs
$r$ such that for all $k\geq 0$, either 
%
%
\begin{enumerate}
\item 
$\pi_1^{\rob}(r[0..k])=
\tuple{\Delta,\bot_1}$ and $ r[k+1]\in 
\w{\delta}_{\jd}(r[k],\tuple{\Delta,\bot_1},\pi_2(r[0..k]))$;
or
\item 
$\pi_1^{\rob}(r[0..k])= \tuple{[\alpha,\beta],a_1}$  for some 
$\beta > \alpha\geq 0$ \\with
$r[k+1]\in  \w{\delta}_{\jd}(r[k],\tuple{\Delta,a_1},\pi_2(r[0..k]))$
for some  $\Delta\in [\alpha,\beta]$.
\end{enumerate}
It can be observed that \[\outcomes(s,\pi_1^{\rob},\pi_2)=
\bigcup_{\pi_1'}\outcomes(\w{s},\pi_1',\pi_2)\]
where $\pi_1'$  ranges over all those (non-robust) player-1 strategies
such that  for runs $r\in \outcomes(\w{s},\pi_1',\pi_2)$ and
 for all $k \geq 0$ we have 
\[\pi_1'(r[0..k]) =
\begin{cases}
 \tuple{\Delta,\bot_1} & \text{if}\quad
  \pi_1^{\rob}(r[0..k]) = \tuple{\Delta,\bot_1}\\
\tuple{\Delta,a_1}\quad & \text{if}\quad\pi_1^{\rob}(r[0..k]) =
\tuple{[\alpha,\beta],a_1}
\end{cases}
\]
for some $\Delta\in[\alpha,\beta]$;
and $\pi_1'$ acts like $\pi_1$ otherwise (note that the runs $r$ and the
strategies $\pi_1'$ are
defined inductively with respect to $k$, with $r[0]=\w{s}$).
Each player-1 strategy $\pi_1'$ in the preceding union is 
region equivalent to $\pi_1$ since $\pi_1^{\rob}$ is region equivalent 
to $\pi_1$ and hence each $\pi_1'$ is a winning
strategy for player~1 by Lemma~\ref{lemma:RegionStrategies}.
Thus, $\outcomes(s,\pi_1^{\rob},\pi_2)=
\bigcup_{\pi_1'}\outcomes(\w{s},\pi_1',\pi_2)$ is a subset of
$\w{\Phi}$, and hence $\pi_1^{\rob}$ is a winning strategy for player~1.
\qed

We now show how to compute the set $\robwintimediv_1^{\A}(\Phi)$.
Given a timed automaton game $\A$, we have the corresponding
enlarged game structure $\w{\A}_3$ which encodes time-divergence
(we use the modified blame variable $\bl_{1,3}$ as in 
Lemma~\ref{lemma:ThreeBlameGames}).
We add another boolean variable to $\w{\A}_3$ to obtain another
game structure $\w{\A}_{\rob}$.
The state space of $\w{\A}_{\rob}$ is $S^{\w{\A}}\times\set{\true,\false}$.
The transition relation $\delta^{\w{\A}_{\rob}}$ is such that 
\[
\delta^{\w{\A}_{\rob}}(\tuple{\w{s},\rb_1},\tuple{\Delta,a_i})=
\tuple{\w{\delta}(\w{s},\tuple{\Delta, a_i}),\rb_1'}\] 
where $\rb_1'=\true$ iff $\rb_1=\true$ and one of the following hold:
\begin{enumerate}[(i)]
\item $a_i \in A_2^{\bot}$; or
\item $a_i =\bot_*$; or
\item $a_i\in A_1^{\bot_1}$ and $s+\Delta$ belongs to an open region of 
$\w{\A}$.
\end{enumerate}
A region $\w{R}$ of $\w{\A}$ is said to be \emph{open} if for all states
$\w{s}\in\w{R}$ we have all the clock values in $\w{s}$ to be non-integral.
Given a location parity index function $\Omega$ on $\A$ of order $d$, we define
another game structure $\wrtd{\A}$ (based on $\w{\A}_{\rob}$) 
with the parity index function $\rotd$
encoding $\timedivbl_{1,3}(\parity(\Omega))\,\wedge\, \Box(\rb_1=\true)$ as
follows.
The state space of $\wrtd{\A}$ is $S^{\wrtd{\A}} = S^{\w{\A}_{\rob}}\times
\set{0,\dots,d-1}$.
The transition relation $\delta^{\wrtd{\A}}$ is specified as follows 
(similar to $\wtd{\A}$).
For $\tuple{\Delta,a_i} \in \Gamma_i^{\A}(s)$, we have
$\delta^{\wrtd{\A}}(\tuple{s,\z,\tick,\bl_1,\rb_1,p}, \tuple{\Delta,a_i}) =$
$\tuple{s', \z',\tick', \bl_1',\rb_1',p'}$ where
\begin{enumerate}[$\bullet$]
\item
$s'=\delta^{\A}(s,\tuple{\Delta,a_i})$.
\item
$\z'= (\z+\Delta)\mod 1$.
\item
$\tick'=\true$ iff $\z+\Delta \geq 1$.
\item
$\bl_1'=\true$ iff $i=1$ (i.e., its a player-1 move).
\item
  $\rb_1'=\left\{
    \begin{array}{ll}
      \true &\text{if } \rb_1=\true, a_i\in A_1\cup\set{\bot_1}\\
      &\text{and all clock values in } 
      C\cup\set{z} \text{ are non-integral}\\
      \true & \text{if } \rb_1=\true \text{ and }
      a_i\in A_2^{\bot} \cup\set{\bot_*}\\
      \false & \text{otherwise}
\end{array}
\right.$
\item
$p'=\left\{
\begin{array}{ll}
\max(p,\Omega(s')) & \text{if } \tick=\false \\
\Omega(s') & \text{if } \tick=\true 
\end{array}
\right.$
\end{enumerate}
The parity index function $\rotd$ is defined as
\[
\rotd(\tuple{s,\z,\tick,\bl_{1,3},\rb_1,p})=
\left\{
\begin{array}{ll}
1 & \text{ if } \rb_1=\false\\
0 & \text{ if } \rb_1=\true \text{ and } \tick=\bl_{1,3}=\false \\
1 & \text{ if } \rb_1=\true \text{ and }\tick=\false, \bl_{1,3}=\true \\
p+2 & \text{ if } \rb_1=\true \text{ and }\tick=\true
\end{array}
\right.
\]

\begin{lem}
\label{lemma:ParityTimeRobDiv}
Let $\A$ be a timed automaton game, $\parity(\Omega)$ 
an objective on $\A$,
and  $\wrtd{\A}$  the corresponding enlarged game
structure with the parity index function $\rotd$.
Then in the structure $\wrtd{\A}$, we have 
$\parity(\rotd)= \timedivbl_{1,3}(\parity(\Omega)) \,\wedge\,\Box(\rb_1=\true)$.
\end{lem}
\proof
The proof follows along similar lines to the proof of 
Lemma~\ref{lemma:ParityTimeDiv}.
We also observe than once $\rb_1$ becomes false, it stays false in $\wrtd{\A}$
and hence the parity also stays odd for all the following states.
Thus, if the maximum of $\infoften(\rotd(\w{r}))$ is even for a run $\w{r}$, 
then we must
have $ \Box(\rb_1=\true)$ in the run.
\qed

\begin{thm}
\label{theorem:Limit-Robust}
Given a state $s$ in a timed automaton game $\A$ and an 
$\omega$-regular location  parity index function $\Omega$,
we have $s\in  \robwintimediv_1^{\A}(\parity(\Omega))$
iff 
\begin{flalign*}
&
\tuple{s,\cdot,\cdot,\cdot,\rb_1=\true,\cdot} 
\in\win_1^{\wrtd{\A}}
\big(\timedivbl_{1,3}(\parity(\Omega))\, \wedge\,
\Box(\rb_1=\true)\big).&
\end{flalign*}

\end{thm}
\proof\hfill
  \begin{enumerate}[$\Rightarrow$]
  \item
    Suppose player-1 has a winning limit-robust receptive strategy
    $\pi_1$ for $\parity(\Omega)$,
    starting from a state $s$ in $\A$.
    We show 
\[\tuple{s,\cdot,\cdot,\cdot,\rb_1=\true,\cdot} 
    \in\win_1^{\wrtd{\A}}
    \big(\timedivbl_{1,3}(\parity(\Omega)) \wedge\,
    \Box(\rb_1=\true)\big).
\]
    We may consider $\pi_1$ to be a strategy in $\wrtd{\A}$.
    Since $\pi_1$  is a limit-robust strategy, player-1 
    proposes limit-robust moves
    at each step of the game.
    Given a state $\w{s}$, and a limit-robust move 
    $\tuple{[\alpha,\beta],a_1}$, there always exists 
    $\alpha < \alpha' <\beta'<\beta$ such that for every
    $\Delta\in [\alpha',\beta']$, we have 
    $\w{s}+\Delta$ belonging to an open region of $\w{\A}$.
    Thus, given the limit-robust strategy $\pi_1$, we can obtain another
    limit-robust strategy $\pi_1'$ in $\w{\A}$,  such that for every run
    $\w{r}$ and every $k\geq 0$,
    \begin{enumerate}[(a)]
    \item if $\pi_1(\w{r}[0..k])=\tuple{\Delta, \bot_*}$, then 
    $\pi_1'(\w{r}[0..k])= \pi_1(\w{r}[0..k])$; and
  \item 
    if $\pi_1(\w{r}[0..k]) = \tuple{[\alpha,\beta], a_1}$, then
    $\pi_1'(\w{r}[0..k]) = \tuple{([\alpha',\beta'], a_1}$ with 
    $[\alpha',\beta'] \subseteq [\alpha,\beta]$, and 
    $\set{\w{r}[k]+\Delta' \mid \Delta' \in [\alpha',\beta']}$ being a 
    subset of an open region of $\wrtd{\A}$.
   \end{enumerate}   
 Thus for any player-2 strategy $\pi_2$, and for any run 
    $\w{r}\in\outcomes(\tuple{s,\cdot,\cdot,\cdot,\true,\cdot},\pi_1',\pi_2)$,
    we have that $\w{r}$ satisfies $\Box(\rb_1=\true)$.
    Since $\pi_1$ was a receptive winning strategy for $\parity(\Omega)$, 
    $\pi_1'$ is also a receptive winning strategy for $\parity(\Omega)$
    as $\outcomes(\w{s},\pi_1,\pi_2) \subseteq \outcomes(\w{s},\pi_1',\pi_2)$
    for any player-2 strategy $\pi_2$.
    Thus, $\pi_1'$ enables player-1 to satisfy
    $\timedivbl_{1,3}(\parity(\Omega))\, \wedge\, \Box(\rb_1=\true)$.

  \item[$\Leftarrow$]
    Suppose $\w{s}=\tuple{s,\cdot,\cdot,\cdot,\rb_1=\true,\cdot} 
    \in\win_1^{\wrtd{\A}}
    (\timedivbl_{1,3}(\parity(\Omega))\, \wedge\, \Box(\rb_1=\true))$.
    We show that player-1 has a limit-robust receptive winning 
    strategy from state $\w{s}$ (and hence from $s$).
    Let $\pi_1$ be a winning region  strategy for player-1
    for the objective 
    $\timedivbl_{1,3}(\parity(\Omega))\, \wedge\, \Box(\rb_1=\true)$ in
    $\wrtd{\A}$.
    Since the strategy ensures $\rb_1=\true$ for all states in all runs
    from $\w{s}$, we have that
    for every run $\w{r}$ starting from state
    $\w{s}$, 
    the strategy $\pi_1$ is such that 
    $\pi_1(\w{r}[0..k]) = \tuple{\Delta^k,a_1^k}$ where either
    $a_1^k=\bot_*$, or $\w{r}[k]+\Delta^k$
    belongs to an open region $\w{R}$ of $\wrtd{\A}$ 
    Since $\w{R}$ is an open region, there always exists some
    $\alpha<\beta$ such that for every $\Delta\in [\alpha,\beta]$, we have
    $\w{r}[k]+\Delta\in \w{R}$.
    Consider the strategy $\pi_1^{\rob}$ that prescribes a limit-robust
    move $\tuple{[\alpha,\beta],a_1^k}$ for the history
    $r[0..k]$ if $\pi_1(r[0..k]) = \tuple{\Delta^k,a_1^k}$ with 
    $a_1^k\neq\bot_*$, and $\pi_1^{\rob}(r[0..k])= \pi_1(r[0..k])$ otherwise.
    The strategy $\pi_1^{\rob}$ is region-equivalent to $\pi_1$,
    and hence is also winning for player-1 by 
    Lemma~\ref{lemma:RegionStrategiesExtension}.
    Since it only prescribes limit-robust moves, it is a limit-robust strategy.
    It is also a receptive strategy as it  is region equivalent to the 
    receptive strategy $\pi_1$.\qed
  \end{enumerate}

\begin{thm}
Let $\A$ be a timed automaton game, and let $\Omega$ be a location
parity index function of order $d$.
The limit-robust winning set 
 $\robwintimediv_1^{\A}(\parity(\Omega)))$ can be
computed in time
$$O\left(
|S^{\wrtd{\A}}_{\reg}|\cdot|\A_{\clkcond}|  + 
\left(|A_1|^* + |A_2|^*\right) \cdot
    \left(8\cdot |S^{\wrtd{\A}}_{\reg}|\right)^{\frac{d+2}{3}+
      \frac{3}{2}} 
\right)
$$
\end{thm}
\proof
Using reductions similar to those in Section~\ref{section:Reduction},
the game on $\wrtd{\A}$ can be solved in time 
$O((|A_1|^* + |A_2|^*) \cdot
    (8\cdot |S^{\wrtd{\A}}_{\reg}|)^{\frac{d+2}{3}+
      \frac{3}{2}} 
)
$ where $S^{\wrtd{\A}}_{\reg}$ is the set of regions of $\wrtd{\A}$, with
$|\wrtd{\A}_{\reg}| = 64\cdot (|C|+1)\cdot d\cdot |S^{\A}_{\reg}|$.
We also need to
 build a list of regions with valid invariants
together with  edge constraints satisfied at the region.
This takes $O(|S^{\wrtd{\A}}_{\reg}|\cdot |\A_{\clkcond}|)$ time.
\qed

We say a timed automaton $\A$ is \emph{open}
if  all the guards and invariants in $\A$ are from $\clkcond^*(C)$.
Note that even though all the guards and invariants are open,
a player might still propose moves to closed regions, e.g., consider
an edge between two locations $l_1$ and $l_2$ with the guard $0<x<2$;
a player might propose a move from $\tuple{l_1,x=0.2} $ to $\tuple{l_2,x=1}$.
The following example shows that player~1 might not have a robust winning strategy
in an open timed automaton.

\begin{exa}
There exists an open  timed automaton game $\A$ such that for a reachability
objective $\Phi$, player~1 has a receptive winning strategy for $\Phi$ from
a state $s$, but does not have a limit-robust receptive strategy for 
$\Phi$ from $s$.

Consider the open timed automaton game $\A$ of 
Figure~\ref{figure:OpenCounterEx}.
The invariants of all the locations are true everywhere.
The objective of player~1 is to reach $l^2$.
The set of player-1 actions is $\set{a_1^0, a_1^1}$ and the set of player-2
actions is $\set{a_2^0, a_2^1, a_2^2}$.
The location $l^3$ and $l^4$ are absorbing locations.
Consider the location $l^1$.
If $x\neq 1$, then player~2 can propose a 0 time duration move to either
$l^4$ or $l^5$, and hence prevent player~1 from reaching $l^2$.
If $x=1$, then player~1 can propose a 0 time duration move to $l^2$, which 
a receptive strategy of player~2 must eventually allow.
Now consider the location $l^0$.
If $x>1$ then player~2 can propose a 0 time duration move to $l^3$.
If $x\leq 1$,  player~1 wins provided that it moves to
the location $l^1$  exactly when $x=1$.
Thus, no limit-robust winning strategy exists from any location (other than 
trivially from $l^2$).
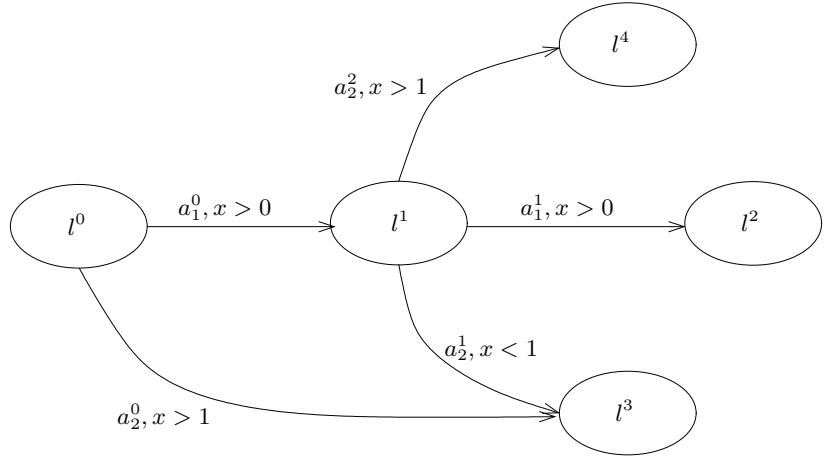
\begin{figure}[t]
\strut\centerline{\setlength{\unitlength}{0.00043745in}
\begingroup\makeatletter\ifx\SetFigFontNFSS\undefined%
\gdef\SetFigFontNFSS#1#2#3#4#5{%
  \reset@font\fontsize{#1}{#2pt}%
  \fontfamily{#3}\fontseries{#4}\fontshape{#5}%
  \selectfont}%
\fi\endgroup%
{\renewcommand{\dashlinestretch}{30}
\begin{picture}(9616,5386)(0,-10)
\put(8798,2751){\ellipse{1620}{990}}
\put(7313,503){\ellipse{1620}{990}}
\put(818,2716){\ellipse{1620}{990}}
\put(4604,2753){\ellipse{1620}{990}}
\put(7313,4868){\ellipse{1620}{990}}
\path(1634,2708)(3839,2708)
\path(3659.000,2648.000)(3839.000,2708.000)(3659.000,2768.000)
\path(5414,2708)(7979,2708)
\path(7799.000,2648.000)(7979.000,2708.000)(7799.000,2768.000)
\path(4604,2258)(4604,2256)(4605,2251)
	(4607,2242)(4609,2228)(4613,2209)
	(4618,2184)(4623,2154)(4630,2119)
	(4638,2080)(4647,2038)(4656,1993)
	(4666,1948)(4677,1902)(4687,1856)
	(4698,1812)(4709,1770)(4721,1729)
	(4732,1690)(4744,1653)(4755,1618)
	(4767,1586)(4780,1555)(4792,1525)
	(4806,1498)(4819,1471)(4834,1445)
	(4849,1421)(4865,1397)(4882,1373)
	(4899,1350)(4918,1327)(4937,1304)
	(4958,1281)(4979,1259)(5002,1236)
	(5026,1213)(5050,1191)(5076,1168)
	(5103,1146)(5131,1124)(5159,1102)
	(5188,1080)(5218,1058)(5249,1037)
	(5279,1016)(5310,996)(5341,976)
	(5372,957)(5403,939)(5434,921)
	(5464,903)(5494,887)(5524,871)
	(5553,856)(5582,841)(5610,827)
	(5637,814)(5665,801)(5692,788)
	(5720,775)(5749,762)(5777,750)
	(5806,738)(5835,726)(5865,714)
	(5896,702)(5929,690)(5962,677)
	(5998,665)(6034,652)(6073,639)
	(6113,625)(6154,611)(6196,598)
	(6238,584)(6280,570)(6320,558)
	(6357,546)(6391,535)(6420,526)
	(6445,518)(6464,512)(6494,503)
\path(6304.350,497.253)(6494.000,503.000)(6338.832,612.192)
\path(4604,3248)(4605,3250)(4606,3256)
	(4610,3265)(4615,3280)(4621,3301)
	(4630,3328)(4641,3360)(4654,3398)
	(4668,3440)(4683,3485)(4700,3533)
	(4717,3581)(4734,3630)(4752,3678)
	(4769,3725)(4786,3770)(4802,3813)
	(4818,3853)(4834,3891)(4849,3927)
	(4864,3960)(4878,3991)(4893,4020)
	(4907,4048)(4921,4073)(4935,4097)
	(4949,4120)(4964,4142)(4979,4163)
	(4997,4186)(5015,4209)(5034,4231)
	(5054,4252)(5074,4273)(5096,4293)
	(5118,4313)(5142,4333)(5166,4351)
	(5191,4370)(5217,4388)(5243,4405)
	(5270,4422)(5297,4438)(5325,4453)
	(5353,4468)(5381,4482)(5410,4496)
	(5438,4508)(5466,4520)(5494,4532)
	(5521,4543)(5549,4553)(5576,4564)
	(5604,4573)(5632,4583)(5657,4592)
	(5684,4601)(5711,4609)(5739,4618)
	(5767,4627)(5798,4636)(5829,4645)
	(5863,4655)(5899,4665)(5936,4676)
	(5976,4687)(6018,4698)(6062,4710)
	(6108,4722)(6155,4735)(6202,4747)
	(6249,4760)(6294,4771)(6337,4782)
	(6376,4793)(6409,4801)(6437,4808)
	(6459,4814)(6494,4823)
\path(6334.614,4720.063)(6494.000,4823.000)(6304.729,4836.282)
\path(824,2213)(825,2211)(827,2208)
	(831,2201)(837,2191)(845,2176)
	(856,2156)(870,2131)(887,2101)
	(907,2067)(929,2028)(954,1985)
	(981,1939)(1010,1890)(1040,1840)
	(1072,1788)(1103,1736)(1135,1685)
	(1168,1633)(1199,1584)(1231,1535)
	(1262,1488)(1293,1444)(1323,1401)
	(1352,1360)(1381,1321)(1409,1285)
	(1437,1250)(1465,1217)(1492,1186)
	(1520,1156)(1547,1128)(1574,1101)
	(1602,1076)(1630,1051)(1658,1028)
	(1687,1005)(1717,983)(1744,964)
	(1772,945)(1800,926)(1830,908)
	(1860,890)(1891,873)(1923,855)
	(1956,839)(1990,822)(2025,806)
	(2061,790)(2098,774)(2136,759)
	(2176,744)(2216,729)(2257,715)
	(2299,701)(2343,687)(2387,674)
	(2432,661)(2478,649)(2524,637)
	(2571,625)(2619,614)(2667,604)
	(2716,594)(2765,584)(2815,575)
	(2865,566)(2915,558)(2965,550)
	(3015,542)(3066,535)(3117,529)
	(3168,522)(3219,516)(3270,511)
	(3322,506)(3374,501)(3426,496)
	(3478,492)(3532,488)(3573,485)
	(3616,482)(3659,479)(3702,477)
	(3747,474)(3792,472)(3838,470)
	(3885,468)(3934,466)(3984,464)
	(4035,463)(4088,461)(4143,460)
	(4199,459)(4257,457)(4318,456)
	(4380,455)(4444,455)(4511,454)
	(4579,453)(4650,453)(4723,452)
	(4798,452)(4875,452)(4954,451)
	(5034,451)(5116,451)(5199,451)
	(5283,451)(5368,452)(5452,452)
	(5536,452)(5619,452)(5701,453)
	(5780,453)(5857,454)(5931,454)
	(6000,455)(6066,455)(6127,455)
	(6183,456)(6233,456)(6278,457)
	(6317,457)(6350,457)(6378,457)
	(6400,458)(6417,458)(6430,458)(6449,458)
\path(6269.000,398.000)(6449.000,458.000)(6269.000,518.000)
\put(6044,2843){\makebox(0,0)[lb]{\smash{{\SetFigFontNFSS{9}{10.8}{\rmdefault}{\mddefault}{\updefault}$a_1^1, x  > 0$}}}}
\put(1994,2843){\makebox(0,0)[lb]{\smash{{\SetFigFontNFSS{9}{10.8}{\rmdefault}{\mddefault}{\updefault}$a_1^0, x  > 0$}}}}
\put(3839,4283){\makebox(0,0)[lb]{\smash{{\SetFigFontNFSS{9}{10.8}{\rmdefault}{\mddefault}{\updefault}$a_2^2, x  > 1$}}}}
\put(689,2618){\makebox(0,0)[lb]{\smash{{\SetFigFontNFSS{9}{10.8}{\rmdefault}{\mddefault}{\updefault}$l^0$}}}}
\put(8654,2663){\makebox(0,0)[lb]{\smash{{\SetFigFontNFSS{9}{10.8}{\rmdefault}{\mddefault}{\updefault}$l^2$}}}}
\put(7169,413){\makebox(0,0)[lb]{\smash{{\SetFigFontNFSS{9}{10.8}{\rmdefault}{\mddefault}{\updefault}$l^3$}}}}
\put(7124,4778){\makebox(0,0)[lb]{\smash{{\SetFigFontNFSS{9}{10.8}{\rmdefault}{\mddefault}{\updefault}$l^4$}}}}
\put(1274,368){\makebox(0,0)[lb]{\smash{{\SetFigFontNFSS{9}{10.8}{\rmdefault}{\mddefault}{\updefault}$a_2^0, x  > 1$}}}}
\put(5144,1178){\makebox(0,0)[lb]{\smash{{\SetFigFontNFSS{9}{10.8}{\rmdefault}{\mddefault}{\updefault}$a_2^1, x  < 1$}}}}
\put(4514,2663){\makebox(0,0)[lb]{\smash{{\SetFigFontNFSS{9}{10.8}{\rmdefault}{\mddefault}{\updefault}$l^1$}}}}
\end{picture}
}}
\caption{An open timed automaton game $\A$ with no player-1 limit-robust winning strategy.}
\label{figure:OpenCounterEx}
\end{figure}
\qed
\end{exa}

\subsection{Winning with Bounded Jitter and Response Time}
\label{subsection:BoundedJitter}
The limit-robust winning strategies described in 
subsection~\ref{subsection:Jitter} did not have a lower bound on
the jitter: player~1 could propose a move 
$\tuple{[\alpha,\alpha+\varepsilon],a_1}$
for arbitrarily small $\alpha$ and $\varepsilon$.
In some cases, the controller may be required to work with a known jitter, and also
a finite \emph{response time}.
Intuitively, the response time is the minimum delay between a discrete action
(of either the controller or the environment)
and a discrete action of the controller.
The response time models the delay between a location change in a timed game, 
and when the controller is allowed to take an action based on the location
change.
We incorporate  the response time in timed automaton games by 
restricting player~1 strategies.
The jitter is modeled by  expanding the set of resulting states to include
all the states which lie in a jitter interval around the proposed player-1 
delay.

\smallskip\noindent\textbf{Strategies compatible for
$\varjit$-jitter $\varres$-response 
  bounded-robust winning.}
Given a finite response time $\varres$, player~2 can always propose pure time
moves of duration $\varres/2$.
Thus, if player~1 is restricted to only playing moves of duration longer than
$\varres$, player~2 can ensure that player-1  moves are never chosen (by
repeatedly playing pure time moves, and action moves of duration less than
$\varres$).
To allow for such blocking player~2 pure time moves, we only  restrict player-1 strategies to
 contain moves that are of duration greater than $\varres$ from  the last time
a non-pure time move was chosen.
In case  either player  proposes moves such that only time advances, 
without any
discrete action being taken, 
we  adjust the remainder of the response time.
%

Let $\varjit \geq 0 $ and $\varres \geq 0$ be given bounded jitter and response
time (we assume both are rational).
Formally, a strategy $\pi_1$ compatible for
\emph{$\varjit$-jitter $\varres$-response bounded-robust winning} of
player~1 proposes a move $\pi_1(r[0..k])=\tuple{\Delta,a_1}$ such that if
\[r[0..k] = s_0,\tuple{m_1^0,m_2^0},s_1,\tuple{m_1^1,m_2^1},\dots,s_k\]
then at least one of the following
conditions holds.
 \begin{enumerate}[$\bullet$]
 \item
   For all $0\leq j < k$, we have
   \begin{enumerate}[$-$]
   \item
     if 
     $\Blfunc_i(s_j, m_1^j,m_2^j, s_{j+1}) = \true$, then 
     $m_i^j = \tuple{\Delta_i^j,\bot_i}$ for $i\in\set{1,2}$.
   \item 
     $\Delta \geq \varres - \sum_{j=0}^{k-1}\Delta_{g(j)}^j$, where
     $g(j) = 
     \begin{cases}
       1 & \text{if } \Blfunc_1(s_j, m_1^j,m_2^j, s_{j+1}) = \true\\
       2 &\text{otherwise.}
     \end{cases}
     $
   \item 
     $\set{m_1^k+\epsilon \mid \epsilon \in [0,\varjit]}\ 
     \subseteq\, \Gamma_1(s_k)$.
   \end{enumerate}
   This corresponds to the case where in the entire run $r[0..k]$,
   (a)~no discrete
   actions have been taken, only simple time moves; 
   (b)~player~1 can take  a non-pure time move only
   after $\varres$ time units from the start of the run; and
   (c)~the moves $m_1^k +\epsilon$ must be legal player-1 moves for all
   $\epsilon \in [0,\varjit]$.
 \item
     There exists a $p$ with $1\leq p\leq k$ such that all of the following
     hold.
   \begin{enumerate}[$-$]
   \item
     $\Blfunc_i(s_{p-1}, m_1^{p-1},m_2^{p-1}, s_{p}) = \true$, for 
     $i\in\set{1,2}$,
     and $m_i^{p-1} = \tuple{\Delta_i^{p-1}, a_i^{p-1}}$ with 
     $a_i^{p-1}\neq \bot_i$ (i.e., the state $r[p]$ arises in the run $r$ due to
     a non pure-time move).
   \item
     For all $p\leq j < k$, 
     if $\Blfunc_i(s_j, m_1^j,m_2^j, s_{j+1}) = \true$
     then we have $m_i^j = \tuple{\Delta_i^j,\bot_i}$ for  $i\in\set{1,2}$
     (i.e., only simple time 
     passage moves are taken after the $p-1$-th stage).
   \item
     $\Delta \geq \varres - \sum_{j=p}^{k-1}\Delta_{g(j)}^j$, where
     $g(j) = 
     \begin{cases}
       1 & \text{if } \Blfunc_1(s_j, m_1^j,m_2^j, s_{j+1}) = \true\\
       2 &\text{otherwise.}
     \end{cases}
     $
   \item 
     $\set{m_1^k+\epsilon \mid \epsilon \in [0,\varjit]}\ 
     \subseteq\, \Gamma_1(s_k)$.
   \end{enumerate}
   This corresponds to the case where in the run $r[0..k]$, 
   (a)~the last non-pure
   move was taken at $r[p-1]$;
   (b)~only simple time 
     passage moves are taken from $r[p-1]$ till $r[k]$;
   (c)~player~1 has to wait $\varres$ time units 
   after the discrete action at $r[p-1]$ to propose a new discrete action at
   $r[k]$; and
   (c)~the moves $m_1^k +\epsilon$ must be legal player-1 moves for all
   $\epsilon \in [0,\varjit]$.
 \end{enumerate}
\smallskip\noindent\textbf{$\varjit$-jitter $\varres$-response 
  bounded-robust winning.}
Given a move $m_1=\tuple{\Delta,a_1}$ of player~1 and a move 
$m_2$ of player~2, the set of $\varjit$-jittered 
states is given by 
$\set{\delta_{\jd}(s,m_1+\epsilon, m_2) \mid \epsilon \in [0,\varjit]}$.
%
Given a player-1 strategy 
$\pi_1$ compatible for $\varjit$-jitter $\varres$-response bounded-robust winning
of player~1, and a  strategy
$\pi_2$ of player~2, the set of possible outcomes in the present
semantics is denoted by $\outcomes_{\jr}(s,\pi_1,\pi_2)$.
Each run $r$ in $\outcomes_{\jr}(s,\pi_1,\pi_2)$ is such that
for all $k\geq 0$, the state $r[k+1]$ belongs to the set of $\varjit$-jittered
states arising due to the moves $\pi_1(r[0..k])$ and $\pi_2(r[0..k])$
of player~1 and player~2 respectively.
We denote the $\varjit$-jitter $\varres$-response bounded-robust
winning set for player~1 for an objective $\Phi$ given
finite $\varjit$ and $ \varres$ by 
$\jrwintimediv^{\A,\varjit,\varres}_1(\Phi)$.

\medskip\noindent\textbf{The  timed automaton $\A^{\varjit,\varres}$ for computing
$\jrwintimediv^{\A,\varjit,\varres}_1(\Phi)$.}
We now show that $\jrwintimediv^{\A,\varjit,\varres}_1(\Phi)$ can be
computed by obtaining a timed automaton $\A^{\varjit,\varres}$ from
$\A$ such that $\wintimediv^{\A^{\varjit,\varres}}_1(\Phi)= 
\jrwintimediv^{\A,\varjit,\varres}_1(\Phi)$.
Given a clock constraint $\varphi$ we make the clocks appearing in 
$\varphi$ explicit by denoting the constraint as $\varphi(\overrightarrow{x})$
 for $\overrightarrow{x}=[x_1,\dots, x_n]$.
Given a real number $\delta$, we let $\varphi(\overrightarrow{x}+\delta)$ 
denote the
clock constraint $\varphi'$ where $\varphi'$ is obtained from $\varphi$ by
syntactically substituting $x_j+\delta$ for every occurrence of
$x_j$ in $\varphi$.
Let $f^{\varjit}: \clkcond(C) \mapsto \clkcond(C)$ be a function
defined by $f^{\varjit}\left(\varphi(\overrightarrow{x})\right) = 
\elimquant\left( \forall\delta\, \left(0\leq \delta\leq \varjit \rightarrow
    \varphi(\overrightarrow{x}+\delta) \right)\right)$, where 
$\elimquant$ is a function
that eliminates quantifiers (this function exists as we are working in
the theory of reals with addition, which admits quantifier elimination).
The formula  $f^{\varjit}(\varphi)$ ensures that $\varphi$ holds at all
the points in $\set{ \overrightarrow{x}+\Delta \mid \Delta\leq \varjit}$.

We now describe the  timed automaton $\A^{\varjit,\varres}$
such that  
\[\wintimediv^{\A^{\varjit,\varres}}_1(\Phi)= 
\jrwintimediv^{\A,\varjit,\varres}_1(\Phi).
\]
The automaton  has an extra clock $z$ in addition to the clocks of $\A$.
\begin{enumerate}[\hbox to8 pt{\hfill}]
\item\noindent{\hskip-12 pt\bf Locations:}\ 
  Corresponding to  each location  $l$ of $\A$ with outgoing player-1 edges
  $e_1^1,\dots, e_1^m$, the automaton $\A^{\varjit,\varres}$ has
  $m+1$ locations: $l, l_{e_1^1}, \dots, l_{e_1^m}$.
  The invariant for $l$ is the same as the invariant for $l$ in $\A$.
  The invariant for $l_{e_1^k} $ is $z \leq \varjit$ for all $k$.
\item\noindent{\hskip-12 pt\bf Actions:}\ 
  The automaton $\A^{\varjit,\varres}$ has the following actions:
  \begin{enumerate}[$\bullet$]
  \item
    The set of actions for player~1 is 
    $\set{\tuple{1,e}\mid e \text{ is a player-1 edge in } \A}$.
  \item
    The set of actions
    for player~2 is $A_2^{\A}\cup 
    \set{\tuple{a_2,e}\mid a_2\in A_2^{\A} \text{ and } 
      e \text{ is a player-1 edge in } \A}  \,\cup\, 
    \set{\tuple{2,e}\mid  e \text{ is a player-1 edge in } \A}$ 
    (we assume
    the unions are disjoint).
  \end{enumerate}
\item\noindent{\hskip-12 pt\bf Edges:}\ 
  The automaton $\A^{\varjit,\varres}$ has the following edges 
  (every edge includes  $z$ in the reset set):
  \begin{enumerate}[$\bullet$]
  \item
     For  $\tuple{l,a_2,\varphi,l',\lambda}$ a player-2 edge of $\A$,
     the automaton $\A^{\varjit,\varres}$ contains the player-2 edge
     $\tuple{l,a_2,\varphi,l',\lambda\cup\set{z}}$.
  \item
    For every player-1 edge $e_j=\tuple{l,a_1^j,\varphi,l',\lambda}$ of $\A$,
    the location $l$ of
    $\A^{\varjit,\varres}$ has the  outgoing
    player-1 edge 
    $\tuple{l,\; \tuple{1,e_j},\;
      f^{\varjit}\left(\gamma^{\A}(l)\right) \wedge (z\geq \varres) 
      \wedge 
      f^{\varjit}(\varphi), \;l_{e_j}, \;
      \lambda\cup\set{z}\;}$.
 \item
    For  $\tuple{l,a_2,\varphi_2,l',\lambda}$ a player-2 edge of $\A$ 
    and $e_j$ a player-1 edge from $l$, the location $l_{e_j}$ of the
    automaton $\A^{\varjit,\varres}$
    has a player-2 edge
    $\tuple{l_{e_j},\tuple{a_2,e_j},\varphi_2,l', \lambda\cup\set{z}}$. 
  \item
    For every player-1 edge $e_j=\tuple{l,a_1^j,\varphi_1,l',\lambda}$ of $\A$,
    the location $l_{e_j}$  of $\A^{\varjit,\varres}$ also has an additional  
    outgoing \emph{player-2} edge 
    $\tuple{l_{e_j}, \tuple{2,e_j}, \varphi_1, l',\lambda\cup\set{z}}$.
  \end{enumerate}
\end{enumerate}
The automaton $\A^{\varjit,\varres}$ as described above contains the rational
constants $\varres$ and $\varjit$. 
We can change the timescale by multiplying every constant by the
least common multiple of the denominators of $\varres$ and $\varjit$
to get a timed automaton with only integer constants.

The role of the different edges in
$\A^{\varjit,\varres}$ is described below.
\begin{enumerate}[$\bullet$]
\item
  A player-2 edge in $\A^{\varjit,\varres}$ labelled with $a_2$ such that
  $a_2 \in A_2^{\A}$ corresponds to the player-2 edge labelled $a_2$ in $\A$.
\item
  Player~1 moving from
  $l$ to $l_{e_j}$ with the edge labelled
  $\tuple{1,e_j}$  indicates the desire of player~1 to pick the edge
  $e_j$ from location $l$ in the game $\A$.
  This is possible in $\A$ iff  the following conditions hold.
 \begin{enumerate}[(a)]
 \item 
 More that $\varres$ time has passed 
  since the last discrete action.
\item  
The edge $e_j$ is enabled for at least
  $\varjit$ more time units.
  \item The invariant of $l$ is satisfied
  for at least $\varjit$ more time units.
\end{enumerate}
  These three requirements are captured by the new guard in 
  $\A^{\varjit,\varres}$, namely 
  $(z\geq \varres) \wedge f^{\varjit}(\varphi) \wedge 
  f^{\varjit}\left(\gamma^{\A}(l)\right)$.
\item
  Consider a location $l_{e_j}$ in $\A^{\varjit,\varres}$.
  If the game is at $l_{e_j}$, then it corresponds to the 
  situation in $\A$ where player~1 has picked the edge labelled  $e_j$ from
  location $l$, and it is up to player~2 to allow it or not.
  The presence of jitter in $\A$ causes uncertainty in when exactly the
  edge $e_j$ is taken.
  This is modeled in $\A^{\varjit,\varres}$ by having the location
  $l_{e_j}$ be controlled entirely by player~2 for a  duration of
  $\varjit$ time units.
  Within $\varjit$ time units, player~2 must either:
  \begin{enumerate}[$-$]
  \item
    propose 
    a move $\tuple{a_2,e_j}$ (corresponding to one of its
    own moves $a_2$ in $\A$), or,
  \item
    allow the action  $\tuple{2,e_j}$ 
    (corresponding to the original
    player-1 edge $e_j$) to be taken.
  \end{enumerate}
\end{enumerate}

\noindent Given a parity function $\Omega^{\A}$ on $\A$, the parity function
$\Omega^{\A^{\varjit,\varres}}$ on $\A^{\varjit,\varres}$ is given by 
$\Omega^{\A^{\varjit,\varres}}(l)=\Omega^{\A^{\varjit,\varres}}(l_{e_j})
=\Omega^{\A}(l)$.
In computing the winning set for player~1, we need to modify $\Blfunc_1$
for technical reasons.
Whenever an  action of the form $\tuple{1,e_j}$ is taken, we blame player~2 
(even though the action is controlled by player~1); and whenever an
action of the form $\tuple{2,e_j}$ is taken, we blame player~1 
(even though the action is controlled by player~2).
Player~2 is blamed as usual for the actions $\tuple{a_2,e_j}$.
This modification is needed because player~1 taking the edge $e_j$ in $\A$
is broken down into two stages in $ \A^{\varjit,\varres}$.
If player~1 were to be blamed for the edge $\tuple{1,e_j}$, then the following
could happen:
\begin{enumerate}[(a)]
\item 
 player~1 takes the edge $\tuple{1,e_j}$ in 
$\A^{\varjit, \varres}$ corresponding
to its intention to take the edge $e_j$ in $\A$, and
\item player~2 then
proposes its own move $\tuple{a_2,e_j}$ from $l_{e_j}$, corresponding to
it blocking the move $e_j$ by $a_2$ in $\A$.
\end{enumerate}
If the preceding scenario happens infinitely often, player~1 gets
blamed infinitely often even though all it  has done is signal its
intentions infinitely often, but its actions have not been chosen.
Hence player~2 is blamed for the edge $\tuple{1,e_j}$.
If player~2 allows the intended player~1 edge by taking $\tuple{2,e_j}$,
then we must blame player~1.
We note that this modification is not required if $\varres >0$, as in this case
player-1 signalling its moves infinitely often via moves of the type
$\tuple{1,e_j}$ can only happen if time progresses by $\varres$ infinitely
often, which implies time divergence.
The construction of $\A^{\varjit,\varres}$ can be simplified if $\varjit=0$
(then we do not need  locations of the form $l_{e_j}$).

\begin{exa}[Construction of $ \A^{\varjit,\varres}$]
\begin{figure}[t]
\strut\centerline{\input Figures/jitter-reduction.eepic}
\caption{The timed automaton game $\A^{\varjit, \varres}$ obtained from $\A$.}
\label{figure:jitter-reduction}
\end{figure}
An example of the construction is given in 
Figure~\ref{figure:jitter-reduction}, corresponding to the timed
automaton of Figure~\ref{figure:jitter}.
For the automaton $\A$, we have $A_1=\set{a_1^1,a_1^2, a_1^3, a_1^4}$ and
$A_2= \set{a_2^1,a_2^2, a_2^3}$.
The invariants of the locations of $\A$ are all
$\true$.
Since $\A$ has at most a single edge from any location $l^j$ to
$l^k$, all edges can be denoted as  $e_{jk}$.
The set of player-1 edges is then $\set{e_{01},e_{02}, e_{20}, e_{10}}$.
The location $l^3$ has been replicated for ease of drawing in
$ \A^{\varjit,\varres}$.
The location $l^3$ is also an absorbing location --- it only has self-loops 
(we omit these self loops in the figures for simplicity).
Observe that $f^{\varjit}(x\leq 1)\, =\, x\leq 1-\varjit$ and
$f^{\varjit}(y>1) \,=\, y>1$.
\qed 
\end{exa}
Given a set of states $\widetilde{S}$ of $\A^{\varjit,\varres}$,
let $\jstates(\widetilde{S})$ denote the projection of states to
$\A$, defined formally by  
$\jstates(\widetilde{S})= 
\set{\tuple{l,\kappa} \mid l\text{ is a location of } \A \text{ and }
\tuple{l,\widetilde{\kappa}} \in 
\widetilde{S} \text{ such that } \kappa(x)=\widetilde{\kappa}(x) 
\text{ for all } x\in C}$, where 
$C$  is the set of clocks of $\A$. 
The next theorem states that the timed automaton game $\A^{\varjit,\varres}$
can be used to compute the winning states in $\A$ from which player~1
has an $\varjit$-jitter $\varres$-response bounded-robust winning strategy.
The proof of correctness of the construction follows from the arguments
given in the description of $\A^{\varjit,\varres}$.

\begin{thm}
Let $\A$ be a timed automaton game,  $\varres\geq 0$
the response time of player~1, and  $\varjit \geq 0$ 
the jitter of player~1 actions such that both
$\varres$ and $\varjit$ are rational constants.
Then, for any $\omega$-regular location objective  
$\parity(\Omega^{\A})$ of $\A$, we have
\[\jstates\left( 
\symb{z=0}\,\cap\,
\wintimediv^{\A^{\varjit,\varres}}_1(\parity(\Omega^{\A^{\varjit,\varres}}))
\right)
= 
\jrwintimediv^{\A,\varjit,\varres}_1(\parity(\Omega^{\A})),
\]
 where
$\jrwintimediv^{\A,\varjit,\varres}_1(\Phi)$ is the winning set in
the jitter-response semantics, $\A^{\varjit,\varres}$ is the timed
automaton with the parity function $\Omega^{\A^{\varjit,\varres}}$
described above,and $\symb{z=0}$ is the set of states of
$\A^{\varjit,\varres}$ with $\widetilde{\kappa}(z)=0$.
\qed
\end{thm}

\begin{thm}
Let $\A$ be a timed automaton game,  $\varres\geq 0$
the response time of player~1, and  $\varjit \geq 0$ 
the jitter of player~1 actions such that both
$\varres$ and $\varjit$ are rational constants.
Then, for any $\omega$-regular location objective  
$\parity(\Omega^{\A})$ of $\A$, the winning set
\[\jrwintimediv^{\A,\varjit,\varres}_1(\parity(\Omega^{\A}))\]
can be computed in time
$$
O\left( \left(|S_{\reg}^{\A,\varjit,\varres,\Omega}|\cdot |\A_{\clkcond}|^2) \right)
  +
  |A_1|\cdot|A_2| \cdot 
  \left(8\cdot |S_{\reg}^{\A,\varjit,\varres,\Omega}|
  \right)^{\frac{d+2}{3} + \frac{3}{2} }
  \right)
$$
where 
    $S_{\reg}^{\A,\varjit,\varres,\Omega}$ is the set of regions of
$\A^{\varjit,\varres}$ with
\[\eqalign{|S_{\reg}^{\A,\varjit,\varres,\Omega}| &= |S_{\reg}^{\A}|\cdot
128\cdot (|C|+1)\cdot (|C|+2) \cdot d \cdot (|A_1|+1)\cdot\cr&\phantom{{}={}} \cdot 
\max\big(\num(\varjit), \num(\varres)\big)\cdot
\big(\lcm(\denom(\varres),\denom(\varjit))\big)^{|C|+1}}
\]
 in which
\begin{enumerate}[$\bullet$]
\item 
  $S_{\reg}^{\A}$ is the set of regions of $\A$;
\item
  $\lcm()$ is the least common multiple function, $\denom()$ and 
  $\num()$ are the  denominator and numerator functions respectively;
\item
  $|\A_{\clkcond}|$ is the length of the clock constraints in 
  $\A$.
\end{enumerate}
\end{thm}
\proof
Let the  timed automaton game $\A$ have $|L|$ locations and $|A_i|$ player-$i$ edges
for $i\in\set{1,2}$.
The automaton $\A^{\varjit,\varres}$ has $|L|\cdot(1 + |A_1|)$ locations, 
$(|A_1| + |A_2| + |A_1|\cdot|A_2|)$ player-2 edges, and
$|A_1|$ player-1 edges.
Given rational constants $\varjit$ and  $\varres$, all the constants in the system 
need to be multiplied by the least common multiple of the denominators of
$\varjit$ and $\varres$.
The timed parity game $\A^{\varjit,\varres}$ can hence be solved
in time 
$$
O\left( 
  |A_1|\cdot|A_2| \cdot 
  \left(8\cdot S_{\reg}^{\A,\varjit,\varres,\Omega}
  \right)^{\frac{d+2}{3} + \frac{3}{2} }
  \right)
$$
For every player-1 edge $e_j=\tuple{l,a_1^j,\varphi,l',\lambda}$
we need to obtain
$f^{\varjit\cdot\lcm(\set{\denom(\varres),\denom(\varjit)})}(\varphi)$.
This takes time $O(|\varphi|^2)$ (see~\cite{Basu}).
We observe that 
$f^{\varjit\cdot\lcm(\set{\denom(\varres),\denom(\varjit)})}(\varphi)$ 
cannot have any constants other than those in 
$\varphi$, and $\varres,\varres$ (this can be seen by 
putting $\varphi$ in a disjunctive normal form and
applying a Fourier-Motzkin like quantifier elimination procedure~\cite{Schrijver}).
Thus, building a  list of regions of $\A^{\varjit,\varres}$  with valid invariants
together with  edge constraints satisfied at the regions takes time
$O(|S_{\reg}^{\A,\varjit,\varres,\Omega}|\cdot |\A_{\clkcond}|^2)$
\qed


%
%
%
\begin{exa}[Differences between various winning modes]
\label{example:Jitter}
Consider the timed automaton $\A$ in Fig.~\ref{figure:jitter}.
Let the objective of player~1 be $\Box(\neg l^3)$, i.e., to avoid $l^3$.
The relevant part of the automaton for this example is the cycle $l^0, l^1$.
The only way to avoid $l^3$ in a time divergent run is to cycle in between
$l^0$ and $l^1$ infinitely often.
In additional player~1 may choose to also cycle in between $l^0$ and $l^2$,
but that does not help (or harm) it.
In our analysis, we omit such $l^0,l^2$ cycles, noting that our receptive
formulation correctly allows any number of  $l^0,l^2$ cycles.
We present an intuitive explanation here, a detailed analysis can be found in the
appendix.
\begin{figure}[t]
\strut\centerline{\setlength{\unitlength}{0.00043745in}
\begingroup\makeatletter\ifx\SetFigFontNFSS\undefined%
\gdef\SetFigFontNFSS#1#2#3#4#5{%
  \reset@font\fontsize{#1}{#2pt}%
  \fontfamily{#3}\fontseries{#4}\fontshape{#5}%
  \selectfont}%
\fi\endgroup%
{\renewcommand{\dashlinestretch}{30}
\begin{picture}(12052,2764)(0,-10)
\put(825,802){\blacken\ellipse{128}{128}}
\put(825,802){\ellipse{128}{128}}
\put(1905,2036){\blacken\ellipse{128}{128}}
\put(1905,2036){\ellipse{128}{128}}
\put(3615,2017){\blacken\ellipse{128}{128}}
\put(3615,2017){\ellipse{128}{128}}
\put(5325,2017){\blacken\ellipse{128}{128}}
\put(5325,2017){\ellipse{128}{128}}
\put(7080,2017){\blacken\ellipse{128}{128}}
\put(7080,2017){\ellipse{128}{128}}
\put(825,262){\ellipse{128}{128}}
\put(2342,2457){\ellipse{128}{128}}
\put(4245,2467){\ellipse{128}{128}}
\put(6135,2493){\ellipse{128}{128}}
\put(8813,2017){\blacken\ellipse{128}{128}}
\put(8813,2017){\ellipse{128}{128}}
\put(10500,2017){\blacken\ellipse{128}{128}}
\put(10500,2017){\ellipse{128}{128}}
\put(8025,2467){\ellipse{128}{128}}
\put(9960,2467){\ellipse{128}{128}}
\path(330,1477)(330,2737)
\path(2130,1477)(2130,2737)
\path(3930,1477)(3930,2737)
\path(5730,1477)(5730,2737)
\path(7530,1477)(7530,2737)
\path(9375,1477)(9375,2737)
\path(11130,1477)(11130,2737)
\thicklines
\path(7080,2017)(7080,892)
\thinlines
\path(6315.000,952.000)(6135.000,892.000)(6315.000,832.000)
\path(6135,892)(7080,892)
\path(6900.000,832.000)(7080.000,892.000)(6900.000,952.000)
\path(4425.000,952.000)(4245.000,892.000)(4425.000,832.000)
\path(4245,892)(5325,892)
\path(5145.000,832.000)(5325.000,892.000)(5145.000,952.000)
\thicklines
\path(4245,2408)(4245,892)
\path(2355,2408)(2355,892)
\path(3615,2017)(3615,892)
\thinlines
\path(2535.000,952.000)(2355.000,892.000)(2535.000,832.000)
\path(2355,892)(3615,892)
\path(3435.000,832.000)(3615.000,892.000)(3435.000,952.000)
\thicklines
\path(5325,2017)(5325,892)
\path(8813,2017)(8813,892)
\path(10500,2017)(10500,892)
\path(6135,2408)(6135,892)
\path(8025,2408)(8025,892)
\thinlines
\path(8205.000,952.000)(8025.000,892.000)(8205.000,832.000)
\path(8025,892)(8813,892)
\path(8633.000,832.000)(8813.000,892.000)(8633.000,952.000)
\thicklines
\path(330,1477)(12030,1477)
\path(9960,2408)(9960,892)
\thinlines
\path(10140.000,952.000)(9960.000,892.000)(10140.000,832.000)
\path(9960,892)(10500,892)
\path(10320.000,832.000)(10500.000,892.000)(10320.000,952.000)
\put(6405,397){\makebox(0,0)[lb]{\smash{{\SetFigFontNFSS{12}{14.4}{\familydefault}{\mddefault}{\updefault}$\alpha^2$}}}}
\put(4515,397){\makebox(0,0)[lb]{\smash{{\SetFigFontNFSS{12}{14.4}{\familydefault}{\mddefault}{\updefault}$\alpha^1$}}}}
\put(8250,397){\makebox(0,0)[lb]{\smash{{\SetFigFontNFSS{12}{14.4}{\familydefault}{\mddefault}{\updefault}$\alpha^3$}}}}
\put(9960,397){\makebox(0,0)[lb]{\smash{{\SetFigFontNFSS{12}{14.4}{\familydefault}{\mddefault}{\updefault}$\alpha^4$}}}}
\put(960,667){\makebox(0,0)[lb]{\smash{{\SetFigFontNFSS{12}{14.4}{\familydefault}{\mddefault}{\updefault}$a_1^1$}}}}
\put(960,127){\makebox(0,0)[lb]{\smash{{\SetFigFontNFSS{12}{14.4}{\familydefault}{\mddefault}{\updefault}$a_1^2$}}}}
\put(2625,397){\makebox(0,0)[lb]{\smash{{\SetFigFontNFSS{12}{14.4}{\familydefault}{\mddefault}{\updefault}$\alpha^0$}}}}
\put(2130,1252){\makebox(0,0)[lb]{\smash{{\SetFigFontNFSS{9}{10.8}{\rmdefault}{\mddefault}{\updefault}1}}}}
\put(3975,1207){\makebox(0,0)[lb]{\smash{{\SetFigFontNFSS{9}{10.8}{\rmdefault}{\mddefault}{\updefault}2}}}}
\put(5730,1207){\makebox(0,0)[lb]{\smash{{\SetFigFontNFSS{9}{10.8}{\rmdefault}{\mddefault}{\updefault}3}}}}
\put(7530,1162){\makebox(0,0)[lb]{\smash{{\SetFigFontNFSS{9}{10.8}{\rmdefault}{\mddefault}{\updefault}4}}}}
\put(9330,1207){\makebox(0,0)[lb]{\smash{{\SetFigFontNFSS{9}{10.8}{\rmdefault}{\mddefault}{\updefault}5}}}}
\put(11040,1207){\makebox(0,0)[lb]{\smash{{\SetFigFontNFSS{9}{10.8}{\rmdefault}{\mddefault}{\updefault}6}}}}
\put(15,1162){\makebox(0,0)[lb]{\smash{{\SetFigFontNFSS{9}{10.8}{\rmdefault}{\mddefault}{\updefault}Time  $\longrightarrow$}}}}
\end{picture}
}}
\caption{General timeline of a run in the game of Fig.~\ref{figure:jitter} (decreasing sequence of timegaps $\alpha^j$).}
\label{figure:GeneralTimeline}
\end{figure}
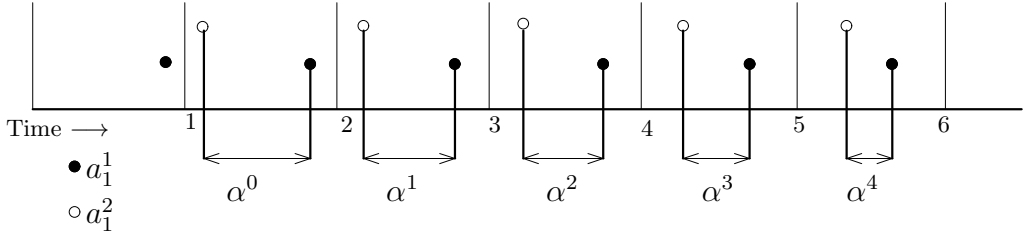

Let the game start from the location $l^0$.
In a run $r$, let $t_1^j$ and $t_2^j$ be the times when the 
$a_1^1$ transition and the $a_1^2$ transitions respectively are
taken for the $j$-th time.
The timeline is depicted in Figure~\ref{figure:GeneralTimeline}.
The guards $x\leq 1$ on $a_1^1$ and $ y> 1$ on $a_1^2$ ensure that
the distance between the $j+1$-th $a_1^1$ transition, and the $j$-th
$a_1^2$ transition keeps on strictly decreasing with increasing $j$.
To see this, observe that $t_1^{j+1}-t_1^{j} \leq 1 $ because of the guard
$x\leq 1$, and  $t_2^j-t_2^{j-1}>1$ because of the guard $y >1$.
Rearranging, we get
$t_1^{j+1}-  t_2^j < t_1^{j}-t_2^{j-1}$.
Consider the time gap sequence $\alpha^{j+1}= t_1^{j+1}-  t_2^j$, 
i.e., the sequence of time
gaps between the $j+1$-th $a_1^1$ transition, and the $j$-th
$a_1^2$ transition. 
For any $\epsilon$-jitter strategy of player~1 with $\epsilon >0$, this
sequence must decrease by more than $\epsilon$ for each step, which clearly cannot
happen infinitely often since $\alpha^j$ must be positive for all $j$, as
the $j+1$-th $a_1^1$ transition must always 
happen after the $j$-th
$a_1^2$ transition.
Thus, player~1 has no  $\epsilon$-jitter bounded-robust 
winning  strategy from $l^0$.
Player~1 does however have a limit-robust strategy, as  a limit-robust
strategy can be such that the time gap sequence $\alpha^j$ decreases, but by
a smaller and smaller amount at each step, ensuring that
$\alpha^j$ stays positive for all $j$.
This shows that  $\epsilon$-jitter bounded-robust winning 
strategies for $\epsilon > 0$ are strictly less
powerful than limit-robust strategies.

To see that limit-robust strategies are strictly less powerful than general
receptive strategies, observe that player~1 does not have a winning limit-robust
strategy from
$\tuple{l^0, x=y=1}$ as it would have to take the first $a_1^1$ transition 
immediately.
 It can be shown that there exists a winning player-1 receptive strategy
from $\tuple{l^0, x=y=1}$.
\qed
\end{exa}

The following theorem follows from the fact that bounded-robust winning 
strategies
are no more powerful than limit-robust strategies, which are in turn no more
powerful than general receptive strategies.
The strictness of the inclusions can be observed in Example~\ref{example:Jitter}.

\begin{thm}
Let  $\A$ be a timed automaton and  $\Phi$ an objective.
For all $\varjit >0$ and $\varres \geq 0$, we have
$\jrwintimediv_1^{\varjit,\varres}(\Phi) \subseteq \robwintimediv_1(\Phi)
\subseteq \wintimediv_1(\Phi)$.
All the subset inclusions are strict in general.
\qed
\end{thm}

\smallskip\noindent{\bf Sampling semantics.}
Instead of having a response time for actions of player~1, we can have a model
where player~1 is only able to take actions in an $\varjit$ interval around
sampling times, with a given time period $\varsam$.
A timed automaton can be constructed along similar lines to that of 
$\A^{\varjit,\varres}$ to obtain the winning set.

\bibliographystyle{alpha}
\bibliography{robust}

\newpage

\section{Appendix}
\subsection{Proofs of Section~\ref{section:TimedGames}}
\noindent\textbf{Proof of Lemma~\ref{lemma:ParityTimeDiv}}.
\proof
Consider a run $\w{r}$ of $\wtd{\A}$.
We show that the maximum index visited infinitely often is even iff
the run $\w{r}$ satisfies 
$\left((\Box\Diamond \tick \rightarrow \parity(\Omega))\ \wedge\ 
  (\neg\Box\Diamond\tick \rightarrow \Diamond\Box \neg\bl_1)\right)$
\begin{enumerate}[$\Rightarrow$]
\item
  Suppose the run $\w{r}$ satisfies 
  $\left((\Box\Diamond \tick \rightarrow \parity(\Omega))\ \wedge\ 
    (\neg\Box\Diamond\tick \rightarrow \Diamond\Box \neg\bl_1)\right) $
  We show the maximum index visited infinitely often is even.
  The following cases can arise.
  \begin{enumerate}[(a)]
  \item
    The run $\w{r}$ satisfies  $\neg\Box\Diamond\tick$ and $
    \Diamond\Box \neg\bl_1$,
    i.e., $\w{r}[j]$ has $\tick=\bl_1 =false$ for all $j\geq n$ for
    some $n$.
    In this case the parity seen infinitely often is 0 (even).
  \item
    The run $\w{r}$ satisfies  $\Box\Diamond\tick$ and also belongs to
    $\parity(\Omega)$.
    The component  $p$ of the state in $\wtd{\A}$ remembers the maximum
    $\Omega$ index seen since the state following the 
    last occurrence of $\tick=\true$.
    That is, in a run $\w{r}$, if $\w{r}[j]$ has $\tick=\true$, and
    $\w{r}[j+1], \dots \w{r}[j+m]$ all have $\tick=\false$ (except
    possibly for $\w{r}[j+m]$), then for $j+1\leq k\leq j+m$ the
    value of $p$ in 
    $\w{r}[k]$ is equal to $
    \max\set{\Omega(\w{r}[i]) \mid j+1\leq i\leq k }$.
    Since $\tick$ is true infinitely often, the maximum $\otd$ index
    seen infinitely often is $m+2$ where
    $m$ is the maximum $\Omega$ index seen infinitely often in $\w{r}$.
    Since $\w{r}$ belongs to $\parity(\Omega)$, $m$ is even.
    Thus the parity seen infinitely often is $m+2$ (even).
  \end{enumerate}
\item[$\Leftarrow$]
  Suppose the run $\w{r}$ does not satisfy 
  $\left((\Box\Diamond \tick \rightarrow \parity(\Omega))\ \wedge\ 
    (\neg\Box\Diamond\tick \rightarrow \Diamond\Box \neg\bl_1)\right) $.
  We show the maximum index visited infinitely often is odd.
  The following cases can arise.
  \begin{enumerate}[(a)]
  \item
    The run $\w{r}$ satisfies  $\neg\Box\Diamond\tick$ and 
    $\Box\Diamond\bl_1$.
    In this case $\tick=false$ for all $j\geq n$ for
    some $n$; and $\bl_1 =\true$ infinitely often. 
    Thus, the maximum index seen infinitely often is 1.
  \item
    The run $\w{r}$ satisfies  $\Box\Diamond\tick$ and also belongs to
    $\parity(\Omega)$.
    As above,  the maximum $\otd$ index seen infinitely often in
    $\w{r}$ is $m+2$ where
    $m$ is the maximum $\Omega$ index seen infinitely often in $\w{r}$.
    Since $\w{r}$ does not belong to $\parity(\Omega)$, $m$ is odd.
    Thus the parity seen infinitely often is $m+2$ (odd).\qed
   \end{enumerate}
\end{enumerate}

\medskip\noindent\textbf{Proof of Lemma~\ref{lemma:RegionStrategies}.}
\proof
  Consider the $\mu$-calculus formula $\varphi$ for describing the winning set
  $\win_1^{\wtd{\A}}\!(\parity(\otd))$.
  The formula contains the $\CPre_1$ operator.
  The set $\CPre_1(Z)$ remains unchanged (for $Z$ a union of regions of 
  $\wtd{\A}$)  if player~1 is restricted to use only memoryless  strategies.
  Suppose  $\w{s}\in \CPre_1(Z)$, and let $m_1^{\w{s}}$ be the winning move
  of player~1 from $\w{s}$ such that no matter what player~2 does, the next state 
  lies in $Z$.
  Let $\w{R}_1= \reg(\delta(\w{s},m_1^{\w{s}}))$, and let the 
  available moves of player~2 from $\w{s}$ be to regions 
  $\w{R}_2^1,\dots, \w{R}_2^n$.
  We have that from \emph{any} state in $\reg(\w{s})$, player~1 has a move
  to $\w{R}_1$, and that  player~2 can only take moves to 
  $\w{R}_2^1,\dots, \w{R}_2^n$.
  From 
  Lemma~\ref{lemma:RegionsBeatRegions}, 
  it then follows that if player~1 proposes a move to $\w{R}_1$ from any state
  in $\reg(\w{s})$, then no matter what player~2 does,
  the resulting state will lie in $Z$.
  Thus, memoryless region strategies suffice as winning strategies.
  Moreover, again from Lemma~\ref{lemma:RegionsBeatRegions}, \emph{any}
  move of player~1 to the region $\w{R}_1$ is a winning move.
  Thus, we have that there is a memoryless region winning strategy $\pi_1$ from
  winning states, and that
  any strategy region equivalent to $\pi_1$ is also a  winning strategy.
\qed

\medskip\noindent\textbf{Proof of Proposition~\ref{proposition:MoveIndependent}.}
\proof
Intuitively, in the structure $\A$, we want player~1 to be able to infer 
the values corresponding to $\z,\tick, \bl_1, p$. 
If player~1 can do this, then it can maintain the structure $\wtd{\A}$ in memory,
and thus it can use a winning memoryless  strategy of $\wtd{\A}$ (by 
Lemma~\ref{lemma:RegionStrategies}   memoryless  strategies
suffice in $\wtd{\A}$).
This strategy will then be move-independent.
The values of $\z$ and $\tick$ can be inferred from the value of the global
clock $z$.
And given a state $r[k]$ in a run $r$, and a move $\tuple{\Delta_1,a_1}$ of
player~1, the $\bl_1$ component will be true iff
$(\runtime(r[k+1]) - \runtime(r[k]))= \Delta_1$ and
$\delta(r[k], \tuple{\Delta_1, a_1}) =r[k+1]$.
The value of the component $p$ can be inferred from the parity values of $\Omega$,
and the values of $\tick,\bl_1$.
\qed

\medskip\noindent\textbf{Proof of 
Lemma~\ref{lemma:ExpandedMemorylessPlayerTwo}.}
\proof
  Consider the $\CPre_1$ operator in the $\mu$-calculus formula for describing the 
  winning set $\win_1^{\wtd{\A}}(\parity(\otd))$.
    The set $\CPre_1(Z)$ remains unchanged (for $Z$ a union of regions of 
  $\wtd{\A}$)  if player~2 is restricted to playing only move-independent region
  strategies. 
  This is because from it can be shown from Lemma~\ref{lemma:RegionsBeatRegions}
  that from any state $\w{s}$, the ability of player~2 to prevent
  player~1 from reaching $Z$ in the next step depends only on $\reg(\w{s})$,
  the regions in  $Z$,
  and the \emph{current} move of player~1.
  Moreover, if player~2 can prevent  player~1 from reaching $Z$ from $\w{s}$ for
  all player~1 moves, then there is a unique region $\w{R}^*$ such that
  for all $\w{s}\,'\in\reg(\w{s})$, against any player~1 move 
  $m_1^{\w{s}\,'}$, 
  player~2 has a counter-move $m_2^{\w{s}\,',m_1^{\w{s}\,'}}$ with
  $\delta^{\wtd{\A}}(\w{s}\,',m_2^{\w{s}\,',m_1^{\w{s}\,'}})\in \w{R}^*$ such that
  the move $m_2^{\w{s}\,',m_1^{\w{s}\,'}}$ prevents the player-1 move
  $m_1^{\w{s}\,'}$ from reaching $Z$.
  Thus, move-independent region strategies of player~2 suffice as
  spoiling strategies.
  \qed

\subsection{Proofs of Section~\ref{section:Reduction}}

We start with the statement of a classical result of~\cite{AlurD94} that the region equivalence 
relation induces a time abstract bisimulation on the regions.

\begin{lem}[\cite{AlurD94}]
\label{lemma:Bisimulation}
Let $Y,Y'$ be  regions in the timed game structure
$\A$.
Suppose player~$i$ has a move from $s_1\in Y$
to $s_1'\in Y'$, for $i\in\set{1,2}$.
Then, for any $s_2\in Y$,
player~$i$ has a move from $s_2$ to some $s_2'\in Y'$.
\qed
\end{lem}

\medskip\noindent\textbf{Proof of Lemma~\ref{lemma:RegionsBeatRegions}.}
\proof
From Lemma~\ref{lemma:Bisimulation}, if player~$i$ has a move from some
$s_1\in Y$
to $s_1'\in Y'$, for $i\in\set{1,2}$.
Then, for any $s_2\in Y$,
player~$i$ has a move from $s_2$ to some $s_2'\in Y'$.

Consider the case when $Y_1'\neq Y_2'$.
The proof follows from the fact that each region has a unique first
\emph{time-successor} region.
Thus, if $Y_1'$ is ``closer'' to $Y$ than $Y_2'$, then the move of
player~1 wins, otherwise, the move of player~2 wins. 
A region $R'$ is a first time-successor of $R\neq R'$ if for all 
states $s\in R$, 
there exists $\Delta >0$ such that $s+\Delta \in R'$ and for all 
$\Delta' < \Delta$, we have $s+\Delta' \in R\cup R'$.
The time-successor of $\tuple{l,h,\parti(C)}$ is
$\tuple{l,h',\parti'(C)}$ when
\begin{enumerate}[$\bullet$]
\item
$h=h'$, 
$\parti(C) = \tuple{C_{-1},C_0\neq \emptyset, C_1,\dots, C_n}$, and 
$\parti'(C) = \tuple{C_{-1},C_0'=\emptyset, C_1',\dots, C_{n+1}'}$ where $C_i'=C_{i-1}$, 
and $h(x) < c_x$ for every $x\in C_0$.
\item
$h=h'$, 
$\parti(C) = \tuple{C_{-1},C_0\neq \emptyset, C_1,\dots, C_n}$, and 
$\parti'(C) = \tuple{C_{-1}'=C_{-1}\cup C_0,C_0'=\emptyset, C_1,\dots, C_n}$,
and $h(x) \geq  c_x$ for every $x\in C_0$.
\item
$h=h'$, 
$\parti(C) = \tuple{C_{-1},C_0\neq \emptyset, C_1,\dots, C_n}$, and 
$\parti'(C) = \tuple{C_{-1}',C_0'=\emptyset, C_1',\dots, C_{n+1}'}$ where 
$C_i'=C_{i-1}$ for $i\geq 2$, 
$h(x) < c_x$ for every $x\in C_1'\subseteq C_0$, and 
$h(x)\geq  c_x$ for every $x\in C_0\setminus C_1'$, and 
$C_{-1}'=C_{-1}\cup C_0\setminus C_1'$.
\item 
$\parti(C) = \tuple{C_{-1},C_0= \emptyset, C_1,\dots, C_n}$, 
$\parti'(C) = \tuple{C_{-1},C_0'=C_n, C_1,\dots, C_{n-1}}$, and 
 $h'(x)=h(x)+1 \leq c_x$ for every $x\in C_n$, and $h'(x) = h(x)$ otherwise.
\item 
$\parti(C) = \tuple{C_{-1},C_0= \emptyset, C_1,\dots, C_n}$, 
$\parti'(C) = \tuple{C_{-1}'=C_{-1}\cup C_n,C_0, C_1,\dots, C_{n-1}}$, and 
 $h'(x)=h(x)=c_x$ for every $x\in C_n$, and $h'(x) = h(x)$ otherwise.
\item 
$\parti(C) = \tuple{C_{-1},C_0= \emptyset, C_1,\dots, C_n}$, 
$\parti'(C) = \tuple{C_{-1}'=C_{-1}\cup C_n\setminus C_0',C_0', C_1,\dots, 
C_{n-1}}$, and 
$h'(x)= h(x)+1 \leq c_x$ for every $x\in C_1'\subseteq C_n$,
$h'(x)=h(x)=c_x$ for every $x\in C_n\setminus C_1'$, 
and $h'(x) = h(x)$ otherwise.

\end{enumerate}

\noindent In case $Y_1'=Y_2'$, then player~2 can pick the same time to
elapse as player~1, and ensure that the conditions of the lemma hold.
\qed

\medskip\noindent\textbf{Completion of proof of 
Lemma~\ref{lemma:ThreeRegionWinningStrategies}.}
\proof
(Continued).\\
We constructed the 3-region strategy $\pi_1^*$ from $\pi_1$, and we claimed
$\pi_1^*$ was a winning strategy.
We were proving it by contradiction.
We assumed a spoiling strategy $\pi_2^*$ for $\pi_1^*$, and we constructed a 
player-2 strategy $\pi_2$ that we claimed was spoiling for $\pi_1$.
We were showing by induction that 
there exists a run $\w{r}_3 \in \outcomes(\w{s},\pi_1, \pi_2)$ such that
$\w{r}_3[k] = \w{r}_3^*[k]$ for all $ k\geq 0$ ($\w{r}_3^*$ was the
run used in defining $\pi_2$, and is such that
$\w{r}_3^*\in\outcomes(\w{s}, \pi_1^*,\pi_2^*)$ and
$\w{r}_3^*\notin \timedivtbl_1(\parity(\Omega))$).
We present  the details of the induction proof.
The proof of the above claim is by induction on $k$.
For $k=0$ the claim is trivially true.
Suppose the claim is true for all $j\leq k$.
Thus, we have a run $\w{r}_3$ such that $\w{r}_3[j] = \w{r}_3^*[j]$ for all 
$j\leq k$.
We show that the run $\w{r}_3[0..k]$ can be extended to $\w{r}_3[0..k+1]$ according
to $\pi_1,\pi_2$ such that $\w{r}_3[k+1]=\w{r}_3^*[k+1]$.
We have the following cases:
\begin{enumerate}[(1)]
\item
  The following hold.
  \begin{enumerate}[$\bullet $]
  \item 
  $\pi_1(\w{r}_3[0..k]) =  \tuple{\Delta,a_1} $, and 
\item $a_1=\bot_*$.
\end {enumerate}
  Since $\pi_1$ is a memoryless strategy of $\wtdbl{\A}$,
  we have $\pi_1(\w{r}_3^*[0..k]) =  \tuple{\Delta,\bot_*}$.
  Suppose $\pi_2^*(\w{r}_3^*[0..k])$ has the form $\tuple{\Delta_2,a_2}$.
  By definition of $\pi_2$, we have 
  $\pi_2(\w{r}_3[0..k]) = \pi_2^*(\w{r}_3[0..k]) =
  \pi_2^*(\w{r}_3^*[0..k]) = \tuple{\Delta_2,a_2}$ ($\pi_2^*$ is a 
  memoryless  strategy).
  Hence, we have 
  $\w{r}_3^*[k+1]=\delta^{\wtdbl{\A}}(\w{r}_3^*[k], \tuple{\Delta_2,a_2}) =
  \delta^{\wtdbl{\A}}(\w{r}_3[k], \tuple{\Delta_2,a_2}) =
  \w{r}_3[k+1]$.
  
\item
 The following hold.
  \begin{enumerate}[$\bullet $]
  \item 
  $\pi_1(\w{r}_3[0..k]) =  \tuple{\Delta,a_1} $,
\item 
  $a_1\neq \bot_*$, and
\item 
   $\succr_3^{\wtd{\A}}(\w{r}_3[k], \Delta) = \true$
\end {enumerate}
  Since $\pi_1$ is a memoryless strategy, we have 
  $\pi_1(\w{r}_3^*[0..k]) =  \tuple{\Delta,a_1} $.
  Combining this with $\succr_3^{\wtd{\A}}(\w{r}^*_3[k], \Delta) = \true$,
  yields $\pi_1^*(\w{r}_3^*[0..k]) =  \pi_1(\w{r}_3^*[0..k])= 
  \tuple{\Delta,a_1}$.
  Let $\pi_2^*(\w{r}_3^*[0..k]) = \tuple{\Delta_2,a_2}$.
  By definition of $\pi_2$, we have 
  $\pi_2(\w{r}_3[0..k]) = \pi_2^*(\w{r}_3[0..k]) =
  \pi_2^*(\w{r}_3^*[0..k]) = \tuple{\Delta_2,a_2}$ ($\pi_2^*$ is a 
  memoryless  strategy).
  Thus, 
\[\eqalign{\delta^{\wtdbl{\A}}_{\jd}(\w{r}_3[k], \pi_1(\w{r}_3[0..k]), \pi_2(\w{r}_3[0..k]))&=
  \delta^{\wtdbl{\A}}_{\jd}(\w{r}_3[k], \tuple{\Delta,a_1},
  \tuple{\Delta_2,a_2})\cr
 &=
  \delta^{\wtdbl{\A}}_{\jd}(\w{r}_3^*[k], \tuple{\Delta,a_1}, \tuple{\Delta_2,a_2})\cr
 &=\delta^{\wtdbl{\A}}_{\jd}(\w{r}_3^*[k], \pi_1^*(\w{r}^*_3[0..k]),
  \pi_2^*(\w{r}^*_3[0..k])).
  }
\]
  Hence $\w{r}_3[0..k]$ can be extended to $\w{r}_3[0..k+1]$ according to 
  $\pi_1,\pi_2$ such that $\w{r}_3[0..k+1]=\w{r}_3^*[0..k+1]$.

\item  The following hold.
  \begin{enumerate}[$\bullet $]
  \item 
    $\pi_1(\w{r}_3[0..k]) = 
    \tuple{\Delta,a_1}, a_1\neq \bot_*$, 
  \item 
    $ \succr_3^{\wtd{\A}}\!\!(\w{r}_3[k], \Delta) = \false$, 
  \item 
    $\pi_1^*(\w{r}_3[0..k]) = \tuple{\Delta',\bot_1}$, and
  \item 
  $\pi_2^*(\w{r}_3[0..k]) = \tuple{\Delta^*,a_2^*}$  with 
  $\Delta^* < \Delta'$.
\end{enumerate}
  We have $\pi_2(\w{r}_3[0..k])=\pi_2^*(\w{r}_3[0..k]) = 
  \tuple{\Delta^*,a_2^*}$.
  Since $\Delta' < \Delta$ (by definition of $\pi_1^*$), we have
  $\Delta^* < \Delta$, hence
  \[
  \w{r}_3[k+1]\in\delta^{\wtdbl{\A}}_{\jd}\!(\w{r}_3[k], \tuple{\Delta,a_1}, 
  \tuple{\Delta^*,a_2}) = 
  \set{\delta^{\wtdbl{\A}}\!(\w{r}_3[k], \tuple{\Delta^*,a_2})}.
\]
  Since $\pi_1^*,\pi_2^*$ are memoryless, we have
\[\eqalign{\pi_1^*(\w{r}^*_3[0..k]) &= \pi_1^*(\w{r}_3[0..k])=
  \tuple{\Delta',\bot_1}\cr
  \pi_2^*(\w{r}^*_3[0..k]) &= \pi_2^*(\w{r}_3[0..k])=
  \tuple{\Delta^*,a_2}.
  }
\]
  Thus, 
  $\w{r}^*_3[k+1]$ belongs to
\[\delta^{\wtdbl{\A}}_{\jd}(\w{r}^*_3[k], \tuple{\Delta',\bot_1},
  \tuple{\Delta^*,a_2}) = 
  \set{\delta^{\wtdbl{\A}}(\w{r}^*_3[k], \tuple{\Delta^*,a_2})}
   = \set{\delta^{\wtdbl{\A}}(\w{r}_3[k], \tuple{\Delta^*,a_2})}.
\]
  
\item   The following hold.
\begin{enumerate}[$\bullet $]
  \item 
 $\pi_1(\w{r}_3[0..k]) = 
 \tuple{\Delta,a_1}, a_1\neq \bot_*$, 
\item 
 $ \succr_3^{\wtd{\A}}\!\!(\w{r}_3[k], \Delta) = \false$, 
\item 
  $\pi_1^*(\w{r}_3[0..k]) = \tuple{\Delta',\bot_1}$, and
\item 
  $\pi_2^*(\w{r}_3[0..k]) = \tuple{\Delta^*,a_2^*}$  with 
  $\Delta^* > \Delta'$.
\end{enumerate}
  We have $\pi_2(\w{r}_3[0..k]) = \tuple{\Delta', \bot_2}$ by definition.
  Since $\Delta' < \Delta$(by definition of $\pi_1^*$),  
  we have
  \[
  \w{r}_3[k+1]\in\delta^{\wtdbl{\A}}_{\jd}(\w{r}_3[k], \tuple{\Delta,a_1}, 
  \tuple{\Delta',\bot_2}) = 
  \set{\delta^{\wtdbl{\A}}(\w{r}_3[k], \tuple{\Delta',\bot_2})}.
  \]
  Also, since $\pi_1^*$ and $\pi_2^*$ are memoryless, we have
  \begin{align*}
    \w{r}_3^*[k+1]  & \in
    \delta^{\wtdbl{\A}}_{\jd}(\w{r}_3^*[k], \tuple{\Delta',\bot_1}, 
    \tuple{\Delta^*,a_2}) \\
    & = 
    \set{\delta^{\wtdbl{\A}}(\w{r}_3^*[k], \tuple{\Delta',\bot_1})} \\
    &=
    \set{\delta^{\wtdbl{\A}}(\w{r}_3[k], \tuple{\Delta',\bot_1})}.
  \end{align*}
  Thus, $\w{r}_3[k+1]$ and $\w{r}_3^*[k+1]$ are the same except for perhaps the
  $\tbl_1$ component.
  Since $\succr_2^{\wtd{\A}}(\w{r}_3[k], \Delta') = \false$ (by
  definition of $\pi_1^*$), we must have $\tbl_1=\false$ in both 
  $\w{r}_3[k+1]$ and $\w{r}_3^*[k+1]$.
  Hence $\w{r}_3[k+1]=\w{r}_3^*[k+1]$.

\item  The following hold.
\begin{enumerate}[$\bullet $]
  \item 
  $\pi_1(\w{r}_3[0..k]) = \tuple{\Delta,a_1}, a_1\neq \bot_*$, 
\item 
  $\succr_3^{\wtd{\A}}(\w{r}_3[k], \Delta) = \false$,
\item 
  $\pi_1^*(\w{r}_3[0..k]) = \tuple{\Delta',\bot_1}$,
\item 
  $\pi_2^*(\w{r}_3[0..k]) = \tuple{\Delta^*,a_2^*}$ with 
  $\Delta^* = \Delta'$, 
\item 
  $\delta^{\wtdbl{\A}}(\w{r}_3^*[k], \pi_2^*(\w{r}^*_3[0..k])) = 
  \w{r}_3^*[k+1]$, 
  and
\item 
  $\delta^{\wtdbl{\A}}(\w{r}_3^*[k], \pi_1^*(\w{r}^*_3[0..k])) \neq \w{r}_3^*[k+1]$.
\end{enumerate}
  We have $\pi_2(\w{r}_3[0..k]) = \pi_2^*(\w{r}_3[0..k]) = 
  \tuple{\Delta^*,a_2^*}$ 
  by definition.
  Also $\pi_1^*(\w{r}^*_3[0..k]) = \pi_1^*(\w{r}_3[0..k]) = 
  \tuple{\Delta',\bot_1}$; and
  $\pi_2^*(\w{r}^*_3[0..k]) = \pi_2^*(\w{r}_3[0..k]) = 
  \tuple{\Delta^*,a_2^*}$ since $\pi_1^*$ and $\pi_2^*$ are both memoryless.
  Now, $\w{r}_3[k+1]$ belongs to
$\delta^{\wtdbl{\A}}_{\jd}(\w{r}_3[k],\pi_1(\w{r}_3[0..k]),\pi_2(\w{r}_3[0..k]))$, and  
\[\eqalign{\delta^{\wtdbl{\A}}_{\jd}(\w{r}_3[k],\pi_1(\w{r}_3[0..k]),\pi_2(\w{r}_3[0..k]))&=
  \delta^{\wtdbl{\A}}_{\jd}(\w{r}_3[k], \tuple{\Delta,a_1}, \tuple{\Delta^*,a_2^*})\cr
  &=\set{\delta^{\wtdbl{\A}}(\w{r}_3[k],\tuple{\Delta^*,a_2^*})}\cr
  &=\set{\delta^{\wtdbl{\A}}(\w{r}_3^*[k],\tuple{\Delta^*,a_2^*})}\cr
  &=\set{\w{r}_3^*[k+1]}.
  }
\]
  Thus, we have $\w{r}_3[k+1]=\w{r}_3^*[k+1]$.

\item The following hold.
\begin{enumerate}[$\bullet $]
  \item 
  $\pi_1(\w{r}_3[0..k]) = \tuple{\Delta,a_1}, a_1\neq \bot_*$,
\item 
  $\succr_3^{\wtd{\A}}(\w{r}_3[k], \Delta) = \false$,
\item 
  $\pi_1^*(\w{r}_3[0..k]) = \tuple{\Delta',\bot_1}$,
\item 
  $\pi_2^*(\w{r}_3[0..k]) = \tuple{\Delta^*,a_2^*}$ with 
  $\Delta^* = \Delta'$; and
\item 
  $\delta^{\wtdbl{\A}}(\w{r}_3^*[k], \pi_1^*(\w{r}^*_3[0..k])) =
  \w{r}_3^*[k+1]$.
\end{enumerate}
  We have $\pi_2(\w{r}_3[0..k]) = \tuple{\Delta^*,\bot_2}$ by definition.
  Also $\pi_1^*(\w{r}^*_3[0..k]) = \pi_1^*(\w{r}_3[0..k]) = 
  \tuple{\Delta',\bot_1}$; and
  $\pi_2^*(\w{r}^*_3[0..k]) = \pi_2^*(\w{r}_3[0..k]) = 
  \tuple{\Delta^*,a_2^*}$ since $\pi_1^*$ and $\pi_2^*$ are both memoryless.
  Then, $\w{r}_3[k+1] \in 
  \delta^{\wtdbl{\A}}_{\jd}(\w{r}_3[k],\pi_1(\w{r}_3[0..k]),\pi_2(\w{r}_3[0..k]))$
  and
\[\eqalign{
   \delta^{\wtdbl{\A}}_{\jd}(\w{r}_3[k],\pi_1(\w{r}_3[0..k]),\pi_2(\w{r}_3[0..k]))
   &= 
  \delta^{\wtdbl{\A}}_{\jd}(\w{r}_3[k], \tuple{\Delta,a_1}, \tuple{\Delta^*,\bot_2}) 
  \cr
  &=\set{\delta^{\wtdbl{\A}}(\w{r}_3[k],\tuple{\Delta^*,\bot_2})}\cr
  &=\set{\delta^{\wtdbl{\A}}(\w{r}_3^*[k],\tuple{\Delta^*,\bot_1})}
  }\]
  since $\succr_3^{\wtd{\A}}(\w{r}_3^*[k], \Delta) = 
\succr_3^{\wtd{\A}}(\w{r}_3[k], \Delta) =
\false$, and $\Delta'=\Delta^*$
and \\ $\succr_3^{\wtd{\A}}(\w{r}_3^*[k], \Delta') = \true$.

\noindent We have , 
\[\w{r}_3^*[k+1]
  =\delta^{\wtdbl{\A}}(\w{r}_3^*[k],\tuple{\Delta^*,\bot_1})
=  \delta^{\wtdbl{\A}}(\w{r}_3^*[k],\tuple{\Delta',\bot_1})
 \]
  and 
\[\eqalign{
\delta^{\wtdbl{\A}}(\w{r}_3^*[k],\tuple{\Delta^*,\bot_1})  & \in
\delta^{\wtdbl{\A}}_{\jd}(\w{r}_3^*[k],\pi_1^*(\w{r}_3^*[0..k]),\pi_2^*(\w{r}_3^*[0..k]))
\cr
&  =
  \delta^{\wtdbl{\A}}_{\jd}(\w{r}_3^*[k], \tuple{\Delta',\bot_1},\tuple{\Delta^*,a_2^*})
\cr
&  =
  \delta^{\wtdbl{\A}}_{\jd}(\w{r}_3[k], \tuple{\Delta',\bot_1},\tuple{\Delta^*,a_2^*})
\cr
&\owns  \w{r}_3[k+1]
}
\]
  Thus,  the run $\w{r}_3[0..k]$ can be extended such that 
$\w{r}_3[k+1]=\w{r}_3^*[k+1]$.
\end{enumerate}
Thus, in all cases, we have that $\w{r}_3[0..k]$ can be extended to 
$\w{r}_3[0..k+1]$ according to $\pi_1,\pi_2$ such that 
$\w{r}_3[0..k+1]=\w{r}_3^*[0..k+1]$.
Hence, we have $\w{r}_3\in \outcomes(\w{s},\pi_1,\pi_2)$ and 
$\w{r}_3\notin\timedivtbl_1(\parity(\Omega))$ as 
$\w{r}^*_3\notin\timedivtbl_1(\parity(\Omega))$, 
a contradiction since $\pi_1$ was assumed to be a winning strategy.
Hence, we cannot have the existence of the strategy $\pi_2^*$ from
which $\w{r}_3^*$ and $\pi_2$ were derived, i.e., 
$\pi_1^*$ is a winning strategy for player~1 from $\w{s}$.
\qed

\medskip\noindent\textbf{Completion of proof of 
  Lemma~\ref{lemma:ThreeRegionSpoilingStrategies}.}
\proof
  (Continued).\\
  We continue to show that if $\w{s}\in \CPre_{1,3}(Z)$, then
   $\w{s}\in \CPre_{1}(Z)$.
   Suppose $\w{s}\in  \CPre_{1,3}(Z)$.

   We can have the following cases.
   \begin{enumerate}[(1)]
   \item
     $\set{\delta(\w{s}, \tuple{\Delta,a_2}) \mid \tuple{\Delta,a_2} \in
       \Gamma_2(\w{s}) \text{ and } \succr_3(\w{s},\Delta)=\true}
   \subseteq Z$. \\
   Consider  the cardinality of the set
   $B=\set{\reg(\delta(\w{s}, \tuple{\Delta,\bot_2}) \mid 
     \tuple{\Delta, \bot_2} \in \Gamma_2(\w{s}))}$.
   \begin{enumerate}[$\bullet$]
   \item
     If $|B|\leq 3$ then we have 
     \begin{align*}
     &  \set{\delta(\w{s}, \tuple{\Delta,a_2}) \mid \tuple{\Delta,a_2} \in
       \Gamma_2(\w{s}) \text{ and } \succr_3(\w{s},\Delta)=\true}\\
     = & \set{\delta(\w{s}, \tuple{\Delta,a_2}) \mid \tuple{\Delta,a_2} \in
       \Gamma_2(\w{s})}.
     \end{align*}
     Hence $\w{s} \in \CPre_1(Z)$ in this case (when $|B|\leq 3$).
   \item
     Suppose $|B| > 3$.
     Then, there exists $\Delta$ such that
     $\tuple{\Delta,a_2}\in \Gamma_2(\w{s})$ (and hence 
     $\tuple{\Delta,\bot_1}\in \Gamma_1(\w{s})$) with 
     $|\set{\reg(\w{s} + \Delta')\mid \Delta'\leq \Delta}| =3$.

     Consider the player-1 move  $\tuple{\Delta,\bot_1}$.
     Since $\tuple{\Delta,a_2}\in \Gamma_2(\w{s})$, we must have
     $\tuple{\Delta,\bot_2}\in \Gamma_2(\w{s})$.
     By assumption, we have 
     \begin{align*} 
      &  \delta(\w{s}, \tuple{\Delta,\bot_2})
     \in Z \\
     & \qquad \text{(the assumption being }
     \left\{\delta(\w{s}, \tuple{\Delta,a_2}) \left\vert
     \begin{array}{l}
      \tuple{\Delta,a_2} \in
       \Gamma_2(\w{s}) \text{ and } \\
       \succr_3(\w{s},\Delta)=\true
     \end{array}
     \right.
   \right\}
     \subseteq Z).
         \end{align*}
     Since $\Delta$ is such that
     $\succr_2(\w{s},\Delta) = \false$, we have
     $\delta(\w{s}, \tuple{\Delta,\bot_2}) = 
     \delta(\w{s}, \tuple{\Delta,\bot_1})$ (the $\tbl_1$ component is false in
     both cases).
     Hence, $\delta(\w{s}, \tuple{\Delta,\bot_1})\in Z$.
     Also, since 
     \begin{enumerate}[$\star$]
     \item 
     $\set{\delta(\w{s}, \tuple{\Delta',a_2}) \mid \tuple{\Delta',a_2} \in
       \Gamma_2(\w{s}) \text{ and } \succr_3(\w{s},\Delta')=\true}
     \subseteq Z$ and
   \item 
     $|\set{\reg(\w{s} + \Delta')\mid \Delta'\leq \Delta}| =3$
     \end{enumerate}
     we have
     \[
     \set{\delta(\w{s},\tuple{\Delta',a_2}) \mid \Delta' \leq \Delta,
       \text{ and }\tuple{\Delta',a_2} \in
       \Gamma_2(\w{s})}
     \subseteq Z .
     \]
     Hence $\w{s} \in \CPre_1(Z)$ in this case by the second condition
     of $\CPre_1(Z)$ (the winning move of player~1 being 
     $\tuple{\Delta,\bot_1}$).
   \end{enumerate}
 \item
   There exists $\tuple{\Delta,a_1}\in  \Gamma_1(\w{s})$ with 
   $a_1\neq \bot_*$ such that
   \begin{enumerate}[(a)]
   \item
     $\delta(\w{s},\tuple{\Delta,a_1})\in Z$, and 
   \item
     $\set{\delta(\w{s},\tuple{\Delta',a_2}) \mid \Delta' \leq \Delta,\ 
       \succr_3(\w{s},\Delta')=\true, 
       \text{ and }\tuple{\Delta',a_2} \in
       \Gamma_2(\w{s})}
     \subseteq Z $.
   \end{enumerate}
   Consider  the cardinality of the set
   $D=\set{\reg(\w{s} +\Delta') \mid \Delta' \leq \Delta}$.
   \begin{enumerate}[$\bullet$]
   \item
     If $|D|\leq 3$ then   
     \begin{align*}
     & \set{\delta(\w{s},\tuple{\Delta',a_2}) \mid \Delta' \leq \Delta,\ 
       \succr_3(\w{s},\Delta')=\true, 
       \text{ and }\tuple{\Delta',a_2} \in
       \Gamma_2(\w{s})}\\
      = & \set{\delta(\w{s},\tuple{\Delta',a_2}) \mid \Delta' \leq \Delta,
       \text{ and }\tuple{\Delta',a_2} \in
       \Gamma_2(\w{s})}.
     \end{align*}
     Hence $\w{s} \in \CPre_1(Z)$ in this case (when $|D|\leq 3$) by the
     second condition of $\CPre_1(Z)$.
   \item
     Suppose $|D| > 3$.
     Since $\tuple{\Delta,a_1}\in  \Gamma_1(\w{s})$, we must have
     that there exists $\Delta'< \Delta$ such that
     $\tuple{\Delta,\bot_1}\in  \Gamma_1(\w{s})$ and
     $|\set{\reg(\w{s} +\Delta'') \mid \Delta'' \leq \Delta'}| = 3$.

     Consider the player-1 move $\tuple{\Delta',\bot_1}$.
     As $\tuple{\Delta,a_1}\in  \Gamma_1(\w{s})$, we must have
     $\tuple{\Delta',\bot_2}\in  \Gamma_2(\w{s})$.
     Also, since
     \[
     \set{\delta(\w{s},\tuple{\Delta'',a_2}) \mid \Delta'' \leq \Delta,\ 
       \succr_3(\w{s},\Delta'')=\true, 
       \text{ and }\tuple{\Delta'',a_2} \in
       \Gamma_2(\w{s})} \subseteq Z, \]
     we must have
     $\delta(\w{s},\tuple{\Delta',\bot_2}) \in Z$.
     Since $\Delta'$ is such that
     $\succr_2(\w{s},\Delta) = \false$, we have
     $\delta(\w{s}, \tuple{\Delta',\bot_2}) = 
     \delta(\w{s}, \tuple{\Delta',\bot_1})$ (the $\bl_1$ component is false in
     both cases).
     Hence $\delta(\w{s},\tuple{\Delta',\bot_1}) \in Z$.
     Also, since 
     \begin{enumerate}[$\star$]
       \item
$\set{\delta(\w{s}, \tuple{\Delta'',a_2}) \mid \Delta''\leq \Delta,\ \tuple{\Delta'',a_2} \in
       \Gamma_2(\w{s}) \text{ and } \succr_3(\w{s},\Delta'')=\true}
     \subseteq Z$
     and
   \item 
     $|\set{\reg(\w{s} + \Delta'')\mid \Delta''\leq \Delta'}| =3$,
\end{enumerate}
     we have
     \[
     \set{\delta(\w{s},\tuple{\Delta'',a_2}) \mid \Delta'' \leq \Delta',
       \text{ and }\tuple{\Delta'',a_2} \in
       \Gamma_2(\w{s})}
     \subseteq Z .
     \]
     Hence $\w{s} \in \CPre_1(Z)$ in this case by the second condition
     of $\CPre_1(Z)$ (the winning move of player~1 being 
     $\tuple{\Delta',\bot_1}$).
     
     \end{enumerate}
   \end{enumerate}
   Thus, in all cases $\w{s}\in\CPre_1(Z)$ whenever $\w{s}\in\CPre_{1,3}(Z)$.
\qed

\medskip\noindent\textbf{Proof of Lemma~\ref{lemma:TwoToThreeGame}.}

\proof
Recall that by Equation~\ref{equation:PreOne} mentioned in 
Theorem~\ref{theorem:Reduction},
given $X=X_1\times\set{1}\cup X_2\times\set{2}\subseteq S^f$, we have
\[
 \regstates\left(\Pre_1^{\A^f}\left(\Pre_1^{\A^f}\left(X\right)\right)\right) = 
 \regstates\left(\Pre_1^{\A^f}\left(\Pre_1^{\A^f}\left(X_1\times \set{1}\right)\right)\right) 
\]
Similarly for the structure ${\A^f}^*$, given
$X^*=X_{1_a}^*\times\set{1_a}\, \cup\, X_{1_c}^*\times\set{1_c}\, \cup\,
X_2^*\times\set{2}\subseteq {S^f}^*$, we have
\begin{equation}
  \label{equation:PreOneLinear}
  \regstates\left(\Pre_1^{{\A^f}^*}\!\left(\Pre_1^{{\A^f}^*}\!\left(\Pre_1^{{\A^f}^*}\!\left(
          X^*\right)\right)\right)\right)\! =
  \regstates\left(\Pre_1^{{\A^f}^*}\!\left(\Pre_1^{{\A^f}^*}\!\left(\Pre_1^{{\A^f}^*}\left(
          X_{1_a}^*\right)\right)\right)\right) 
 \end{equation}
Thus, it suffices to show that if $X_1\times\set{1}\subseteq S^f$ and 
$X_{1_a}^*\times\set{1_a}\subseteq {S^f}^*$ are such that
$\regstates(X_1\times\set{1}) = \regstates(X_{1_a}^*\times\set{1})$, then
\[\regstates\left(\Pre_1^{\A^f}\!\left(\Pre_1^{\A^f}\!\!\left(
X_1\times\set{1}\right)\right)\right)\! =
\regstates\!\left(\Pre_1^{{\A^f}^*}\!\left(\Pre_1^{{\A^f}^*}\!
\left(\Pre_1^{{\A^f}^*}\!\!\left(
X^*_{1_a}\times\set{1_a}\right)\right)\right)
\right).
\]

Note that $\regstates(X_1\times\set{1}) = \regstates(X_{1_a}^*\times\set{1})$
implies $X_1=X_{1_a}^* \subseteq \w{S}_{\reg}$, where
$\w{S}_{\reg}$ is the set of regions of $\wtdbl{\A}$.
\begin{enumerate}[($\Rightarrow$)]
\item  For
\[s^f= \tuple{\w{R},1}\in \Pre_1^{\A^f}\left(\Pre_1^{\A^f}\left(
      X_1\times\set{1}\right)\right)
\]
  we show
\[{s^f}^* =
  \tuple{\w{R},1_a}\in \Pre_1^{{\A^f}^*}\left(\Pre_1^{{\A^f}^*}
    \left(\Pre_1^{{\A^f}^*}\left(
        X_{1_a}^*\times\set{1_a}\right)\right)\right).
\]
  Let player~1 choose to go to $\tuple{\w{R},j,a_1,2}$ from $s^f$.
  We can have have two cases:
  \begin{enumerate}[$-$]
  \item
    $a_1=\bot_*$.\\
    If this is the move of player~1 for demonstrating that
    $s^f \in\Pre_1^{\A^f}\left(\Pre_1^{\A^f}\left(
        X_1\times\set{1}\right)\right)$, it means that
    every player-2 move from $\tuple{\w{R},j,a_1,2}$ goes to
    $X_1\times\set{1}$.
    For this to happen,
    we must have that in the game structure $\wtdbl{\A}$,  for every 
    state $\w{s}\in \w{R}$,
    for every player-2 move $\tuple{\Delta,a_2}$ such that
    $\succr^{\wtdbl{\A}}_3(\w{s},\Delta)=\true$ and 
    $\tuple{\Delta,a_2}\in \Gamma_2(\w{s})$, we have 
    $\delta(\w{s}, \tuple{\Delta,a_2}) \in \w{R}\,'$ with
    $\tuple{\w{R}\,',1}\in X_1\times\set{1}$.

    Consider ${s^f}^* = \tuple{\w{R},1_a}$ in ${S^f}^*$.
    Let player~1 choose to go to ${s^f}^*_2 =  \tuple{\w{R},2}$.
    Because of the restriction of player-2 moves from $\w{R}$ proved above (that
    every player-2 move within two successor regions must lead to a state such that
    the region corresponds to a region in $X_1$ and hence in $X_{1_a}^*$),
    we must have that in ${\A^f}^*$ every player-2 edge from ${s^f}^*_2$ leads to 
    a state $\tuple{\w{R}\,',1_c^{\dum}}$ such that
    $\w{R}\,' \in X_1=X_{1_a}^*$.
    Let $X_{1_c}^*=X_1$.
    We now have that every player~2 edge from ${s^f}^*_2$ leads to a state in
    $X_{1_c}^* \times\set{1_c^{\dum}}$.
    The next state in the game will obviously be in $X_{1} \times\set{1_a} =
    X_{1_a}^* \times\set{1_a}$.
    Thus, ${s^f}^* =
    \tuple{\w{R},1_a}\in \Pre_1^{{\A^f}^*}\left(\Pre_1^{{\A^f}^*}
      \left(\Pre_1^{{\A^f}^*}\left(
          X_{1_a}^*\times\set{1_a}\right)\right)\right)$ if the first move
    of player~1 corresponds to $\bot_*$.

  \item
    $a_1\neq\bot_*$.\\
    This case can have two subcases according to the available move for
    player~2.
    \begin{enumerate}[$-$]
    \item
      This subcase corresponds to the case where player~2 does not
      ``allow'' the player-1 move in a corresponding stage in $\wtdbl{\A}$.
      We must have \underline{\texttt{fact 1}}: that for every 
      $\w{s}\in \w{R}$, for every
      player-2 move $\tuple{\Delta,a_2}$ such that
      $\tuple{\Delta,a_2}\in \Gamma_2^{\wtdbl{\A}}(\w{s})$ and 
      $|\set{\reg{\w{s}+\Delta'}\mid \Delta'\leq \Delta}| \leq j+1$,
      we have that $\delta^{\wtdbl{\A}}(\w{s}, \tuple{\Delta,a_2}) 
      \in \w{R}\, '$ 
      such that $\reg( \w{R}\, ')\in X_1$.
    \item
      This subcase corresponds to the case where player~2 ``allows'' the player-1
      move.
      We must have \underline{\texttt{fact 2}}: that for every 
      $\w{s}\in \w{R}$, there exists a  player-1
      move $\tuple{\Delta,a_1}$ such that
      $\tuple{\Delta,a_1}\in \Gamma_1^{\wtdbl{\A}}(\w{s})$ and
      $|\set{\reg{\w{s}+\Delta'}\mid \Delta'\leq \Delta}| = j+1$,
      we have that $\delta^{\wtdbl{\A}}(\w{s}, \tuple{\Delta,a_1}) \in 
      \w{R}\, '$ 
      such that $\reg( \w{R}\, ')\in X_1$.
    \end{enumerate}
    
    Let $X^*_{1_c}= X_1 = X^*_{1_a}$.
    Consider ${s^f}^* = \tuple{\w{R},1_a}$ in ${S^f}^*$.
    Let player~1 choose to go to $\tuple{\w{R},j,2}$.
    we show player~1 can ensure going to $X_{1_a}^*\times\set{1_{1_a}}$ with
    this choice.
    Player~2 can do two things from $\tuple{\w{R},j,2}$.
    \begin{enumerate}[$\bullet$]
    \item
      Player~2 decides to ``not allow'' the player-1 move 
      in a corresponding stage in $\wtdbl{\A}$.
      This corresponds to player~2 taking an edge to 
      a state $\tuple{\w{R}\,', 1_c^{\dum}}$ in ${\A^f_{\Omega}}^*$ 
      from ${s^f}^*$
      which can happen if
      for every $\w{s}\in \w{R}$, there exists a player-2 move
      $\tuple{\Delta,a_2}$ such that
      $\tuple{\Delta,a_2}\in \Gamma_2^{\wtdbl{\A}}(\w{s})$ and 
      $|\set{\reg{\w{s}+\Delta'}\mid \Delta'\leq \Delta}| \leq j+1$.
      By \texttt{fact 1},  we must have that for every such  $\tuple{\Delta,a_2}$,
      we must have $\delta^{\wtdbl{\A}}(\w{s}, \tuple{\Delta,a_2}) 
      \in \w{R}\, '$ 
      such that $\reg( \w{R}\, ')\in X_1$.
      Hence, every ``not allow'' player-2 edge in ${\A^f_{\Omega}}^*$ 
      from ${s^f}^*$
      leads a state in $X^*_{1_c}\times \set{1_c^{\dum}}$, the next state from 
      which will lie in $X^*_{1_a} \times \set{1_a}$.
    \item
      Player~2 decides to ``allow'' the player-1 move 
      in a corresponding stage in $\wtdbl{\A}$.
      This corresponds to player~2 taking the edge from ${s^f}^*$ in 
      ${\A^f_{\Omega}}^*$ 
      to 
      the state $\tuple{\w{R},j,1_c}$.
      From $\tuple{\w{R},j,1_c}$, because of \texttt{fact 2}, player~1
      can pick the edge corresponding to the action $a_1$ which will
      lead it to a state in $X_{1} \times \set{1_a} = 
      X^*_{1_a} \times \set{1_a}$.
    \end{enumerate}
    Thus, if $a_1\neq \bot_*$, player1 has a strategy to go from ${s^f}^*$ to
    $X^*_{1_a} \times \set{1_a}$ in three steps.
    ${s^f}^* =
    \tuple{\w{R},1_a}\in \Pre_1^{{\A^f}^*}\left(\Pre_1^{{\A^f}^*}
      \left(\Pre_1^{{\A^f}^*}\left(
          X_{1_a}^*\times\set{1_a}\right)\right)\right)$.
\end{enumerate}
Hence in both cases, we have that 
\[{s^f}^* =
    \tuple{\w{R},1_a}\in \Pre_1^{{\A^f}^*}\left(\Pre_1^{{\A^f}^*}
\left(\Pre_1^{{\A^f}^*}\left(
          X_{1_a}^*\times\set{1_a}\right)\right)\right).
\]

\item[($\Leftarrow$)]
  For 
\[{s^f}^* =
  \tuple{\w{R},1_a}\in 
  \Pre_1^{{\A^f}^*}\left(\Pre_1^{{\A^f}^*}\left(\Pre_1^{{\A^f}^*}\left(
        X_{1_a}^*\times\set{1_a}\right)\right)\right)
\]
  we show
\[s^f= \tuple{\w{R},1}\in \Pre_1^{\A^f}\left(\Pre_1^{\A^f}\left(
      X_1\times\set{1}\right)\right).
\]
  The following cases arise depending on the move of player~1 from ${s^f}^*$
  which witnesses 
\[{s^f}^* \in  \Pre_1^{{\A^f}^*}
  \left(\Pre_1^{{\A^f}^*}\left(\Pre_1^{{\A^f}^*}\left(
        X_{1_a}^*\times\set{1_a}\right)\right)\right).
\]
  \begin{enumerate}[$\bullet$]
  \item
    Player~1 moves to $ \tuple{\w{R},2}$ (this corresponds to a move
    $a_1=\bot_*$ in $\A$).\\
    This means that every player~2 move from 
    $\tuple{\w{R},2}$ goes to $X_{1_a}^*\times\set{1_a}$.
    Since $X_{1_a}^* = X_1$, we have
    $\tuple{\w{R},1}\in \Pre_1^{\A^f}(\Pre_1^{\A^f}(
        X_1\times\set{1}))$ if the first move of player~1 from
    ${s^f}^*$ corresponds
    to 
    $\bot_*$.
  \item
    Player~1 moves to $\tuple{\w{R},j,2}$.
    This case has two subcases.
    \begin{enumerate}[$-$]
    \item
      Suppose we have \underline{\texttt{fact 3}}: 
      every player~2 move from $\tuple{\w{R},j,2}$ to 
      $\tuple{\w{R}\,', 1_c^{\dum}}$ is such that
      $\w{R}\,' \in X_{1_a}^*$.\\
      Consider the situation in $\A^f$ where
      player~1 moves to some $\tuple{\w{R},j,a_1,2}$ with $a_1\neq \bot_*$ and
      player~2 takes an edge corresponding to it not allowing player~1 in
      $\A$.
      Because of \texttt{fact 3}, all such edges must go to $X_1\times\set{1}$, no
      matter the action $a_1$.
    \item
      Suppose we have \underline{\texttt{fact 4}}: 
      every player~2 move from $\tuple{\w{R},j,2}$
        to $\tuple{\w{R},j,1_c}$ is such that
        $\tuple{\w{R},j,1_c} \in \Pre_1^{{\A^f}^*}(X_{1_a}^*\times\set{1_a})$.
        This means that  in $\A$, from any state in $\w{R}$, if player~1 is allowed to
        let time elapse to get to the $j$-th successor from $\w{R}$, then it can
        take a
        discrete action  $a_1$ such that the resultant state will be in some
        region in $X_1=X_{1_a}^*$.
        Thus, there must exist a player-1 action $a_1$ such that from
        the state $\tuple{\w{R},j,a_1,2}$ in $\A^f$, all player-2 edges, which
        correspond to it allowing a move in $\A$, end up in $X_1\times\set{1}$.

    \end{enumerate}
    Because of the two above subcases, if the witness for 
\[{s^f}^* \in  \Pre_1^{{\A^f}^*}\left(\Pre_1^{{\A^f}^*}\left(\Pre_1^{{\A^f}^*}\left(
        X_{1_a}^*\times\set{1_a}\right)\right)\right)
\]
  in the structure
  ${\A^f}^*$ is a move to  $\tuple{\w{R},j,2}$, then in the structure $\A^f$,
    there exists a move of player~1 to
    a state $ \tuple{\w{R},j,a_1,2}$ such that
    $ \tuple{\w{R},j,a_1,2} \in \Pre_1^{\A^f}(X_1\times\set{1})$.
  \end{enumerate}
  Thus in both cases of player-1 winning moves in ${\A^f}^*$ from ${s^f}^*$,
  we have the existence of a player-1 move from $s^f$
  which witnesses that 
  $s^f= \tuple{\w{R},1}\in \Pre_1^{\A^f}\left(\Pre_1^{\A^f}\left(
    X_1\times\set{1}\right)\right)$.
  
\end{enumerate}
\qed

\subsection{Proofs of Section~\ref{section:Robust}}
\noindent\textbf{Completion of analysis of Example~\ref{example:Jitter}}\\
\noindent\textit{Example~\ref{example:Jitter} (continued)}.
\hspace*{5mm}Let the game start from the location $l^0$.
In a run $r$, let $t_1^j$ and $t_2^j$ be the times when the 
$a_1^1$-th transition and the $a_1^2$-th transitions respectively are
taken for the $j$-th time.
The constraints are $t_1^j- t_1^{j-1} \leq 1$ and $t_2^j-t_2^{j-1} >1$.
If the game cycles infinitely often in between $l^0$ and $l^1$ we must also
have that for all $j\geq 0,\, t_1^{j+1} \geq t_2^j \geq t_1^j$.
Conversely, we  also have
that if this condition holds then we can construct an infinite time
divergent cycle of $l^0, l^1$ for some suitable  initial 
clock values.
Observe that $t_i^j = t_i^0 + (t_i^1-t_i^0) + (t_i^2-t_i^1)+\dots+ 
(t_i^j-t_i^{j-1})$ for $i\in \set{1,2}$.
We need 
\[t_1^{m+1} -  t_2^m = 
(t_1^{m+1} - t_1^m) + 
\sum_{j=1}^m\left\{(t_1^j - t_1^{j-1}) -  (t_2^j - t_2^{j-1})\right\} 
+ (t_1^0 - t_2^0) \geq 0 \text{ for all }m\geq 0.
\]
Rearranging, we get the requirement
\[\sum_{j=1}^m\left\{(t_2^j - t_2^{j-1}) -  (t_1^j - t_1^{j-1})\right\}
\leq
 (t_1^{m+1} - t_1^m) + (t_1^0 - t_2^0).\]
Consider the initial state $\tuple{l^0,x=y=0}$.
Let $t_1^0 =1, t_2^0=1.1, t_1^j - t_1^{j-1} = 1, t_2^j-t_2^{j-1}= 
1+10^{-(j+1)}$.
We have 
\[
\sum_{j=1}^m\left\{(t_2^j - t_2^{j-1}) -  (t_1^j - t_1^{j-1})\right\}
\leq \sum_{j=1}^\infty 10^{-(j+1)} = 10^{-2}*\frac{1}{0.9}
\,\leq\, 1-0.1=(t_1^{m+1} - t_1^m) + (t_1^0 - t_2^0).\]
Thus, we have an infinite time divergent trace with the given values.
Hence $\tuple{l^0, x=y=0} \in \wintimediv_1^{\A}(\Box(\neg l^3))$.
It can also be similarly seen that 
$\tuple{l^0, x=y=1} \in \wintimediv_1(\Box(\neg l^3))$ (taking
$t_1^0=0$ and $t_2^0=0.1$).
We also have  $\tuple{l^0, x=y=1} \notin \robwintimediv_1(\Box(\neg l^3))$
as player~1 would have to take the first $a_1^1$ transition 
immediately from this state.

We now show $\tuple{l^0, x=y=0} \in \robwintimediv_1(\Box(\neg l^3))$.
Consider
\begin{align*}
t_1^0 \ & \in [0.9,1], t_1^j-t_1^{j-1}\in [1-10^{-(j+1)},1] \\
t_2^0 \ & \in [1.05,1.1]\\
t_2^j-t_2^{j-1}\ & \in   [1+0.5*10^{-(j+1)}, 1+10^{-(j+1)}].
\end{align*}
We have 
\begin{align*}
\sum_{j=1}^m\left\{(t_2^j - t_2^{j-1}) -  (t_1^j - t_1^{j-1})\right\}
& \leq \sum_{j=1}^m 10^{-(j+1)}  - (- 10^{-(j+1)})\\
&  \leq 
2*\sum_{j=1}^\infty 10^{-(j+1)}\\
& = 2*10^{-2}*\frac{1}{0.9}.
\end{align*}
We also have $(t_1^{m+1} - t_1^m) + (t_1^0-t_2^0) \geq 1-10^{-(m+2)} +(0.9-1.1)
\geq 0.7$.
Thus, we have 
\begin{align*}
\sum_{j=1}^m\left\{(t_2^j - t_2^{j-1}) -  (t_1^j - t_1^{j-1})\right\} &
< 2*10^{-2}*\frac{1}{0.9} \\
& < 0.7 \\
& \leq (t_1^{m+1} - t_1^m) + (t_1^0-t_2^0).
\end{align*}
This shows that we can construct an infinite cycle in between $l^0$ and $l^1$
for all the values in our chosen intervals, and hence that 
 $\tuple{l^0, x=y=0} \in \robwintimediv_1(\Box(\neg l^3))$.

We next show that $\tuple{l^0, x=y=0} \notin 
\jrwintimediv_1^{\varjit,\varres}(\Box(\neg l^3))$ for any
$\varjit  >0$.
Observe that for any objective $\Phi$, we have
$\jrwintimediv_1^{\varjit,\varres}(\Phi)\subseteq 
\jrwintimediv_1^{\varjit,0}(\Phi)$.
Let $\varjit = \epsilon$ and let $\varres=0$.
Consider  any player-1 strategy $\pi_1$, for $\epsilon$-jitter 0-response time 
bounded-robust winning,
that makes the game cycle in between $l^0$ and $l^1$.
Player~2 then has a strategy which ``jitters'' the player-1 moves
by $\epsilon$.
Thus, the player-1 strategy $\pi_1$  can only propose $a_1^1$ moves 
with the value of $x$ being
less than or equal to $1-\epsilon$ (else the jitter would make the move 
invalid).
Thus, player~2 can ensure that $t_1^j-t_1^{j-1} \leq 1-\epsilon$ for all
$j$ for some run
(since $x$ has the value $t_1^j-t_1^{j-1}$ when $a_1^1$ is taken
for the $j$-th time for $j>0$).
We then have that for any player-1 strategy that is a candidate for 
$\epsilon$-jitter 0-response time bounded-robust winning,
 player~2 has a strategy such that for some resulting run, we have 
$t_1^j-t_1^{j-1} \leq 1-\epsilon$ and $t_2^j-t_2^{j-1} > 1$.
Thus, 
$\sum_{j=1}^m\left\{(t_2^j - t_2^{j-1}) -  (t_1^j - t_1^{j-1})\right\}
> m*\epsilon$, which can be made arbitrarily large for a sufficiently
large $m$ for any $\epsilon$ and hence greater than
$(t_1^{m+1} - t_1^m) + (t_1^0-t_2^0) \leq 1+(t_1^0-t_2^0) $ for any
initial values of $t_1^0$ and $t_2^0$.
This violates the requirement for an infinite $l^0, l^1$ cycle.
Thus, $\tuple{l^0, x=y=0} \notin 
\jrwintimediv_1^{\epsilon,0}(\Box(\neg l^3))$ for any $\epsilon >0$.
\qed


\end{document}